# Structural Change in the Multilayer World Trade Web: An Economic Dominance Theory Approach

**Working Paper**

October 2026

Didier Lebert
Unité d'Économie Appliquée, ENSTA – Institut Polytechnique de Paris
i3-CRG, École Polytechnique – CNRS (UMR 9217)
didier.lebert@ensta.fr

**Abstract**
Most studies of the World Trade Web (WTW) describe its topology: degrees, strengths, centralities, communities, or the similarity between product networks and years. This paper presents Economic Dominance Theory (EDT), initiated by Lantner (1974), and positions it as an alternative theory of structural dependence in directed flow networks rather than an additional toolbox. EDT normalizes bilateral flows into a Ghosh matrix of directed propagation and measures circular interdependence through its determinant, which aggregates the feedback circuits by which influence returns to its origin. Through the log-determinant and its trace expansion, a single quantity decomposes exactly by feedback length, arc and node deletion, product layer, network partition and, in the dynamic extension developed here, time. Structural change can therefore be measured without defining a distance between graphs, and layers and regions become exact components of one object rather than separate descriptions. EDT differs from centrality and community detection, which describe positions and densities rather than feedback closure; from economic complexity, which addresses capabilities rather than dependence; and from input-output and value-chain analysis, whose algebra it shares but which locate positions along chains where EDT measures their closure. An application to CEPII BACI data (1995-2023, 96 HS2 layers) shows what this adds. Circular interdependence does not grow monotonically with trade: the diffusion determinant rises from 0.12 in 1995 to 0.21 in 2006, falls to 0.16 in 2013 and reaches 0.19 in 2023. Its long-run change is cross-product, with more than 90% of order-two and about 99% of order-three feedback genuinely multilayer, and its geography depends on feedback length: short reciprocal loops remain regional, while long feedback is overwhelmingly transregional.

**How to read this paper**

This paper is self-contained but long, and readers meeting the Economic Dominance Theory (EDT) for the first time are advised to proceed as follows.

Begin with **Annex A.1 to A.5**, a five-country worked example in which the core identities used in the empirical analysis can be verified by hand. Then read **Sections 2.1 to 2.5**, which construct the normalized matrix, the Ghosh inverse, the determinant, and the two exact representations of the determinant. These five subsections contain everything required to follow the empirical results. Sections 2.6 to 2.14 develop the suppression, partition, multilayer and temporal apparatus, and may be consulted as each is first used.

**Section 3** is a replication protocol for CEPII BACI. It is written for readers who intend to rebuild the annual matrices and may be skipped entirely on a first reading.

For the empirical argument, read **Section 4.2** on the aggregate trajectory, **Section 4.5** on the 2017 diagnostic, which illustrates the full chain of EDT operations from the global determinant down to a single bilateral flow, **Sections 5.1 to 5.3** on the multilayer result, and **Section 6.5** on the interaction between feedback length and geography. The remaining subsections refine these results.

**Section 7** situates EDT relative to network analysis, economic complexity, input-output analysis, global value chains, multilayer network theory and community detection. It is written as a methodological reference rather than as a linear argument. Section 7.9, on scope conditions and limitations, should be read by anyone intending to apply the framework.

A note on terminology is necessary at the outset. The quantity $D$ is Lantner's *diffusion* index: it *rises* when circular interdependence *weakens*. This sign convention is inherited from Lantner (1974) and is a recurrent source of misreading; it is restated at each empirical result. An open chain can transmit influence even when the determinant equals one. The index measures circular closure, not the total amount or geographical reach of propagation.

## Contents

## 1. Introduction

International trade is commonly described as increasingly interconnected. The long expansion of cross-border exchanges, the fragmentation of production, the growth of global value chains, and the multiplication of traded product categories all seem to point in the same direction: a world economy in which countries depend increasingly on one another. Yet this broad proposition hides a more difficult question. What exactly is meant by *structural interdependence*, and how should its change through time be measured?

The World Trade Web (WTW) literature has shown why this question cannot be reduced to the growth of trade volumes. Once bilateral trade is represented as a directed and weighted network, the organization of links matters in its own right. Connectivity, clustering, assortativity, reciprocity, centrality, core-periphery organization, and community structure reveal properties that are invisible in aggregate trade statistics (Fagiolo et al., 2008, 2009, 2010; De Benedictis et al., 2013). Commodity-specific and multilayer approaches further show that the aggregate WTW is composed of product networks with distinct topologies and patterns of interaction (Barigozzi et al., 2010). Other research traditions extract different forms of structure from related data. Economic-complexity approaches use country-product relations to infer productive capabilities and diversification opportunities (Hidalgo et al., 2007; Hidalgo & Hausmann, 2009; Hausmann et al., 2014), while input-output and global-value-chain approaches reconstruct production requirements, value-added flows, and positions along international production chains (Johnson & Noguera, 2012; Timmer et al., 2014).

These literatures have greatly expanded what can be learned from trade data, but they also make clear that *structure* is not a single object. A graph can be structurally dense without generating strong recurrent dependence. Two countries can be central under a geodesic or eigenvector criterion without playing the same role in circular feedback. Two product layers can be individually important without interacting in the same way when they are superimposed. A network can become more geographically dispersed while some of its feedback mechanisms become more regionally concentrated. More generally, an observed change in a trade network can be summarized by a distance between two graphs, but such a distance does not by itself specify which propagation mechanisms changed or why the structural effect has the observed sign.

This article approaches the problem from the Economic Dominance Theory (EDT), initiated by Lantner (1974) and recently extended to multilayer trade structures by Lebert and Poindron (2025). EDT distinguishes directed dependence from circular interdependence. Normalized bilateral coefficients allow influence to propagate along open chains. When a chain returns to its starting country, it also generates feedback. The inverse matrix represents both forms of propagation; the determinant studied here summarizes their circular component.

Let $X_t = (x_{ij,t})$ denote the annual matrix of bilateral merchandise exports, with domestic flows excluded. For country $i$, total exports and imports are $E_{i,t}$ and $M_{i,t}$, and the trade-balance normalization uses

$$q_{i,t} = \max(E_{i,t}, M_{i,t}).$$

The normalized coefficient matrix is

$$\Phi_t = Q_t^{-1} X_t,$$

where $Q_t = \mathrm{diag}(q_{1,t}, \dots, q_{n,t})$. Because each row sum is at most one, the normalization gives the matrix a propagation interpretation and represents a trade deficit as structural leakage. Under the spectral condition established in Section 2, the associated Ghosh inverse is

$$(I - \Phi_t)^{-1} = I + \Phi_t + \Phi_t^2 + \cdots.$$

The powers of $\Phi_t$ therefore accumulate directed walks of increasing length. EDT summarizes the global feedback architecture through the determinant

$$D_t = \det(I - \Phi_t).$$

With zero domestic flows, the diagonal product of the Ghosh matrix is equal to one, so $1 - D_t$ is the corresponding measure of circular structural interdependence. A lower determinant means that a larger part of the normalized system is locked into recurrent feedback; a higher determinant means greater diffusion and weaker circular interdependence. This terminology is inherited from EDT and should not be confused with the ordinary-language use of "diffusion" as the geographical spread of trade.

The determinant is useful here because it is not treated as a black-box scalar. It has two exact representations that expose its structural content. Its Leibniz expansion expresses it in terms of simple directed circuits and families of vertex-disjoint circuits. Its logarithm has the trace representation

$$\log D_t = -\sum_{k=2}^{\infty} \frac{\mathrm{Tr}\left(\Phi_t^k\right)}{k},$$

where $\mathrm{Tr}\left(\Phi_t^k\right)$ is the total weight of closed walks of order $k$. Changes in the measured circular interdependence can consequently be decomposed as follows:

$$\log D_{t_1} - \log D_{t_0} = -\sum_{k=2}^{\infty} \frac{\mathrm{Tr}\left(\Phi_{t_1}^k\right) - \mathrm{Tr}\left(\Phi_{t_0}^k\right)}{k}.$$

This identity is central to the article. It decomposes a change in circular interdependence without first selecting a distance between graphs. The same determinant-based criterion can be examined by feedback length, product layers, geographical partitions, nodes, or arcs. It does not characterize every network transformation: distinct trade matrices can have the same determinant, and changes in different feedback components can offset one another.

The empirical application uses CEPII's BACI database in the HS92 classification over 1995–2023. The raw bilateral-product records are first converted into annual country-by-country flow matrices. A fixed universe of 238 country positions is used throughout the period; countries that are inactive in a given year contribute zero rows and columns and therefore do not mechanically change the determinant. Product-specific matrices are constructed at HS6 and then aggregated exactly to 96 observed HS2 chapters. A crucial feature of the multilayer construction is that the normalization vector is calculated once from aggregate trade in each year and applied identically to every product layer. Hence, if $g$ indexes HS2 categories,

$$\Phi_t = \sum_g \Phi_t^{(g)}.$$

The layers are therefore additive components of the same normalized economy rather than separately rescaled networks. This makes it possible to compare the aggregate determinant with the determinant of the block-diagonal structure in which product layers remain separated.

The article makes a first contribution by developing a fully dynamic reading of EDT. The static determinant and its circuit interpretation are well established within the theory, and Lebert and Poindron (2025) provide the multilayer extension. Here, time is introduced as a sequence of comparable EDT structures. The exact trace identity decomposes changes in the log-determinant by feedback order. The multilayer representation then yields an exact temporal separation between changes occurring inside individual product networks and changes created specifically by the combination of product categories. If $D_t^{\circ} = \prod_g D_{g,t}$ denotes the determinant of the separated-layer structure, then

$$\log \frac{D_{t_1}}{D_{t_0}} = \log \frac{D_{t_1}^{\circ}}{D_{t_0}^{\circ}} + \left[ \log \frac{D_{t_1}}{D_{t_1}^{\circ}} - \log \frac{D_{t_0}}{D_{t_0}^{\circ}} \right].$$

The first term is the within-layer component of structural change; the second is genuinely multilayer. No additional index is introduced. The decomposition follows directly from existing EDT objects.

A second formal contribution applies the theorem of the partition dynamically. Countries are assigned to a fixed geographical partition – Africa, the Americas, Asia, Europe, and Oceania – defined independently of the trade network. Suppressing all inter-regional arcs produces a block-diagonal window-induced structure whose determinant is the product of the regional determinants. Comparing its log-determinant with the global one separates feedback that closes within broad regions from feedback that requires at least one regional boundary to be crossed. The same temporal change can therefore be opened simultaneously along a product dimension and a geographical dimension while remaining anchored in the same structural function.

The empirical results provide a markedly different picture from a simple narrative of continuously rising interdependence. At the aggregate level, the WTW does not move monotonically toward stronger circular dependence. The determinant rises from 0.1205 in 1995 to a sample maximum of 0.2137 in 2006, indicating a substantial weakening of circular feedback. It then falls to 0.1619 in 2013, rises again through 2016, and remains comparatively stable after 2018 apart from pronounced movements in 2022 and 2023. Over the full period, $D_t$ increases to 0.1895. The long-run direction is therefore toward weaker circular interdependence in the EDT sense, but that endpoint is produced by a sequence of reversals rather than by a trend. The chronology is endogenous to the feedback structure and does not simply reproduce the chronology of trade-value expansion and contraction.

The trace decomposition adds an important qualification. Bilateral reciprocal feedback is significant, but it does not explain the full transformation. For the 1995–2006 phase, order-two feedback accounts for only about one third of the exact log-determinant change; longer recurrent chains are required to reconstruct the rest. In later phases, structural change becomes more concentrated in lower feedback orders. The result does not imply that the WTW has become a collection of short elementary circuits, because traces count closed walks, including repeated traversals and combinations of shorter cycles. It shows more precisely that the *difference between successive structures* has become increasingly concentrated in shorter recurrent mechanisms.

The product decomposition reveals the strongest result of the paper. The aggregate WTW is structurally very different from the juxtaposition of its product networks. The separated-layer determinant remains close to 0.92 outside the isolated 2017 episode, while the aggregate determinant is generally between 0.12 and 0.21. More than 90% of order-two closed-walk mass is already generated by cross-product terms, about 99% at order three, and virtually all feedback at higher orders is multilayer. Correspondingly, the major temporal phases are overwhelmingly changes in cross-product feedback. Between 1995 and 2006, the aggregate log-determinant rises by 0.5724, of which only 0.0030 comes from within-layer change and 0.5695 from the multilayer component. Between 2006 and 2013, the reversal is again primarily multilayer. Over 1995–2023 as a whole, the within-layer component moves slightly in the opposite direction to the aggregate change: the entire net increase in diffusion is generated by a weakening of cross-product feedback strong enough to offset a small strengthening of circularity inside individual product networks.

The identity of the layers that structure this feedback system also changes. Machinery, electrical machinery, vehicles, and mineral fuels remain important throughout, but mineral fuels acquire the largest marginal structuring contribution by the end of the period. Pharmaceuticals, precious stones and metals, and ores also gain, whereas several paper, textile, wood, and material categories lose structural weight. These changes are not statements about export shares. They concern the extent to which removing a layer breaks the recurrent architecture of the aggregate WTW. Pairwise interactions also change: the positive interaction between the suppression effects of energy and major manufacturing layers weakens sharply, while separate removals within the manufacturing and chemical complexes increasingly overlap in their effects on diffusion. The transformation therefore concerns both the marginal role of individual layers and the architecture of their interactions.

The geographical decomposition adds a different perspective. Outside the isolated 2017 episode, roughly 35–42% of the absolute aggregate log-determinant is generated by feedback that crosses at least one continental boundary. The long-run change is much less dominated by geography than by products: approximately 59% of the 1995–2023 increase in the log-determinant comes from weaker feedback inside the broad regions and 41% from weaker cross-regional feedback. The composition nevertheless changes across phases. Cross-regional feedback accounts for about 60% of the major movements before 2013, while the 2013–2016 recovery is predominantly internal to regions. From 2018 to 2023, the two geographical components move in opposite directions: within-region circularity weakens while cross-region circularity strengthens, partially offsetting one another.

Feedback length changes the meaning of geography. At order two, only about 15–20% of closed-walk mass crosses continental boundaries. At order three the cross-regional share approaches one half, and at longer orders it rapidly becomes dominant. In 2023, more than 90% of order-eight feedback and about 98% of order-twelve feedback are transregional. This result contrasts sharply with the product dimension, where cross-layer feedback is already dominant at order two. Put differently, the WTW is *multilayer before it is transregional*: product heterogeneity is embedded in elementary reciprocal feedback, whereas geographical globalization becomes increasingly visible as recurrent chains lengthen.

One exceptional observation illustrates the diagnostic value of this framework. The aggregate determinant falls sharply in 2017 and returns almost completely in 2018. Rather than deleting the observation or treating it as an unexplained outlier, the EDT operations allow the anomaly to be opened progressively. Layer suppression locates almost the entire movement in HS89, ships and boats; finer disaggregation identifies HS890392, motorboats; the order-two trace locates the mechanism in reciprocal feedback; and bilateral inspection identifies flows linking Seychelles with Cayman Islands and the British Virgin Islands. The same disturbance is therefore monolayer in product space but cross-regional in geographical space. A configuration representing less than one percent of gross world trade can have a large structural effect if it closes exceptionally heavy normalized feedback loops. The episode illustrates why structural importance and trade value should not be conflated.

These findings motivate a broader methodological argument. EDT should not be presented as a replacement for established approaches to the WTW. Network analysis, economic complexity, input-output analysis, global-value-chain accounting, multilayer network theory, and community detection each preserve different information and answer different questions. The discussion in Section 7 therefore compares the approaches at the level of formal objects, propagation mechanisms, analytical purposes, and substantive interpretation rather than ranking their indicators. EDT is distinctive because global structure, local contribution, layer interaction, partition, and temporal change remain linked to the same determinant-based theory of feedback dependence.

The article is organized as follows. Section 2 develops the complete EDT framework used in the paper, beginning with the trade-balance normalization and deriving the Ghosh inverse, the determinant, its circuit and trace representations, suppression results, the theorem of the partition, and the multilayer extension. It then introduces the exact temporal decompositions used in the empirical analysis. Section 3 describes the BACI data and gives a fully reproducible procedure for constructing the aggregate and product-specific matrices from the raw files. Section 4 examines the aggregate structural trajectory of the WTW from 1995 to 2023 and decomposes its changes by feedback order. Section 5 opens the system by product layers, distinguishing within-layer from genuinely multilayer structural change and analyzing the marginal contribution and pairwise interaction of HS2 categories. Section 6 applies the theorem of the partition to the geographical organization of the WTW. Section 7 places the results in relation to the established WTW, economic-complexity, input-output, GVC, multilayer, and community-detection literatures, and states the scope conditions and limitations of the approach. Section 8 concludes.

## 2. A structural analysis of the World Trade Web with the Economic Dominance Theory

The Economic Dominance Theory (EDT), initiated by Lantner (1974), is a structural theory of directed flows. Its central question is not how large a flow is in isolation, nor how many partners a node has, but how a system of directed influences is organized once direct and indirect feedbacks are taken into account. Applied to international trade, the theory treats countries as poles of a directed flow system and bilateral exports as the elementary links through which influence can circulate. A simple distinction guides the analysis: exports from A to B and from B to C transmit influence along an open chain. Adding a link from C back to A creates circular feedback. Both structures transmit influence, but only the second contains a circuit contributing to the determinant gap.

This section develops the complete formal framework used in the article. The exposition is deliberately progressive. We begin with raw bilateral flows, introduce the normalization that transforms them into structural coefficients, construct the Ghosh matrix and its inverse, and then show how the determinant of the Ghosh matrix summarizes circular interdependence. Two complementary representations of the determinant are then derived: a combinatorial representation in terms of vertex-disjoint circuits, and a trace representation in terms of closed walks of increasing length. These representations make it possible to establish the monotonic effects of suppressing arcs and nodes, to derive exact marginal sensitivities, and to state the theorem of the partition. We then move to the multilayer formulation of Lebert and Poindron (2025), in which product-specific trade matrices are additive layers of the same aggregate structure. Finally, we introduce time and decompose changes in determinant-based circular interdependence exactly, while distinguishing that measured change from other transformations of the network.

### 2.1 From bilateral trade flows to a normalized structure of dependence

Consider a set of countries $N = \{1, \dots, n\}$. Let

$$X = \left(x_{ij}\right)_{i,j \in N}, \qquad x_{ij} \geq 0,$$

where $x_{ij}$ is the value of exports from country $i$ to country $j$. Domestic trade is excluded throughout the article,

$$x_{ii} = 0 \qquad \forall i \in N.$$

The directed weighted graph associated with $X$ has countries as nodes and a directed arc $i \to j$ whenever $x_{ij} > 0$. Raw trade values, however, cannot be used directly as structural coefficients. Their scale changes over time with prices, inflation, exchange rates, and the general expansion or contraction of world trade. More importantly, the same export flow does not have the same structural meaning for a country that is externally balanced and for a country whose exports are sustained by much larger imports. EDT therefore starts by normalizing the flow matrix.

For every country $i$, define total exports and imports as

$$E_i = \sum_{j=1}^{n} x_{ij}, \qquad M_i = \sum_{j=1}^{n} x_{ji}.$$

The trade-balance normalization used by Lebert and Poindron (2025) defines

$$q_i = \max(E_i, M_i)$$

and the normalized flow coefficient

$$\phi_{ij} = \frac{x_{ij}}{q_i}.$$

Let

$$Q = \mathrm{diag}(q_1, \dots, q_n).$$

The normalized flow matrix can then be written compactly as

$$\Phi = Q^{-1}X.$$

Before using these coefficients in EDT, it is useful to separate three issues that are easily conflated: the choice of a country-level trade margin, the level at which that margin is computed, and the direction in which influence is propagated. At the industry level, Balassa (1966) normalizes the trade imbalance by total trade, while Grubel and Lloyd (1975) use its complement; Aquino (1978), Bergstrand (1983), and Glejser (1983) subsequently modify the treatment of the aggregate imbalance. Here, however, the export and import totals defined above are country totals over all products. The quantities introduced below therefore measure the normalized external balance of a country; they are not, in this aggregate application, measures of intra-industry trade. Define total trade, the normalized trade balance and its absolute value as

$$V_i = E_i + M_i, s_i = \frac{E_i - M_i}{V_i}, b_i = |s_i|.$$

*Proposition — Minimum orientation-independent common trade margin*. For an active country, impose three requirements: the denominator is unchanged when exports and imports exchange roles; both normalized totals are bounded by one; and the smallest denominator satisfying these conditions is selected. Under these postulates, the denominator is the maximum of exports and imports.

$$\frac{E_i}{q_i} \leq 1, \qquad \frac{M_i}{q_i} \leq 1 \Leftrightarrow q_i \geq \max(E_i, M_i).$$

*Proof*. The two boundedness conditions require the denominator to be at least as large as exports and at least as large as imports, hence at least as large as their maximum. Any larger symmetric margin would also satisfy boundedness; minimality selects the maximum. Conversely, the maximum of exports and imports satisfies both inequalities and therefore attains the lower bound.

$$q_i = \max(E_i, M_i) = \frac{E_i + M_i + |E_i - M_i|}{2} = \frac{V_i}{2}(1 + b_i).$$

The last identity locates the selected margin within the classical family of trade-normalization statistics, but it does not by itself characterize the choice. The pair of export and import totals contains exactly the same information as total trade together with the signed trade difference, or equivalently total trade together with the signed normalized balance. The absolute normalized balance retains only the magnitude of that imbalance, not its direction. The identity therefore shows that the common margin depends on total trade and on the magnitude of the external imbalance, while remaining invariant to whether that imbalance is a surplus or a deficit. The proposition therefore characterizes the maximum conditional on these measurement postulates. The accounting data alone do not require this choice of margin.

This also clarifies the relation to the intra-industry-trade literature. That literature often treats the aggregate imbalance as something to be adjusted or purged in order to isolate within-industry overlap. The present construction deliberately uses the imbalance in the opposite way: it is retained as information in the country margin. Once a direction of propagation is chosen, the imbalance can therefore affect the amount of influence transmitted through the network rather than being removed before the network is analyzed.

$$\phi_{ij} = \frac{x_{ij}}{q_i} = \frac{2}{1 + b_i} \frac{x_{ij}}{V_i}.$$

Thus the bilateral EDT coefficient can be written as a conventional flow share relative to total external trade, multiplied by a country-balance factor. This places the coefficient in a familiar measurement family without turning it into an intra-industry-trade index. The economic definition of the common margin comes first; sub-stochasticity is a property that follows once the export-oriented propagation matrix is constructed.

This normalization has two immediate properties. First,

$$\sum_j \phi_{ij} = \frac{E_i}{\max(E_i, M_i)} \leq 1.$$

If $E_i \geq M_i$, country $i$ is balanced or in surplus and its row sum is equal to one. If $E_i < M_i$, the country is in deficit and

$$\sum_j \phi_{ij} = \frac{E_i}{M_i} < 1.$$

Under the chosen export-oriented rule, a deficit produces a normalized row sum below one, interpreted as structural leakage. This is a property of the propagation model. It does not mean that some recorded trade value disappears, or that the deficit measures a causal loss of productive capacity.

Crucially, the common margin itself is symmetric in exports and imports, whereas leakage in the present matrix is not. The asymmetry is introduced by the Ghosh orientation: rows are exporters. The ratio of the smaller to the larger side of the trade position is

$$\frac{\min(E_i, M_i)}{\max(E_i, M_i)} = \frac{1 - b_i}{1 + b_i}.$$

This ratio becomes the row sum only for deficit countries. Surplus countries have a row sum equal to one. In the dual import-oriented representation the asymmetry would be reversed. The Balassa–Grubel–Lloyd genealogy therefore helps situate the symmetric margin; the one-sided leakage is produced by the direction of propagation chosen for the present analysis.

Second, the normalization removes a pure scale effect. If every bilateral flow is multiplied by the same positive scalar $a$, then exports, imports and denominators are multiplied by the same scalar, so that

$$\frac{a x_{ij}}{\max(aE_i, aM_i)} = \frac{x_{ij}}{\max(E_i, M_i)}.$$

Consequently, $\Phi$ describes the organization of trade rather than its common nominal scale.

Because $\Phi$ is non-negative and every row sum is at most one,

$$\rho(\Phi) \leq \| \Phi \|_\infty = \max_i \sum_j \phi_{ij} \leq 1,$$

where $\rho(\Phi)$ denotes the spectral radius. A strict inequality is required for the inverse and logarithmic expansions used below. If $\Phi$ is irreducible, the existence of at least one strictly sub-stochastic row is sufficient to imply

$$\rho(\Phi) < 1.$$

More generally, if the graph is reducible, the same conclusion holds provided that every final strongly connected class contains at least one node whose normalized row sum is strictly below one. Economically, no closed autonomous subsystem can then recycle its full outgoing mass indefinitely. In the empirical application, the condition $\rho(\Phi_t) < 1$ is checked directly for every year.

The normalized matrix should therefore be understood as a matrix of structural coefficients. The coefficient $\phi_{ij}$ measures the direct exposure of $j$ to the outgoing influence of $i$, relative to the larger side of country $i$'s external trade position. The object is directional: in general, $\phi_{ij} \neq \phi_{ji}$. This is the exposure assigned by the model, not an estimated causal response.

### 2.2 The Ghosh matrix, indirect propagation and Lantner centrality

We use the term Ghosh matrix in its EDT sense (Ghosh, 1958), for the identity matrix minus the normalized coefficient matrix:

$$G = I - \Phi.$$

Since $\phi_{ii} = 0$, its diagonal is equal to one,

$$g_{ii} = 1,$$

while every off-diagonal coefficient is non-positive,

$$g_{ij} = -\phi_{ij} \leq 0 \qquad (i \neq j).$$

Under $\rho(\Phi) < 1$, the Neumann series converges and gives

$$G^{-1} = (I - \Phi)^{-1} = I + \Phi + \Phi^2 + \Phi^3 + \cdots.$$

This identity gives a direct network interpretation to the Ghosh inverse. Matrix multiplication implies that

$$(\Phi^2)_{ij} = \sum_k \phi_{ik}\,\phi_{kj},$$

which is the total weight of all directed walks of length two from $i$ to $j$. More generally,

$$(\Phi^k)_{ij} = \sum_{i_1,\ldots,i_{k-1}} \phi_{ii_1}\,\phi_{i_1 i_2} \cdots \phi_{i_{k-1}j},$$

so $(\Phi^k)_{ij}$ is the total weight of all directed walks of length $k$ from $i$ to $j$, where the weight of a walk is the product of the coefficients carried by its successive arcs.

Hence

$$(G^{-1})_{ij} = \delta_{ij} + \sum_{k=1}^{\infty} (\Phi^k)_{ij}$$

contains direct and indirect propagation through walks of every possible length. For $i \neq j$, it measures total influence from $i$ to $j$ through the structure. For $i = j$, it includes the unit identity term plus the influence that leaves $i$, circulates through the network, and returns to $i$.

This diagonal quantity is the Lantner centrality of node $i$,

$$L_i = (G^{-1})_{ii}.$$

Since the identity term contributes one,

$$L_i = 1 + \sum_{k=1}^{\infty} (\Phi^k)_{ii}.$$

Because $\phi_{ii} = 0$, the $k = 1$ term vanishes. Thus $L_i - 1$ is entirely generated by feedback walks passing through $i$.

The same centrality admits a determinant representation. Let $\Phi_{-i}$ denote the matrix obtained by deleting row and column $i$, and define

$$D = \det(I - \Phi), \qquad D_{-i} = \det(I - \Phi_{-i}).$$

By the cofactor formula for the inverse,

$$(G^{-1})_{ii} = \frac{D_{-i}}{D}.$$

Therefore,

$$L_i = \frac{D_{-i}}{D}.$$

This identity already reveals one of the characteristic features of EDT: a local quantity, the centrality of one country, is defined by comparing the determinant of the entire structure with the determinant of the structure in which that country is absent. Local position and global structure are therefore not separate objects.

### 2.3 The determinant as a global measure of diffusion and circular interdependence

Lantner's global structural quantity is the determinant of the Ghosh matrix,

$$D = \det(G) = \det(I - \Phi).$$

Lantner (1974) calls $D$ the diffusion index. Under the assumptions above,

$$0 < D \leq 1.$$

The terminology can be read directly from the role of feedback. A structure containing strong circular feedbacks tends to retain and amplify an influence within recurrent chains. Such feedback lowers $D$. Conversely, weakening or breaking feedback circuits increases $D$ and therefore increases diffusion in Lantner's terminology.

In the more general formulation of EDT, the product of the diagonal elements of the Ghosh matrix is denoted

$$U = \prod_i g_{ii}.$$

The part of the determinant gap associated with structural interdependence is

$$R = U - D.$$

In the present application, domestic self-flows are excluded and $g_{ii} = 1$ for every country. Hence

$$U = 1$$

and the measure of circular interdependence becomes simply

$$\boxed{R = 1 - D.}$$

The two quantities therefore carry the same information with opposite orientation. An increase in $D$ means greater diffusion and weaker circular interdependence; an increase in $1 - D$ means stronger circular interdependence.

This relation is especially useful for interpretation because the diagonal baseline is fixed. There is no ambiguity as to the source of the gap between one and the determinant: it is entirely generated by the off-diagonal architecture of directed trade.

<u>A three-country illustration:</u>

Consider the simplest directed cycle involving three countries,

$$\Phi = \begin{pmatrix} 0 & a & 0 \\ 0 & 0 & b \\ c & 0 & 0 \end{pmatrix}, \qquad 0 \leq a, b, c < 1.$$

The only elementary circuit is

$$1 \to 2 \to 3 \to 1,$$

with weight $abc$. The Ghosh matrix is

$$G = \begin{pmatrix} 1 & -a & 0 \\ 0 & 1 & -b \\ -c & 0 & 1 \end{pmatrix},$$

and its determinant is

$$D = 1 - abc.$$

Hence

$$1 - D = abc.$$

The interpretation is transparent. If any one of the three directed links disappears, the cycle is broken and $D$ returns to one. If the three links strengthen, the product $abc$ rises, $D$ falls, and circular interdependence increases. The relevant object is not the sum $a + b + c$ but the multiplicative closure of the chain. After a link is removed, the remaining open chain can still transmit direct and indirect influence. A determinant equal to one therefore signals absence of circular feedback, not absence of dependence.

**2.4 Why the determinant is a circuit quantity**

The circuit interpretation follows directly from the Leibniz expansion of the determinant. For an $n \times n$ matrix $G$,

$$\det(G) = \sum_{\sigma \in S_n} \operatorname{sgn}(\sigma) \prod_{i=1}^{n} g_{i\sigma(i)},$$

where $S_n$ is the set of permutations of $\{1, \dots, n\}$.

Every permutation decomposes uniquely into disjoint cycles. A fixed point $\sigma(i) = i$ contributes the diagonal factor $g_{ii} = 1$. A non-trivial cycle

$$(i_1 \; i_2 \; \dots \; i_r)$$

contributes the off-diagonal product

$$g_{i_1 i_2} g_{i_2 i_3} \cdots g_{i_r i_1} = (-1)^r \phi_{i_1 i_2} \phi_{i_2 i_3} \cdots \phi_{i_r i_1}.$$

The sign of a permutation cycle of length $r$ is $(-1)^{r-1}$. The combined sign is therefore

$$(-1)^{r-1}(-1)^r = -1.$$

Thus every non-trivial directed circuit contributes negatively, irrespective of its length.

Define the weight of a simple directed circuit $c = (i_1, \dots, i_r, i_1)$ as

$$\omega(c) = \prod_{s=1}^{r} \phi_{i_s i_{s+1}}, \qquad i_{r+1} = i_1.$$

Let $F$ be a family of pairwise vertex-disjoint directed circuits, and define

$$\Omega(F) = \prod_{c \in F} \omega(c).$$

Because a permutation cannot assign one node to two different cycles, only vertex-disjoint circuits can appear together in the same determinant term. If $|F|$ is the number of circuits in the family, the determinant becomes

$$\boxed{D = 1 + \sum_{F \neq \varnothing} (-1)^{|F|} \, \Omega(F).}$$

A family containing one circuit enters negatively, a family containing two disjoint circuits positively, a family containing three negatively, and so on. This alternating structure explains why the determinant cannot be interpreted as the simple sum of circuit weights. Nevertheless, every departure from the diagonal baseline is generated by circuits.

The three-country example above contains only one circuit and no pair of disjoint circuits, so the formula reduces immediately to $D = 1 - abc$.

The combinatorial representation is fundamental because it tells us exactly what kind of structure the determinant sees. A directed chain that never closes does not enter the determinant as a non-trivial term. A feedback loop does. EDT therefore assigns a special role to closure: dependence becomes interdependence when influence can return to its source.

### 2.5 The trace representation: closed walks, length and the log-determinant

The circuit expansion is exact but difficult to enumerate in a large network. An equivalent representation based on matrix traces is more convenient for empirical analysis.

Since $\rho(\Phi) < 1$, the matrix logarithm admits the convergent power series

$$\log(I - \Phi) = -\sum_{k=1}^{\infty} \frac{\Phi^k}{k}.$$

Under the maintained spectral condition, the trace of this matrix logarithm equals the real logarithm of the positive determinant:

$$\log\det(I - \Phi) = \mathrm{Tr}[\log(I - \Phi)].$$

Taking the trace term by term therefore gives

$$\log D = -\sum_{k=1}^{\infty} \frac{\mathrm{Tr}(\Phi^k)}{k}.$$

Because $\phi_{ii} = 0$,

$$\mathrm{Tr}(\Phi) = 0,$$

so the series starts at order two:

$$\boxed{\log D = -\sum_{k=2}^{\infty} \frac{\mathrm{Tr}(\Phi^k)}{k}.}$$

Equivalently,

$$\boxed{D = \exp\left[-\sum_{k=2}^{\infty} \frac{\mathrm{Tr}(\Phi^k)}{k}\right].}$$

This expression, used in Lebert and Poindron (2025) from the matrix-logarithm result presented by Hall (2015), is especially valuable because all trace terms are non-negative when $\Phi$ is non-negative. The alternating signs of the direct determinant expansion disappear after taking the logarithm.

To interpret the trace, expand a diagonal element of $\Phi^k$:

$$(\Phi^k)_{ii} = \sum_{i_2,\dots,i_k} \phi_{ii_2}\,\phi_{i_2i_3} \cdots \phi_{i_ki}.$$

Summing over $i$ yields

$$\mathrm{Tr}(\Phi^k) = \sum_{i_1,\dots,i_k} \phi_{i_1i_2}\,\phi_{i_2i_3} \cdots \phi_{i_ki_1}.$$

Thus $\mathrm{Tr}(\Phi^k)$ is the total weight of all closed directed walks of length $k$.

A distinction of terminology is important. The Leibniz determinant expansion involves simple cycles organized into vertex-disjoint families. By contrast, a closed walk counted in $\mathrm{Tr}(\Phi^k)$ may revisit nodes or repeat a shorter circuit several times. The two representations are algebraically equivalent at the level of the determinant, but the objects they enumerate are not identical. We therefore use *circuit* for a simple directed cycle and *closed walk* or *feedback of order* $k$ for the terms entering $\mathrm{Tr}(\Phi^k)$.

The factor $1/k$ has a simple combinatorial role. A simple circuit of length $k$ is counted $k$ times in the trace, once for each possible starting node. Dividing by $k$ removes this rotational multiplicity. In the three-country example,

$$\mathrm{Tr}(\Phi^3) = 3abc,$$

so its order-three contribution to $\log D$ is

$$-\frac{\mathrm{Tr}(\Phi^3)}{3} = -abc.$$

Longer feedbacks are also attenuated by the explicit factor $1/k$, in addition to the multiplicative attenuation produced by coefficients smaller than one. This is why EDT gives greater structural weight, all else equal, to short feedback loops.

The trace representation also provides an immediate proof of the bounds on $D$. Every term in the sum is non-negative, so

$$\log D \leq 0,$$

hence

$$0 < D \leq 1.$$

Moreover, adding or strengthening a directed coefficient, while holding the others fixed and preserving the spectral condition, cannot decrease any trace term. It therefore cannot increase $\log D$ or $D$. This monotonicity does not apply automatically to a raw-flow change followed by recomputation of the country margins.

### 2.6 Structural interventions: arc deletion, node deletion and exact marginal sensitivity

The monotonicity just described is the basis of the structural interventions used throughout EDT.

Arc deletion:

Let $\Phi'$ be obtained from $\Phi$ by reducing one or more non-negative coefficients, with

$$0 \le \Phi' \le \Phi$$

entry by entry. Since all entries are non-negative,

$$(\Phi')^k \le \Phi^k$$

entry by entry for every $k \ge 1$, and therefore

$$\mathrm{Tr}[(\Phi')^k] \le \mathrm{Tr}(\Phi^k).$$

The log-determinant formula then implies

$$\log \det(I - \Phi') \ge \log \det(I - \Phi),$$

hence

$$\boxed{\det(I - \Phi') \ge \det(I - \Phi).}$$

Suppressing or weakening an arc cannot decrease diffusion. The increase is strict only when the change destroys or weakens feedback; deleting an arc on no circuit leaves the determinant unchanged. The remaining coefficients and normalization margins are held fixed.

<u>Node deletion:</u>

Let $\Phi_{-i}$ be obtained by deleting node $i$, together with all incoming and outgoing arcs. Every closed walk remaining in $\Phi_{-i}$ was already present in $\Phi$, while all closed walks visiting $i$ disappear. Hence

$$\mathrm{Tr}\left(\Phi_{-i}^{k}\right) \leq \mathrm{Tr}(\Phi^{k})$$

for every $k$, and therefore

$$\boxed{D_{-i} \geq D.}$$

This is Lantner's node-suppression theorem. Combined with

$$L_i = \frac{D_{-i}}{D},$$

it gives $L_i \geq 1$ and makes explicit that Lantner centrality measures the marginal structural role of node $i$ in the system of feedbacks.

The same reasoning applies to any subset $A \subseteq N$:

$$D_{-A} = \det(I - \Phi_{-A}) \geq D.$$

<u>Jacobi's formula and the exact sensitivity to one arc:</u>

The same monotonicity can be obtained locally from Jacobi's formula. Let

$$G = I - \Phi, \qquad D = \det(G).$$

For an infinitesimal perturbation $dG$,

$$dD = D\,\mathrm{Tr}(G^{-1}dG).$$

Since $g_{ij} = -\phi_{ij}$ for $i \neq j$,

$$\frac{\partial D}{\partial \phi_{ij}} = -D(G^{-1})_{ji} \qquad (i \neq j).$$

Equivalently, using cofactors,

$$\frac{\partial D}{\partial \phi_{ij}} = -\mathrm{cof}_{ij}(G),$$

where $\mathrm{cof}_{ij}(G)$ is the $(i,j)$ cofactor of $G$. Under $\rho(\Phi) < 1$, $G^{-1}$ is non-negative and $D > 0$, so

$$\frac{\partial D}{\partial \phi_{ij}} \leq 0.$$

Thus an increase in any bilateral structural coefficient can never increase diffusion; it can only leave $D$ unchanged or lower it. The derivative is strictly negative whenever the reverse structural propagation from $j$ to $i$, measured by $(G^{-1})_{ji}$, is positive. This gives a precise feedback interpretation to the marginal effect of an arc: the arc $i \to j$ matters for $D$ to the extent that the remaining network provides a path by which influence can return from $j$ to $i$.

Jacobi's formula also gives an exact finite-change decomposition. Consider two structures $\Phi_0$ and $\Phi_1$, and the straight path

$$\Phi(s) = \Phi_0 + s(\Phi_1 - \Phi_0), \qquad 0 \leq s \leq 1.$$

Let $G(s) = I - \Phi(s)$ and $\Delta\phi_{ij} = \phi_{ij,1} - \phi_{ij,0}$. Then

$$D_1 - D_0 = -\sum_{i \neq j} \Delta\, \phi_{ij} \int_0^1 \mathrm{cof}_{ij}\, [G(s)]\, ds.$$

This identity attributes the total change in the determinant to changes in individual links along a specified continuous path between the two structures. The total is exact. The separate link contributions depend on the path chosen when several coefficients change simultaneously, which is why the empirical analysis below gives priority to path-independent circuit and trace decompositions. The Jacobi representation is nevertheless useful for interpreting local sensitivity and for proving monotonicity results.

**2.7 Partitioning a directed flow structure**

EDT also provides exact results for structures obtained by dividing the set of countries into economically meaningful groups. This is the theorem of the partition developed by Lantner and Lebert (2013) and restated in Lebert and Poindron (2025). It is important for the WTW because interdependence can be generated by feedback internal to regions or blocs, or by feedback crossing their boundaries.

Let

$$\mathcal{P} = (N_1, \ldots, N_P)$$

be a partition of $N$, so that

$$\bigcup_{r=1}^{P} N_r = N, \qquad N_r \cap N_s = \varnothing \quad (r \neq s).$$

After ordering countries by group, write $\Phi$ in block form,

$$\Phi = \begin{pmatrix} \Phi_{11} & \Phi_{12} & \cdots & \Phi_{1P} \\ \Phi_{21} & \Phi_{22} & \cdots & \Phi_{2P} \\ \vdots & \vdots & \ddots & \vdots \\ \Phi_{P1} & \Phi_{P2} & \cdots & \Phi_{PP} \end{pmatrix}.$$

The block $\Phi_{rs}$ contains flows from nodes in $N_r$ toward nodes in $N_s$.

<u>The window-induced structure and the theorem of the partition:</u>

The window-induced matrix associated with the partition retains only within-group flows and deletes every inter-group arc:

$$\Phi^\circ = \operatorname{diag}(\Phi_{11}, \dots, \Phi_{pp}).$$

Because the matrix is block diagonal,

$$D^\circ = \det(I - \Phi^\circ) = \prod_{r=1}^{P} \det(I - \Phi_{rr}).$$

Writing

$$D_r = \det(I - \Phi_{rr}),$$

we therefore have

$$D^\circ = \prod_{r=1}^{p} D_r.$$

The inequality with the complete structure follows directly from the trace representation. For every $k \geq 2$,

$$\mathrm{Tr}[(\Phi^\circ)^k] = \sum_{r=1}^{p} \mathrm{Tr}\,(\Phi_{rr}^k).$$

These terms include exactly the closed walks that remain entirely within one partition block. The complete trace $\mathrm{Tr}(\Phi^k)$ contains these walks plus all closed walks that cross at least one block boundary. Since all weights are non-negative,

$$\mathrm{Tr}[(\Phi^\circ)^k] \leq \mathrm{Tr}(\Phi^k).$$

Therefore,

$$\log D^\circ = -\sum_{k=2}^{\infty} \frac{\mathrm{Tr}[(\Phi^\circ)^k]}{k} \geq -\sum_{k=2}^{\infty} \frac{\mathrm{Tr}(\Phi^k)}{k} = \log D,$$

and hence

$$\boxed{\prod_{r=1}^{P} D_r = D^\circ \geq D.}$$

This is the theorem of the partition. Suppressing all inter-group connections cannot decrease diffusion because it removes every feedback mechanism that requires more than one block. Equivalently, the gap between the determinant of the window-induced structure and the determinant of the complete structure is generated by cross-boundary interdependence.

The theorem is particularly useful because it separates two conceptually different forms of closure. A circuit may close entirely inside one region, or it may leave a region and return through other regions. The former survives in $\Phi^\circ$; the latter does not.

Aggregation-induced structure:

A complementary operation collapses each group $N_r$ into a single aggregate node. For $r \neq s$, the flow from aggregate node $N_r$ to aggregate node $N_s$ is obtained by summation,

$$x_{N_r N_s} = \sum_{i \in N_r} \sum_{j \in N_s} x_{ij}.$$

Collapsing countries also requires a rule for the denominator and for within-group flows. To define a specific common-scale matrix $\Phi^{\bullet}$, retain the original country margins and sum them within each group:

$$\phi^{\bullet}_{rs} = \frac{x_{N_r N_s}}{q^{\bullet}_r} \quad (r \neq s), \quad q^{\bullet}_r = \sum_{i \in N_r} q_i.$$

Within-group trade is excluded from the aggregate coefficients by setting their diagonal to zero, while its contribution to the original country margins is retained. An inactive group is assigned a zero row. The associated determinant is

$$D^{\bullet} = \det(I - \Phi^{\bullet}).$$

This construction has non-negative row sums no larger than one; its spectral radius must still be checked before using a logarithmic expansion. In contrast to the window operation, node merging can create return paths. The inequality stated as a node-aggregation corollary in Lebert and Poindron (2025) cannot be used as a general consequence of non-negativity and sub-stochasticity. The effect of merging must be established for the specified aggregation rule.

For example, take four countries with only two unit flows, from A to B and from C to D. Every original margin equals one and the graph is acyclic, so its determinant is one. Merge A with D and B with C. Under the rule above, the two aggregate coefficients are one half in each direction, giving a determinant of three quarters. Merging creates a circuit. Section 6 instead uses country-level windows throughout, so its decomposition relies only on the window theorem.

This distinction will also clarify the multilayer model. A partition of nodes separates feedback according to its geographical support; a partition into product layers separates feedback according to its commodity composition.

**2.8 Multilayer EDT: product-specific flows as additive components of the same structure**

Lebert and Poindron (2025) extend the Lantner model to a multilayer network. Let

$$\mathcal{W} = \{1, \dots, \ell\}$$

be a set of product categories. For each category $g$, let

$$X^{(g)} = \left(x_{ij}^{(g)}\right)$$

be the matrix of bilateral trade in that category. The aggregate raw trade matrix is

$$X = \sum_{g=1}^{w} X^{(g)}.$$

The crucial point is that the product layers are not normalized independently. The denominator $q_i = \max(E_i, M_i)$ is computed once from aggregate trade and then applied identically to every layer. Thus

$$\Phi^{(g)} = Q^{-1} X^{(g)}.$$

It follows by linearity that

$$\sum_{g=1}^{\ell} \Phi^{(g)} = Q^{-1} \sum_{g=1}^{\ell} X^{(g)} = Q^{-1} X = \Phi.$$

Hence

$$\boxed{\Phi = \sum_{g=1}^{\ell} \Phi^{(g)}.}$$

This exact additivity is essential. Each product matrix is a component of the same normalized economy, not a separately rescaled network. If each layer were normalized with its own denominator, the sum of the layers would no longer reproduce the aggregate structure and the structural meaning of layer aggregation would be lost.

There are no inter-layer arcs in this formulation. A layer identifies the product carried by an arc between countries. The multilayer representation can therefore be written as a block-diagonal matrix,

$$\Phi^{\diamond} = \bigoplus_{g=1}^{\ell} \Phi^{(g)}.$$

Its Ghosh matrix is

$$I - \Phi^{\diamond} = \bigoplus_{g=1}^{w} \left( I - \Phi^{(g)} \right).$$

Let

$$D_g = \det\left( I - \Phi^{(g)} \right)$$

be the determinant of layer $g$. The determinant of the block-diagonal multilayer structure is then

$$\boxed{D^{\diamond} = \det(I - \Phi^{\diamond}) = \prod_{g=1}^{\ell} D_g \,.}$$

By contrast, the determinant of aggregate trade is

$$D = \det\left(I - \sum_{g=1}^{\ell} \Phi^{(g)}\right).$$

These two determinants answer different questions. $D^{\diamond}$ allows a closed walk to remain only within one product layer. $D$ allows successive arcs of the same closed walk to belong to different product categories.

Consider, for example,

$$A \overset{g}{\to} B \overset{h}{\to} C \overset{r}{\to} A,$$

with $g$, $h$ and $r$ distinct. No individual layer contains a closed walk. The aggregate network does. The difference between the separated and aggregate representations therefore isolates a specifically multilayer source of circular interdependence.

<u>Why layer aggregation lowers diffusion:</u>

The result can be proved directly from traces. Since

$$\log D^{\diamond} = \sum_{g=1}^{\ell} \log D_g,$$

we have

$$\log D^{\diamond} = -\sum_{k=2}^{\infty} \frac{\sum_{g=1}^{\ell} \mathrm{Tr}\left[\left(\Phi^{(g)}\right)^{k}\right]}{k}.$$

For the aggregate matrix,

$$\log D = -\sum_{k=2}^{\infty} \frac{\mathrm{Tr}\left[\left(\sum_{g} \Phi^{(g)}\right)^{k}\right]}{k}.$$

Expanding the power gives

$$\left(\sum_{g} \Phi^{(g)}\right)^{k} = \sum_{g_1,\dots,g_k} \Phi^{(g_1)}\,\Phi^{(g_2)} \dots \Phi^{(g_k)}.$$

All entries are non-negative. The terms with

$$g_1 = g_2 = \cdots = g_k = g$$

sum to

$$\sum_{g} \left(\Phi^{(g)}\right)^{k}.$$

All remaining terms are additional non-negative cross-layer products. Therefore

$$\sum_{g} \mathrm{Tr}\left[\left(\Phi^{(g)}\right)^{k}\right] \le \mathrm{Tr}\left[\left(\sum_{g} \Phi^{(g)}\right)^{k}\right].$$

Term-by-term comparison of the logarithmic series gives

$$\boxed{D^{\diamond} \geq D.}$$

This is the layer-aggregation inequality of Lebert and Poindron (2025). Aggregating layers cannot increase diffusion: it retains all within-layer closed walks and adds non-negative terms whose successive arcs use different product categories.

The associated heterogeneity quantity is

$$\boxed{\kappa = D^{\diamond} - D \geq 0.}$$

The quantity $\kappa$ records the determinant effect of combining layers on the common scale. It is not a distance between their flow patterns. Even identical layers can yield a positive value, because superposition strengthens existing circuits and creates mixed-label terms in the trace expansion. Dissimilar layers may additionally close circuits absent from each layer separately.

**2.9 Exact separation of within-layer and genuinely multilayer feedback**

The log-determinant representation permits a sharper decomposition than the level difference $\kappa$. Define the within-layer contribution of order $k$ as the exact trace component

$$C_k^{\text{within}} = \sum_{g=1}^{\ell} \operatorname{Tr}\left[\left(\Phi^{(g)}\right)^k\right].$$

Define the genuinely multilayer component as the remainder of the aggregate trace,

$$C_k^{\text{multi}} = \operatorname{Tr}(\Phi^k) - C_k^{\text{within}}.$$

No new index is introduced here: these two quantities are simply an exact partition of the trace term already present in the EDT identity. By construction,

$$C_k^{\text{multi}} = \sum_{\substack{g_1,\dots,g_k \\ \text{at least two distinct layer labels}}} \operatorname{Tr}\left(\Phi^{(g_1)} \cdots \Phi^{(g_k)}\right).$$

Therefore,

$$\log D = -\sum_{k=2}^{\infty} \frac{C_k^{\text{within}} + C_k^{\text{multi}}}{k}.$$

Since

$$\log D^{\diamond} = -\sum_{k=2}^{\infty} \frac{C_k^{\text{within}}}{k},$$

subtracting gives

$$\boxed{\log D - \log D^{\diamond} = -\sum_{k=2}^{\infty} \frac{C_k^{\text{multi}}}{k}.}$$

Thus the difference between the aggregate and separated log-determinants is exactly the cumulative contribution of closed walks containing at least two different product categories.

At order two, a cross-layer term is

$$\operatorname{Tr}\left(\Phi^{(g)}\Phi^{(h)}\right), \qquad g \neq h.$$

Expanding the trace,

$$\operatorname{Tr}\left(\Phi^{(g)}\Phi^{(h)}\right) = \sum_{i,j} \phi_{ij}^{(g)}\, \phi_{ji}^{(h)}.$$

It measures bilateral feedback in which country $i$ exports product $g$ to $j$ while $j$ returns product $h$ to $i$. At order three, products may combine around a three-country chain,

$$\mathrm{Tr}\left(\Phi^{(g)}\Phi^{(h)}\Phi^{(r)}\right) = \sum_{i,j,k} \phi_{ij}^{(g)}\, \phi_{jk}^{(h)} \phi_{ki}^{(r)}.$$

The same logic extends to every order. The multilayer character of the WTW is therefore not defined by the mere coexistence of many products. It is defined by the extent to which different product categories are jointly required to close international feedback mechanisms.

### 2.10 Layer suppression and the structuring role of product categories

The monotonicity of EDT applies directly to layers. Let

$$\Phi_{-g} = \sum_{h\neq g} \Phi^{(h)} = \Phi - \Phi^{(g)}.$$

Since

$$0 \leq \Phi_{-g} \leq \Phi,$$

we have for every $k$,

$$\mathrm{Tr}\left(\Phi_{-g}^{k}\right) \leq \mathrm{Tr}(\Phi^{k}),$$

and consequently

$$\boxed{D_{-g} = \det\left(I - \Phi_{-g}\right) \geq D.}$$

Lebert and Poindron (2025) call the difference

$$\boxed{D_{-g} - D}$$

the diffusion intermediacy or structuring impact of layer $g$ in the aggregate structure. It answers a counterfactual question internal to the EDT: by how much would diffusion increase if every flow belonging to product category $g$ were suppressed, while all other normalized flows were kept unchanged?

A large value means that layer $g$ participates strongly in the circular architecture of aggregate trade. This may occur because the layer itself contains strong within-product feedback, because it combines with other layers to close feedbacks, or both. The quantity is therefore not equivalent to the share of product $g$ in world trade.

For a set of layers $J \subseteq \mathcal{W}$, define

$$\Phi_{-J} = \sum_{g \notin J} \Phi^{(g)}$$

and

$$D_{-J} = \det\left(I - \Phi_{-J}\right).$$

Then

$$D_{-J} \geq D$$

and the set function

$$\mu(J) = D_{-J} - D$$

is grounded,

$$\mu(\varnothing) = 0,$$

and monotone with set inclusion. This provides the bridge between EDT and the theory of capacities used by Lebert and Poindron.

### 2.11 Structural interaction between product layers

Consider a reciprocal link whose outward and return legs belong to different layers. Both layers are needed for the circuit to exist. Yet removing either one breaks it, so their separate removal effects count the same loss of feedback twice. This distinction between joint presence and overlapping suppression effects guides the interpretation below. For categories $g$ and $h$, the simplified pairwise index used by Lebert and Poindron (2025), following Murofushi and Soneda (1993) and Grabisch (1997, 2016), is the second-order discrete difference of the removal capacity $\mu$:

$$\Gamma(g,h) = \mu(\{g,h\}) - \mu(\{g\}) - \mu(\{h\}) + \mu(\varnothing).$$

Since $\mu(\varnothing) = 0$,

$$\boxed{\Gamma(g,h) = D_{-\{g,h\}} - D_{-g} - D_{-h} + D.}$$

The expression compares the increase in diffusion after joint removal with the sum of the increases after the two separate removals, each measured from the complete network. All other coefficients are held fixed. In the reciprocal-link example, the joint effect equals either separate effect, making the interaction negative although both layers are required for closure.

If

$$\Gamma(g,h) > 0,$$

joint suppression raises diffusion by more than the sum of the separate suppression effects. This is super-additivity of removal effects, termed complementarity for this capacity. It is not a direct statement that the goods are complementary inputs to production.

If

$$\Gamma(g,h) < 0,$$

joint suppression raises diffusion by less than the sum of the separate suppression effects. The effects overlap, termed redundancy or substitution for this capacity. As the reciprocal-link example shows, this is compatible with both layers being necessary for the same circuit.

The connection with cross-layer circuits becomes explicit in a second-order approximation. Since $\mathrm{Tr}(\Phi) = 0$,

$$\det(I - \Phi) = 1 - \frac{1}{2}\mathrm{Tr}(\Phi^2) + o(\| \Phi \|^2)$$

around the null structure. For the following two-layer calculation, only the two displayed layers are present; it is not an expansion around a fixed, non-zero background of other layers. Then

$$\mathrm{Tr}\left[\left(\Phi^{(g)} + \Phi^{(h)}\right)^2\right] = \mathrm{Tr}\left[\left(\Phi^{(g)}\right)^2\right] + \mathrm{Tr}\left[\left(\Phi^{(h)}\right)^2\right] + 2\mathrm{Tr}\left(\Phi^{(g)}\Phi^{(h)}\right).$$

Substituting this expansion into the interaction expression yields

$$\boxed{\Gamma(g,h) = -\mathrm{Tr}\left(\Phi^{(g)}\Phi^{(h)}\right) + o\left(\| \Phi^{(g)} + \Phi^{(h)} \|^2\right).}$$

Thus, to second order, negative interaction is generated by two-product bilateral feedback of the form

$$i \overset{g}{\rightarrow} j \overset{h}{\rightarrow} i.$$

Lebert and Poindron show symmetrically that the corresponding two-layer heterogeneity term is approximately

$$\kappa(g,h) = \mathrm{Tr}\left(\Phi^{(g)}\Phi^{(h)}\right) + o\left(\parallel \Phi^{(g)} + \Phi^{(h)} \parallel^2\right),$$

so heterogeneity and the removal interaction have opposite signs at second order. Cross-layer feedback requires both layers to be present, while its destruction is counted in each separate removal effect. Negative interaction therefore expresses overlap of those effects, not technological substitutability or an ability to reroute trade.

The empirical calculations use the exact determinant-based $\Gamma(g,h)$ in the full network, including all remaining layers. The two-layer expansion provides intuition near the null structure, but is not an error-controlled approximation for a pair embedded in the observed WTW.

**2.12 Introducing time: structural change without an external distance between networks**

We now consider a sequence of annual structures. For every date $t$, let

$$X_t, \quad \Phi_t, \quad G_t = I - \Phi_t, \quad \text{and} \quad D_t = \det(I - \Phi_t)$$

be defined as above.

For two dates $a$ and $b$, the change in determinant-based circular interdependence is summarized by

$$\boxed{\Delta D_{a,b} = D_b - D_a.}$$

This quantity records the signed change in $D$. It is not a distance between adjacency matrices: it can be zero when bilateral coefficients or circuits change. The decomposition identifies which feedback components account for the net change and which components offset one another.

<u>Exact change in the circuit representation:</u>

At each date,

$$D_t = 1 + \sum_{F \neq \varnothing} (-1)^{|F|}\, \Omega_t(F).$$

Embedding the two structures on a common node support when necessary, subtraction gives

$$\boxed{D_b - D_a = \sum_{F} (-1)^{|F|}\, [\Omega_b(F) - \Omega_a(F)].}$$

The change in the diffusion index is therefore exactly the signed change in the circuit families entering the determinant. A zero net change does not imply that each family is unchanged: positive and negative contributions can cancel.

<u>Exact change by feedback length:</u>

For empirical work, the trace representation is more tractable. Define

$$\delta_{a,b} = \log D_b - \log D_a = \log\left(\frac{D_b}{D_a}\right).$$

Using the trace identity at both dates,

$$\boxed{\delta_{a,b} = -\sum_{k=2}^{\infty} \frac{\mathrm{Tr}\left(\Phi_b^k\right) - \mathrm{Tr}\left(\Phi_a^k\right)}{k}.}$$

The contribution of order $k$ is therefore

$$-\frac{\mathrm{Tr}\left(\Phi_b^k\right) - \mathrm{Tr}(\Phi_a^k)}{k}.$$

Positive values indicate that feedback of order $k$ has weakened, contributing to an increase in diffusion. Negative values indicate that feedback of order $k$ has strengthened.

The logarithm is a decomposition device, not a new measure replacing $D$. The exact relation back to the determinant is

$$D_b = D_a e^{\delta_{a,b}},$$

and hence

$$\boxed{D_b - D_a = D_a\left(e^{\delta_{a,b}} - 1\right).}$$

Thus every trace decomposition of $\delta$ maps exactly into the observed change in $D$.

<u>Changing country coverage:</u>

EDT remains comparable when the set of active countries changes. Let

$$N = N_a \cup N_b.$$

If a country is absent at date $a$, represent it by a zero row and zero column in the enlarged flow matrix. The corresponding enlarged Ghosh matrix has the block form

$$\tilde{G}_a = \begin{pmatrix} G_a & 0 \\ 0 & I_z \end{pmatrix},$$

where $z$ is the number of nodes added only to obtain common support. Therefore,

$$\det(\tilde{G}_a) = \det(G_a)\det(I_z) = D_a.$$

The trace representation is invariant as well. The enlarged normalized matrix is

$$\tilde{\Phi}_a = \begin{pmatrix} \Phi_a & 0 \\ 0 & 0 \end{pmatrix},$$

so for every $k$,

$$\mathrm{Tr}(\tilde{\Phi}_a^k) = \mathrm{Tr}(\Phi_a^k).$$

Hence neither $D$ nor its closed-walk decomposition changes when structurally inactive nodes are added. Non-zero flows make a country participate in propagation, but affect the determinant only through feedback and any associated changes in normalized coefficients. Temporal comparison does not require an artificially balanced panel of countries.

**2.13 Dynamic decomposition of the multilayer WTW**

The temporal and multilayer constructions can now be combined without introducing any new structural index.

At every date $t$,

$$\Phi_t = \sum_g \Phi_t^{(g)},$$

with

$$D_t = \det(I - \Phi_t)$$

and

$$D_t^{\diamond} = \prod_g D_{g,t}.$$

For two dates $a$ and $b$,

$$\delta_{a,b} = \log D_b - \log D_a.$$

Add and subtract the two separated multilayer log-determinants:

$$\delta_{a,b} = (\log D_b^{\diamond} - \log D_a^{\diamond}) + [(\log D_b - \log D_b^{\diamond}) - (\log D_a - \log D_a^{\diamond})].$$

Define, purely as labels for the two exact components,

$$\delta_{a,b}^{\text{within}} = \log D_b^{\diamond} - \log D_a^{\diamond}$$

and

$$\delta_{a,b}^{\text{multi}} = (\log D_b - \log D_b^{\diamond}) - (\log D_a - \log D_a^{\diamond}).$$

Then

$$\boxed{\delta_{a,b} = \delta_{a,b}^{\text{within}} + \delta_{a,b}^{\text{multi}}.}$$

The within-layer term can be expanded as

$$\delta_{a,b}^{\text{within}} = \sum_g \left(\log D_{g,b} - \log D_{g,a}\right),$$

so it measures the change in feedback that would be observed if every product network remained structurally separate. The multilayer term is the change in feedback created specifically by combining different product categories within the same closed walks.

The same decomposition can be carried out order by order. For every $k$ and date $t$,

$$\mathrm{Tr}\left(\Phi_t^k\right) = C_{k,t}^{\mathrm{within}} + C_{k,t}^{\mathrm{multi}}.$$

Therefore,

$$\boxed{\delta_{a,b} = -\sum_{k=2}^{\infty} \frac{\Delta C_{k,a,b}^{\mathrm{within}} + \Delta C_{k,a,b}^{\mathrm{multi}}}{k},}$$

where

$$\Delta C_{k,a,b}^{\mathrm{within}} = C_{k,b}^{\mathrm{within}} - C_{k,a}^{\mathrm{within}}$$

and

$$\Delta C_{k,a,b}^{\mathrm{multi}} = C_{k,b}^{\mathrm{multi}} - C_{k,a}^{\mathrm{multi}}.$$

This is the central dynamic extension used in the article. It distinguishes two logically different transformations of the WTW. The first is a change in the feedback architecture internal to individual product categories. The second is a change in the way product categories combine to close international feedback loops. The distinction is exact and exhaustive in log-determinant space.

The temporal evolution of layer structuring effects and interactions follows directly from the static EDT quantities. For a product layer $g$, compare

$$D_{-g,b} - D_b$$

with

$$D_{-g,a} - D_a.$$

A rise means that suppressing the layer has a larger effect on diffusion at date $b$ than at date $a$: the product category has become more structurally embedded in the system of feedbacks.

Similarly, the evolution of pairwise structural interaction is

$$\Delta\Gamma_{a,b}(g,h) = \Gamma_b(g,h) - \Gamma_a(g,h).$$

This is not an independent change index. It is simply the temporal difference of the determinant-based interaction already defined by EDT.

**2.14 What the formalism identifies as structural change**

The preceding results define a hierarchy of structural questions without combining unrelated network indicators.

At the most aggregate level,

$$D_t$$

measures diffusion and

$$1 - D_t$$

measures circular interdependence under the no-self-flow convention used here.

The circuit expansion identifies the elementary configurations from which the determinant is built. The trace expansion reorganizes the same information by the length of closed feedback walks. Arc and node suppression reveal marginal structural contributions, while Jacobi's formula identifies the exact local sensitivity of $D$ to bilateral coefficients. The theorem of the partition separates feedback internal to groups of countries from feedback crossing their boundaries. The multilayer construction separates within-product feedback from feedback that requires combinations of product categories. Layer suppression identifies the categories whose presence contributes most to aggregate circularity, while $\Gamma(g,h)$ identifies super-additivity or overlap of pairwise suppression effects. Finally, the temporal extension expresses observed changes in $D$ as changes in these same objects.

Between dates $a$ and $b$, a difference between $D_b$ and $D_a$ identifies a change in the circular component measured by EDT. Equal determinants do not imply identical trade structures. The analysis therefore examines the components of the determinant, including offsetting changes, rather than treating its scalar value as a complete description of the network. The exact decomposition

$$\log \frac{D_b}{D_a} = -\sum_{k=2}^{\infty} \frac{\mathrm{Tr}\left(\Phi_b^k\right) - \mathrm{Tr}\left(\Phi_a^k\right)}{k}$$

identifies the change by feedback length, while

$$\log \frac{D_b}{D_a} = \delta_{a,b}^{\mathrm{within}} + \delta_{a,b}^{\mathrm{multi}}$$

identifies whether the transformation occurred primarily within product networks or in the way those product networks combine. The same structure can subsequently be opened by country groups, product layers, individual layers, layer pairs, or bilateral arcs, always through operations defined on the same determinant.

This internal consistency is the main methodological reason for using EDT to study the long-run transformation of the WTW. The empirical sections apply these identities to annual BACI trade structures and ask what kind of feedback architecture actually changed between 1995 and 2023.

## 3. Data and construction of the multilayer World Trade Web

The empirical objective of this section is deliberately operational. A reader who has downloaded the raw BACI files should be able to reconstruct, without any additional information, the annual matrices that enter the EDT analysis developed in Section 2. We therefore describe the complete transformation from individual exporter–importer–product records to the aggregate normalized trade matrix and to its product-specific layers. The procedure is deterministic: apart from the choice of the time interval and the product aggregation level, it contains no estimated parameter, threshold, filtering rule based on the results, or ex post network transformation.

The analysis uses CEPII's BACI database (Gaulier & Zignago, 2010), release 202601, in the 1992 revision of the Harmonized System (HS92). BACI is constructed from UN Comtrade declarations and reconciles mirror export and import reports into a unique bilateral trade flow. Import values, originally reported CIF, are first adjusted toward an FOB basis, and the two declarations are combined using information on reporting reliability. The resulting database therefore provides a single directed value for each exporter–importer–product–year combination rather than requiring the analyst to choose between exporter and importer declarations.

The 202601 BACI release contains HS92 data through 2024. The empirical window used here is nevertheless frozen at 1995–2023. This is the interval for which the matrix-construction pipeline and the structural analysis were defined before the present extension of the article. Keeping 2023 as the terminal year avoids changing the sample endpoint ex post. Nothing in the procedure depends on this endpoint, and an additional year can be incorporated by applying exactly the same operations.

### 3.1 Why HS92 and why HS2 layers?

BACI is distributed in several revisions of the Harmonized System. We use HS92 because it provides a single product nomenclature over the entire 1995–2023 period. This avoids a problem that would arise if successive HS revisions were concatenated: product codes would have to be translated through concordance tables, and part of the measured structural change could then be generated by changes in classification rather than by changes in trade itself.

The raw BACI HS92 files identify products at the six-digit level. Let

$$\mathcal{K}$$

denote the set of HS6 products. At this level, BACI contains slightly more than five thousand product categories. The matrix construction described below is first performed at HS6 because this is the native resolution of the data. Product layers used in the structural analysis are then obtained by exact aggregation to the two-digit HS level.

For an HS6 code written as a six-character string,

$$p = c_1 c_2 c_3 c_4 c_5 c_6,$$

the corresponding HS2 chapter is simply

$$h(p) = c_1 c_2.$$

The HS2 level is chosen for the empirical analysis because it yields an economically interpretable multilayer system of approximately one hundred product categories while remaining sufficiently disaggregated to preserve differences in trade structure. In the 1995–2023 files used here, 96 HS2 chapters are observed. Chapter 77 is reserved in HS92 and is not observed. No special residual HS2 category is required in the final sample.

With the common normalization vector computed from aggregate trade, summing HS6 flows to HS2 before or after row normalization gives the same result (Section 3.9). The common denominator is the essential condition.

### 3.2 Raw BACI files and variables

For each year $t$, the raw trade file is
BACI_HS92_Y<t>_V202601.csv.

The accompanying metadata files are

country_codes_V202601.csv

and

product_codes_HS92_V202601.csv.

Each raw annual BACI file contains six variables (Table 3.1). Only four are required to construct the matrices used in this article.

| **BACI field** | **Meaning** | **Use in the present analysis** |
|---|---|---|
| t | Year | Identifies the annual file; not needed once the file is selected |
| i | Exporter, ISO numeric country code | Row index of the directed matrix |
| j | Importer, ISO numeric country code | Column index of the directed matrix |
| k | HS6 product code | Identifies the product layer |
| v | Trade value, thousands of current US dollars | Edge weight used in all matrices |
| q | Quantity, metric tons | Not used |

**Table 3.1.** Variables of the raw BACI files and their use in the analysis.

The product code k must be read as a string. Treating it as an integer would delete leading zeros and would therefore corrupt the mapping from HS6 to HS2. A valid product code is stored as a six-character string, for example 010121, not as the integer 10121.

BACI records only strictly positive trade flows. Zero observations are therefore implicit: if no record exists for exporter $i$, importer $j$, product $p$ and year $t$, the corresponding matrix element is set to zero.

Trade values are expressed in thousands of current US dollars. No conversion to dollars is required because the EDT coefficients are dimensionless. Multiplying every flow in one year by the same positive scalar leaves the normalized matrix unchanged. Formally, if

$$X_t' = c_t X_t, \qquad c_t > 0,$$

then exports, imports and normalization denominators are all multiplied by $c_t$, so

$$\Phi_t' = \Phi_t.$$

Consequently, changing the monetary unit or applying a common annual deflator has no effect on the structure. This scale invariance should not be confused with invariance to relative prices: product- or country-specific price changes can alter relative bilateral values and therefore can alter the normalized structure. The analysis is explicitly an analysis of trade values, not physical quantities.

### 3.3 A fixed country universe and matrix indexing

Temporal comparison is easiest when every annual matrix uses the same ordering of countries. The first preprocessing step is therefore to construct a master country table from country_codes_V202601.csv.

Let

$$\mathrm{N} = \{1, \dots, n\}$$

be the ordered set of country codes retained in the BACI metadata. Each numeric BACI country code is assigned once and for all to a matrix index

$$\iota\text{:country code} \to \{0, \dots, n-1\}.$$

The mapping is kept fixed for every year from 1995 to 2023. The resulting table contains three fields (Table 3.2):

| **Field** | **Content** |
|---|---|
| code_pays | BACI numeric country code |
| pays | Country label or ISO identifier supplied by the metadata |
| matrix_index | Permanent row/column position in every annual matrix |

**Table 3.2.** Fields of the permanent country mapping.

The use of a common universe is computationally convenient but does not impose a balanced sample in the economic sense. A country with no recorded flow in year $t$ simply has a zero row and a zero column. As shown in Section 2, adding such a structurally inactive node contributes an identity block to the Ghosh matrix and leaves both the determinant and all trace terms unchanged. No country has to be deleted merely because it is inactive in some years.

This convention also makes entry and exit transparent. If a country becomes active in a later year, the corresponding row and column simply cease to be zero and the feedback structures in which the country participates enter the EDT calculations automatically.

### 3.4 Cleaning the annual bilateral-product records

For each year $t$, the annual file is reduced to the fields

$$(k, i, j, v).$$

For clarity, define

$$p = k, \qquad e = i, \qquad m = j, \qquad x = v.$$

The following operations are then applied in order.

First, missing values in any of these four fields are removed. Second, product codes are left-padded with zeros until they contain exactly six characters. Third, observations with non-positive trade value are removed; in standard BACI files this step is normally redundant because only positive flows are recorded, but retaining the check makes the construction robust to altered extracts. Fourth, observations satisfying

$$e = m$$

are removed before any matrix is constructed.

This last operation implements the no-self-flow convention of Section 2:

$$x_{ii,t} = 0.$$

It is preferable to remove these observations before matrix construction rather than to construct a matrix and overwrite its diagonal afterwards, because the former guarantees that domestic loops never enter the export and import totals used in normalization.

Finally, exporter and importer country codes are converted into their permanent matrix indices. Records whose country codes are not present in the master table are discarded. With the official BACI metadata and the corresponding annual files, this is a consistency check rather than a substantive filter.

**3.5 HS6 raw-flow matrices**

For each HS6 product $p$ and year $t$, define the $n \times n$ raw-flow matrix

$$W_t^{(p)} = \left(w_{ij,t}^{(p)}\right),$$

where

$$w_{ij,t}^{(p)} = \sum_{u \in \mathcal{U}_{ijpt}} v_u.$$

Here $\mathcal{U}_{ijpt}$ is the set of raw BACI records associated with exporter $i$, importer $j$, product $p$ and year $t$. In a standard BACI annual file the combination is normally unique, so the sum contains a single observation. Writing the matrix element as a sum nevertheless makes the procedure robust to duplicated or pre-aggregated input files.

The diagonal satisfies

$$w_{ii,t}^{(p)} = 0$$

by construction.

The aggregate raw-flow matrix is the sum of all HS6 matrices:

$$W_t = \sum_{p \in \mathcal{K}_t} W_t^{(p)},$$

where $\mathcal{K}_t$ denotes the set of HS6 products actually observed in year $t$.

Equivalently, its elements are

$$w_{ij,t} = \sum_{p \in \mathcal{K}_t} w_{ij,t}^{(p)}.$$

This equality is a first reconstruction identity and should be checked numerically when matrices are produced.

Because the bilateral-product data are sparse, the computational implementation stores $W_t$ and all $W_t^{(p)}$ as compressed sparse row (CSR) matrices. Sparsity affects only storage and speed; it does not change any mathematical operation described below.

### 3.6 Aggregate exports, imports, and the trade-balance denominator

Section 2.1 characterizes the country margin as the minimum common denominator that bounds both sides of the external trade position, and then relates that margin to classical trade-imbalance normalizations. A separate empirical decision concerns the level at which the margin is computed. For the multilayer EDT used here, the denominator must be calculated from aggregate trade, not separately within product layers. For every country and year, compute total exports from the row sums of the aggregate raw-flow matrix,

$$E_{i,t} = \sum_{j=1}^{n} w_{ij,t},$$

and total imports from the column sums,

$$M_{i,t} = \sum_{j=1}^{n} w_{ji,t}.$$

The annual normalization denominator is then

$$q_{i,t} = \max(E_{i,t}, M_{i,t}).$$

Using the aggregate margin here involves a second choice that is distinct from the maximum-margin rule itself. The same maximum rule could, in principle, be applied separately to each product using product-specific exports and imports; such a normalization is meaningful when each product network is studied as an autonomous object. In the present article, by contrast, the denominator is computed once from total trade. Every product flow is therefore expressed relative to the same country-level external-trade scale, which is what makes the layers additive components of one multilayer trade system.

Collect these denominators in the diagonal matrix

$$Q_t = \operatorname{diag}(q_{1,t}, \dots, q_{n,t}).$$

If a country is completely inactive in year $t$, then

$$E_{i,t} = M_{i,t} = q_{i,t} = 0.$$

No division is performed for that row. Its normalized row is defined as a row of zeros. Computationally, define

$$\theta_{i,t} = \begin{cases} 1/q_{i,t}, & q_{i,t} > 0, \\ 0, & q_{i,t} = 0, \end{cases}$$

and let the empirical row-scaling matrix be

$$\Theta_t = \operatorname{diag}(\theta_{1,t}, \dots, \theta_{n,t}).$$

For active countries, $\Theta_t$ coincides with $Q_t^{-1}$. The zero convention for inactive countries avoids attempting to invert a zero diagonal element. All normalized matrices can then be constructed by ordinary left multiplication.

### 3.7 Construction of the annual aggregate EDT matrix

The empirical normalized matrix is

$$B_t = \Theta_t W_t.$$

The computational symbol $B_t$ is used in the archived scripts. It is exactly the matrix denoted $\Phi_t$ in Section 2:

$$B_t \equiv \Phi_t.$$

Element by element,

$$b_{ij,t} = \begin{cases} \dfrac{w_{ij,t}}{\max\left(E_{i,t}, M_{i,t}\right)}, & q_{i,t} > 0, \\ 0, & q_{i,t} = 0. \end{cases}$$

The matrix is non-negative and has a zero diagonal. Its row sum is

$$\sum_j b_{ij,t} = \frac{E_{i,t}}{\max\left(E_{i,t}, M_{i,t}\right)} \leq 1.$$

Thus surplus or balanced countries have normalized row sum one, whereas deficit countries have a row sum strictly below one. The normalization used in the formal model is therefore reproduced exactly from the raw BACI values.

The Ghosh matrix entering EDT calculations is subsequently

$$G_t = I - B_t.$$

At this stage, however, no determinant or other EDT indicator need yet be computed. The purpose of the present section is to establish the matrices that constitute the inputs to the structural analysis.

### 3.8 Construction of HS6 normalized product layers

The critical rule in the multilayer construction is that product layers are not normalized using their own exports and imports. The same global denominator vector $q_t$ computed from $W_t$ must be applied to every HS6 raw-flow matrix.

For product $p$,

$$B_t^{(p)} = \Theta_t W_t^{(p)}.$$

Hence

$$b_{ij,t}^{(p)} = \begin{cases} \dfrac{w_{ij,t}^{(p)}}{q_{i,t}}, & q_{i,t} > 0, \\ 0, & q_{i,t} = 0. \end{cases}$$

Because $\Theta_t$ is common to all layers,

$$\sum_{p \in \mathcal{K}_t} B_t^{(p)} = \sum_p \Theta_t W_t^{(p)} = \Theta_t \sum_p W_t^{(p)} = \Theta_t W_t = B_t.$$

Thus

$$\boxed{B_t = \sum_p B_t^{(p)}}.$$

This identity is not an approximate property of the data. It is an exact consequence of the common normalization. It is also the empirical implementation of the multilayer EDT developed in Section 2.

A layer-specific alternative would be to define, for each product,

$$q_{i,t}^{(p)} = \max\left(E_{i,t}^{(p)}, M_{i,t}^{(p)}\right)$$

and divide the product matrix by this product-specific denominator. This is a legitimate construction if the product network is analyzed in isolation. It places each layer on its own country-specific scale and therefore, in general, does not preserve the additive identity with aggregate trade.

$$\sum_p \Theta_t^{(p)} W_t^{(p)} \neq \Theta_t W_t.$$

Such a construction answers a different economic question. It asks how each country distributes trade inside each product network considered separately. The present multilayer analysis instead asks how product flows contribute to one aggregate system measured on a common country scale. For that purpose, the global denominator must be computed once and stored before any product layer is normalized.

### 3.9 Exact aggregation from HS6 to HS2

Let $\mathcal{K}_h$ denote the set of HS6 products whose first two digits equal HS2 chapter $h$. The raw HS2 matrix is

$$W_t^{(h)} = \sum_{p \in \mathcal{K}_h} W_t^{(p)}.$$

The normalized HS2 matrix is then

$$B_t^{(h)} = \Theta_t W_t^{(h)}.$$

By linearity,

$$B_t^{(h)} = \Theta_t \sum_{p \in \mathcal{K}_h} W_t^{(p)} = \sum_{p \in \mathcal{K}_h} \Theta_t W_t^{(p)} = \sum_{p \in \mathcal{K}_h} B_t^{(p)}.$$

Therefore either of the following two computational routes gives exactly the same HS2 layer:

$$\text{raw HS6} \rightarrow \text{sum to HS2} \rightarrow \text{normalize with } \Theta_t,$$

or

$$\text{raw HS6} \rightarrow \text{normalize with } \Theta_t \rightarrow \text{sum normalized HS6 to HS2.}$$

The archived implementation follows the second route because the HS6 matrices are already stored for other analyses.

Summing the HS2 layers reproduces aggregate trade:

$$\boxed{B_t = \sum_h B_t^{(h)}}.$$

No HS2 matrix is renormalized after aggregation. Renormalizing a chapter would destroy this identity and would convert the chapter into a stand-alone network rather than a layer of the world trade structure.

The product code should remain a string throughout this operation. If an unexpected two-digit prefix such as a special or residual code is present, it should first be retained so that the reconstruction identity remains exact. It can be flagged for inspection and, if substantively justified, treated separately after the matrix has been validated. In the data used here, the final annual system contains 96 standard HS2 chapters and no residual chapter.

### 3.10 Storage of the annual matrices

The complete HS6 representation contains thousands of sparse matrices per year. Storing each matrix as a separate file is unnecessary and makes exact reproduction cumbersome. The implementation therefore stores one HDF5 file per year,

matrices_<year>.h5.

Every sparse matrix is stored in CSR form through three arrays (data, indices, and indptr) together with its shape. The annual HDF5 hierarchy is given in Table 3.3.

| **HDF5 path** | **Content** |
|---|---|
| /meta/m_global | Vector $q_t$ of annual global normalization denominators |
| /meta/row_sum_total_exports | Vector $E_t$ of aggregate exports |
| /meta/col_sum_total_imports | Vector $M_t$ of aggregate imports |
| /meta/products | HS6 products observed in the year |
| /global/W | Aggregate raw-flow matrix $W_t$ |
| /global/B | Aggregate normalized matrix $B_t = \Phi_t$ |
| /products/<HS6>/W | Raw HS6 matrix $W_t^{(p)}$ |
| /products/<HS6>/B | Normalized HS6 matrix $B_t^{(p)}$ |
| /products/<HS6>/row_sum_layer_exports | Raw HS6 export totals |
| /products/<HS6>/col_sum_layer_imports | Raw HS6 import totals |

**Table 3.3.** Hierarchy of the annual HDF5 files.

The HDF5 format is a storage choice, not part of the EDT methodology. A researcher can instead use sparse Matrix Market files, NumPy arrays, R sparse matrices, or another lossless format, provided that the fixed country ordering and the common normalization vector are preserved.

The annual HS2 matrices used in the present article do not need to be stored separately. They can be reconstructed at any time by summing the relevant /products/<HS6>/B matrices according to the first two characters of the HS6 code.

### 3.11 Reconstruction and validation tests

Before any structural statistic is computed, each annual dataset is subjected to a sequence of algebraic checks. These checks are important because the EDT identities of Section 2 assume that the matrices have been constructed exactly as specified.

*Test 1 — Non-negativity and zero diagonal*

For every year and layer,

$$B_t \geq 0, \qquad B_t^{(h)} \geq 0,$$

entry by entry, and

$$\mathrm{diag}(B_t) = 0, \qquad \mathrm{diag}\left(B_t^{(h)}\right) = 0.$$

Any non-zero diagonal element indicates that self-flows have not been removed before normalization.

*Test 2 — Row-substochasticity*

For every country,

$$0 \leq \sum_j b_{ij,t} \leq 1.$$

A row sum larger than one indicates either an incorrect denominator or an inconsistency between the matrix and the export/import totals.

*Test 3 — Raw-flow reconstruction*

The product matrices must satisfy

$$W_t = \sum_p W_t^{(p)}.$$

If HS2 matrices are also constructed in raw values,

$$W_t = \sum_h W_t^{(h)}.$$

*Test 4 — Normalized multilayer reconstruction*

The normalized matrices must satisfy both

$$B_t = \sum_p B_t^{(p)}$$

and

$$\boxed{B_t = \sum_h B_t^{(h)}}.$$

In floating-point arithmetic the residual matrix

$$Z_t = B_t - \sum_h B_t^{(h)}$$

should be zero up to machine precision. The implementation checks

$$\max_{i,j} |Z_{ij,t}| \le 10^{-11}.$$

A Frobenius residual can be recorded as a secondary diagnostic,

$$\varepsilon_t^F = \frac{\| Z_t \|_F}{\| B_t \|_F}.$$

This quantity is a numerical audit statistic only; it is not used as an empirical indicator.

*Test 5 — Spectral condition*

The spectral radius is computed from the aggregate normalized matrix,

$$\rho(B_t) = \max_{\lambda \in \mathrm{spec}(B_t)} |\lambda|.$$

The EDT calculations require

$$\rho(B_t) < 1.$$

Because every HS2 layer and every regional block used below is a non-negative matrix bounded entrywise by the aggregate matrix (a principal submatrix in the regional case), the same strict inequality then holds for every layer and block determinant. This condition is checked rather than imposed by an additional rescaling. If it were violated, the correct response would be to inspect the matrix construction and the economic structure, not to multiply the matrix by an arbitrary constant solely to force convergence.

*Test 6 — Determinant domain*

As a final pre-analysis check,

$$D_t = \det(I - B_t)$$

must have positive sign and, under the maintained EDT assumptions,

$$0 < D_t \leq 1.$$

The determinant is evaluated numerically through its sign and log-absolute value (slogdet) rather than by direct determinant multiplication. This avoids avoidable underflow and provides the $\log D_t$ quantity used later in the trace decomposition.

### 3.12 Complete replication algorithm from raw BACI files

The whole preprocessing chain can be summarized as the following deterministic algorithm.

<u>Algorithm. Construction of annual multilayer EDT matrices:</u>

*Input.* BACI HS92 annual trade files for 1995–2023, country_codes_V202601.csv, and product_codes_HS92_V202601.csv.

*Step 1 — Build the permanent country mapping.* Read the country metadata, sort the retained numeric country codes, and assign each code a unique matrix position that remains unchanged over all years.

*Step 2 — Read one annual file.* Load fields k, i, j, and v; read k as a string and preserve leading zeros.

*Step 3 — Clean the records.* Remove missing observations, non-positive trade values, and records with i == j. Convert exporter and importer codes to their permanent matrix positions.

*Step 4 — Construct HS6 raw matrices.* For each product $p$, place trade value $v$ in cell $(i, j)$ and sum any duplicate records, yielding $W_t^{(p)}$.

*Step 5 — Construct aggregate raw trade.* Sum all HS6 matrices,

$$W_t = \sum_p W_t^{(p)}.$$

*Step 6 — Compute global country totals.* Obtain

$$E_{i,t} = \sum_j w_{ij,t}, \qquad M_{i,t} = \sum_j w_{ji,t},$$

and

$$q_{i,t} = \max\left(E_{i,t}, M_{i,t}\right).$$

*Step 7 — Normalize aggregate trade.* Define $\theta_{i,t} = 1/q_{i,t}$ for $q_{i,t} > 0$ and zero otherwise. Construct

$$B_t = \Theta_t W_t.$$

*Step 8 — Normalize every HS6 layer with the same denominator.* For all $p$,

$$B_t^{(p)} = \Theta_t W_t^{(p)}.$$

Do not recompute a product-specific denominator.

*Step 9 — Aggregate HS6 layers to HS2.* For every two-digit chapter $h$,

$$B_t^{(h)} = \sum_{p:h(p)=h} B_t^{(p)}.$$

Do not renormalize the resulting HS2 matrix.

*Step 10 — Validate exact reconstruction.* Verify

$$B_t = \sum_p B_t^{(p)} = \sum_h B_t^{(h)}$$

to machine precision, verify zero diagonals and row sums not exceeding one, and check $\rho(B_t) < 1$.

*Output.* For each year $t$, the aggregate normalized matrix $B_t$ and the set of normalized HS2 layers

$$\left\{B_t^{(h)} : h \in \mathcal{H}\right\}$$

constitute the complete input required by the structural analysis of Section 2.

### 3.13 Minimal pseudocode implementation

The following language-neutral pseudocode reproduces the same operations. It is included to make clear which quantities are calculated globally and which are calculated by product.

```
COUNTRIES = permanent mapping from BACI country metadata

for year in 1995,...,2023:
    data = read BACI_HS92_Y<year>_V202601.csv
    keep k, i, j, v
    read k as a six-character string
    drop missing rows
    keep v > 0
    drop i == j
    map i and j to permanent matrix indices
```

```
for each HS6 product p:
    W[p] = sparse matrix of bilateral values for p

W_total = sum_p W[p]

exports = row_sums(W_total)
imports = column_sums(W_total)
m_global = elementwise_max(exports, imports)

inv_m[i] = 1 / m_global[i] if m_global[i] > 0 else 0

B_total = row_scale(W_total, inv_m)

for each HS6 product p:
    B[p] = row_scale(W[p], inv_m)

for each HS2 chapter h:
    B_HS2[h] = sum_{p with prefix h} B[p]

assert max_abs(B_total - sum_h B_HS2[h]) <= tolerance
assert diagonal(B_total) == 0
assert max(row_sums(B_total)) <= 1 + tolerance
assert spectral_radius(B_total) < 1

save W_total, B_total, m_global, W[p], B[p]
```

The crucial line is the construction of m_global before the product loop. Moving the denominator calculation inside the product loop would produce a different model.

### 3.14 From the validated matrices to the EDT analysis

After the preceding steps, the empirical objects correspond one-for-one to the notation of Section 2. The correspondence is given in Table 3.4.

| Formal object | Empirical object |
|---|---|
| $X_t$ | Aggregate BACI raw-flow matrix $W_t$ |
| $X_t^{(g)}$ | Raw HS2 matrix $W_t^{(h)}$ |
| $Q_t = \mathrm{diag}\left(q_{i,t}\right)$ | $Q_t = \mathrm{diag}\left(q_{i,t}\right)$; empirical zero-row convention implemented by $\Theta_t$ |
| $\Phi_t$ | Aggregate normalized matrix $B_t$ |
| $\Phi_t^{(g)}$ | Normalized HS2 matrix $B_t^{(h)}$ |
| $I - \Phi_t$ | Annual Ghosh matrix $I - B_t$ |

**Table 3.4.** Correspondence between the formal EDT objects and their empirical counterparts.

The structural analysis can therefore start without any further transformation. For each year,

$$D_t = \det(I - B_t),$$

and the HS2 layers satisfy

$$B_t = \sum_{h \in \mathcal{H}} B_t^{(h)}.$$

The separated multilayer determinant (the block-diagonal determinant of Section 2) is

$$D_t^{\diamond} = \prod_{h \in \mathcal{H}} \det\left(I - B_t^{(h)}\right),$$

layer suppression uses

$$B_{-h,t} = B_t - B_t^{(h)},$$

and pairwise interactions use

$$B_{-\{h,g\},t} = B_t - B_t^{(h)} - B_t^{(g)}.$$

No estimated network, thresholded network, or binary projection intervenes between BACI and these EDT objects. Every coefficient entering the analysis can be traced back to a raw bilateral trade value and to the single country-specific denominator computed from aggregate annual trade.

### 3.15 Reproducibility, archived scripts, and numerical precision

The matrix-construction routine used for the article is archived in a first script. It implements the preceding procedure using pandas, NumPy, SciPy sparse matrices, h5py, and optional pyarrow acceleration. Its default period is 1995–2023. The HS6-to-HS2 aggregation and the first structural calculations are implemented in a second script. Both scripts are available from the author on request.

The code is optimized for a large dataset, but none of the optimizations changes the algebra. Parallelization is performed across years. Sparse CSR matrices are used for storage and row scaling. The country mapping is created once and reused. HDF5 compression reduces disk usage. All calculations of structural coefficients use double-precision floating-point values.

For reproducibility, the following outputs should be retained together with the scripts: the BACI release identifier, the permanent country mapping, the annual vector of normalization denominators, the aggregate raw and normalized matrices, and the normalized HS6 layers. These objects are sufficient to reconstruct every HS2 matrix used in the article and to audit the normalization independently.

A replication from the same BACI release should satisfy the matrix identities to floating-point precision rather than merely reproduce rounded published indicators. In particular, the central validation condition is

$$\left\| B_t - \sum_h B_t^{(h)} \right\|_{\max} \approx 0$$

for every year. In our implementation, the observed reconstruction error is of the order of machine precision. This is the empirical condition that ensures that the multilayer analysis is performed on additive components of one and the same normalized world-trade structure.

The resulting annual sequence

$$\left( B_t, \{B_t^{(h)}\}_{h \in \mathcal{H}} \right)_{t=1995}^{2023}$$

is the dataset used in the remainder of the article. Section 4 uses these matrices to identify the long-run evolution of aggregate diffusion and circular interdependence before progressively opening the observed change by feedback length, product layer, and interaction between layers.

## 4. Aggregate structural evolution of the World Trade Web, 1995–2023

The previous sections established the formal meaning of the EDT determinant and described the construction of the annual normalized matrices. We now turn to the empirical evolution of the aggregate World Trade Web (WTW). The purpose of this section is deliberately restricted. Before opening the network by product layers, we ask what can already be learned from the sequence of aggregate structures $B_t$.

The distinction is important. At this stage, the object of interest is not the value of world trade, the number of bilateral links, or the evolution of any conventional network statistic. It is the determinant

$$D_t = \det(I - B_t),$$

and, under the zero-diagonal convention used throughout the article, its complement

$$R_t = 1 - D_t.$$

As established in Section 2, $D_t$ is Lantner's diffusion index and $R_t$ measures circular structural interdependence. An increase in $D_t$ therefore means that the feedback architecture of world trade has become less constraining and less circular. A decrease in $D_t$ means that feedback has intensified. Because the matrices are normalized year by year by the larger of exports and imports of each country, this movement is not a direct measure of the expansion or contraction of world trade in value. It is a change in the architecture of normalized dependence.

The aggregate analysis proceeds in three steps. We first verify that the trace expansion converges sufficiently to reconstruct the determinant at each date. We then identify the main phases of the 1995–2023 trajectory. Finally, we use the exact log-determinant decomposition to determine whether the observed transformations are carried mainly by short feedbacks or also require longer closed walks. A separate diagnostic is devoted to 2017, because that year contains a large but highly localized discontinuity.

**4.1 Numerical domain and convergence of the EDT decomposition**

For every year from 1995 to 2023, the empirical matrix satisfies the conditions required by the EDT formalism. The global country support contains 238 positions. Countries with no trade in a given year simply generate zero rows and columns and therefore do not affect the determinant, as shown in Section 2.12. The diagonal of $B_t$ is zero by construction, and the spectral radius is strictly below one for every year.

Across the period, the estimated spectral radius ranges from approximately 0.831 to 0.897. The highest value is observed in 1995 and the lowest in 2006. Hence the Neumann and logarithmic series used in the structural decomposition converge throughout the sample:

$$\rho(B_t) < 1 \qquad \forall t \in \{1995, \ldots, 2023\}.$$

The empirical decomposition uses the exact identity

$$\log D_t = -\sum_{k=2}^{\infty} \frac{\mathrm{Tr}\left(B_t^k\right)}{k}.$$

The finite trace reconstruction is checked against the log-determinant evaluated independently by slogdet. The relevant criterion is the residual of the whole partial sum, not merely the size of its last term. If

$$\Lambda_{K,t} = -\sum_{k=2}^{K} \frac{\mathrm{Tr}\left(B_t^k\right)}{k},$$

the absolute truncation residual is

$$\left|\log D_t - \Lambda_{K,t}\right|$$

For the reported stopping orders, which range from 119 to 199 across years, this residual is of the order of $10^{-11}$. These are finite reconstructions of an exact infinite-series identity. The residual measures their achieved numerical accuracy; it is not the machine epsilon of double-precision arithmetic. Replication outputs should retain the stopping tolerance, the number of consecutive checks and the final residual.

The large number of orders required for strict numerical convergence should not be interpreted as evidence that very long feedbacks dominate the WTW. The contribution of order $k$ is penalized by $1/k$ and by repeated multiplication of normalized coefficients smaller than one. High orders become progressively small. Yet because the spectral radius is relatively close to one, many small terms must be accumulated before the residual reaches the reported truncation accuracy. Economically, the relevant question is therefore not the maximum $k$ required for numerical closure but how rapidly the cumulative contribution approaches the observed structural change. We address this point in Section 4.4.

## 4.2 The aggregate trajectory: no monotonic globalization of structural interdependence

Figure 4.1 reports the annual sequence of $D_t$ from 1995 to 2023. The most immediate result is that structural interdependence does not follow a monotonic trajectory.

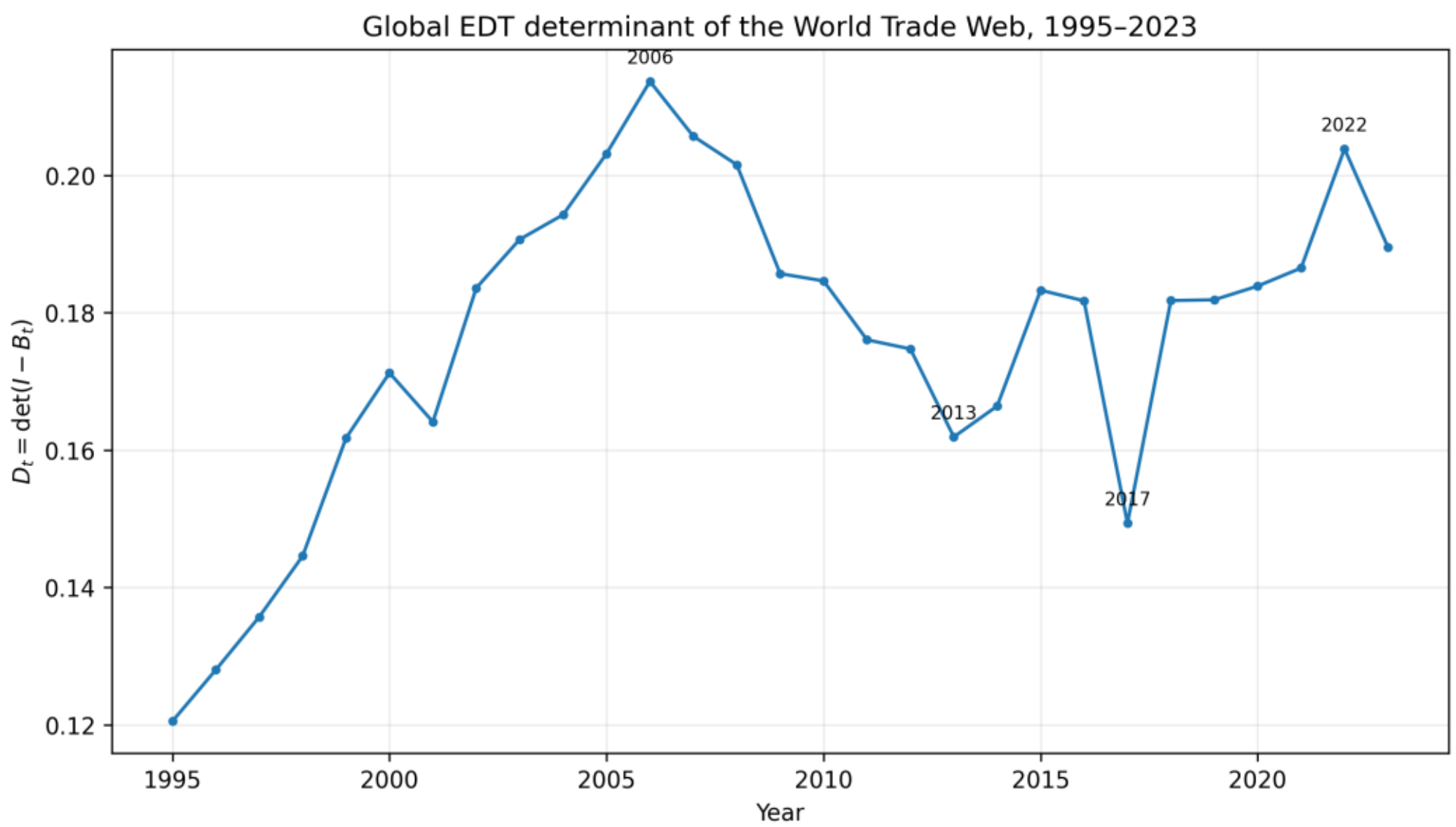

**Figure 4.1.** Global EDT determinant of the World Trade Web, 1995–2023. A higher determinant means greater diffusion and lower circular structural interdependence.

*Source: author's calculations from BACI.*

The period begins with a particularly low determinant. In 1995,

$$D_{1995} = 0.120537,$$

which implies

$$R_{1995} = 1 - D_{1995} = 0.879463.$$

By 2006, the determinant has risen to

$$D_{2006} = 0.213663,$$

and circular interdependence has fallen to

$$R_{2006} = 0.786337.$$

This is the highest value of $D_t$ in the entire sample and therefore the point of minimum circular interdependence. The first major fact emerging from the EDT representation is consequently counterintuitive if the growth of world trade is automatically equated with increasing structural interdependence: the first decade of the sample is characterized by a strong *decline* in circular interdependence.

The movement is not perfectly monotonic. The determinant falls from 0.171240 in 2000 to 0.164113 in 2001, before rising sharply to 0.183598 in 2002. This one-year reversal does not alter the underlying movement. Between 1995 and 2006,

$$\Delta D_{1995,2006} = 0.093126,$$

which is a large change relative to the initial level of the determinant.

The direction reverses after 2006. By 2009,

$$D_{2009} = 0.185680,$$

and by 2013,

$$D_{2013} = 0.161916.$$

Thus, between 2006 and 2013,

$$\Delta D_{2006,2013} = -0.051746.$$

Circular interdependence rises from 0.786337 to 0.838084. The decrease begins before 2009 and continues after it. The 2008–2009 episode is visible — $D$ falls from 0.201535 to 0.185680 — but the EDT trajectory is broader than a single crisis year. The aggregate structure enters a seven-year phase during which feedbacks progressively strengthen.

A second reversal occurs after 2013. The determinant rises to 0.183272 in 2015 and remains close to that level in 2016. The net movement from 2013 to 2016 is

$$\Delta D_{2013,2016} = 0.019805.$$

The 2017 observation then produces an exceptional one-year fall to 0.149331, immediately followed by a return to 0.181764 in 2018. As shown in Section 4.5, this discontinuity does not represent a broad reorganization of the WTW and can be isolated almost entirely within one narrowly defined product configuration. For the analysis of medium-run phases, 2016 and 2018 are therefore remarkably close:

$$D_{2016} = 0.181721, \qquad D_{2018} = 0.181764.$$

From 2018 onward, the aggregate structure is comparatively stable until 2021. The determinant is 0.181876 in 2019, 0.183873 in 2020 and 0.186498 in 2021. The pandemic year therefore does not appear as a major break in the normalized feedback architecture. This does not mean that trade volumes were unaffected. It means that the contraction and reorganization of raw flows did not translate into a discontinuity of the same order in the EDT structure.

The most visible late-sample movement occurs in 2022, when

$$D_{2022} = 0.203849,$$

before falling to

$$D_{2023} = 0.189547.$$

The final year thus partly reverses the 2022 increase. Over the entire 1995–2023 period, however, the determinant remains substantially above its initial value:

$$D_{2023} - D_{1995} = 0.069010.$$

Equivalently,

$$R_{2023} - R_{1995} = -0.069010.$$

The long-run direction of the WTW in the EDT sense is therefore toward lower circular interdependence, but this result is generated by a sequence of reversals rather than by a continuous trend.

Table 4.1 summarizes selected benchmark years.

| Year | $D_t$ | $1 - D_t$ | $\rho(B_t)$ |
|---|---|---|---|
| 1995 | 0.120537 | 0.879463 | 0.896629 |
| 2000 | 0.171240 | 0.828760 | 0.853093 |
| 2006 | 0.213663 | 0.786337 | 0.830688 |
| 2009 | 0.185680 | 0.814320 | 0.846068 |
| 2013 | 0.161916 | 0.838084 | 0.862011 |
| 2016 | 0.181721 | 0.818279 | 0.852602 |
| 2018 | 0.181764 | 0.818236 | 0.854134 |
| 2021 | 0.186498 | 0.813502 | 0.852724 |
| 2022 | 0.203849 | 0.796151 | 0.840734 |
| 2023 | 0.189547 | 0.810453 | 0.852550 |

**Table 4.1.** Selected values of the aggregate EDT structure.

*Source: author's calculations from BACI.*

Two observations should be kept distinct. First, the level of $D_t$ identifies the intensity of circular feedback at each date. Second, the direction of change identifies whether feedbacks are being strengthened or relaxed. A year can therefore remain highly interdependent in level while moving toward greater diffusion, or the reverse. This distinction becomes important when interpreting the structural phases.

**4.3 An endogenous periodization of structural change**

The sequence of determinants suggests four main phases, plus the exceptional 2017 observation. The boundaries are not imposed from an external chronology. They are selected from reversals in the EDT trajectory itself.

The first phase, 1995–2006, is a long movement of structural relaxation. The determinant increases by 0.093126 and its logarithm by

$$\delta_{1995,2006} = \log D_{2006} - \log D_{1995} = 0.572443.$$

Because $D$ rises, the cumulative weight of closed feedback walks falls. The WTW becomes less circular in the precise EDT sense.

The second phase, 2006–2013, reverses approximately half of the preceding change in determinant levels. We obtain

$$\delta_{2006,2013} = -0.277319.$$

The sign change means that the normalized network develops stronger feedback closure. Importantly, this process is gradual. The determinant falls in 2007 and 2008, falls much more strongly in 2009, remains nearly flat in 2010, then continues downward through 2013. The structural change therefore cannot be reduced to a single shock date.

The third phase, 2013–2016, is shorter but again positive:

$$\delta_{2013,2016} = 0.115394.$$

Most of the increase occurs in 2014–2015. The determinant then moves only slightly between 2015 and 2016.

The fourth phase is most clearly read from 2018 onward because of the isolated 2017 discontinuity. Between 2018 and 2023,

$$\delta_{2018,2023} = 0.041932.$$

The net movement is toward greater diffusion, but the endpoint hides substantial late variation: a sharp increase in 2022 is partially reversed in 2023. This final phase is therefore not a stable trend comparable in magnitude to 1995–2006. It is better characterized as a relatively stable structural plateau punctuated by shorter movements.

Over the complete sample,

$$\delta_{1995,2023} = \log D_{2023} - \log D_{1995} = 0.452683.$$

Using the exact relationship established in Section 2,

$$D_{2023} = D_{1995} \exp(0.452683),$$

which reproduces the observed endpoint $D_{2023} = 0.189547$. The logarithmic change is used only because it is exactly additive across feedback orders. The substantive quantity remains the determinant itself.

This periodization has an important methodological implication. If structural change were inferred from the evolution of trade values alone, major episodes would naturally be organized around expansions and contractions of aggregate trade. The EDT instead lets the structure determine its own chronology. Some historically salient years appear clearly in $D_t$; others do not. In particular, 2020 is much less conspicuous structurally than 2009, 2022 or 2023. This difference follows from the scale invariance of the normalization: a common or proportionate contraction of flows does not mechanically generate structural change.

### 4.4 Which feedback orders carry structural change?

The determinant trajectory tells us when circular interdependence rises or falls. The trace decomposition tells us *how* the feedback architecture changes.

For any interval from date $a$ to date $b$,

$$\delta_{a,b} = \log D_b - \log D_a = -\sum_{k=2}^{\infty} \frac{\mathrm{Tr}\left(B_b^k\right) - \mathrm{Tr}\left(B_a^k\right)}{k}.$$

For descriptive purposes, define the finite partial sum

$$\Lambda_K(a,b) = -\sum_{k=2}^{K} \frac{\mathrm{Tr}\left(B_b^k\right) - \mathrm{Tr}\left(B_a^k\right)}{k}.$$

$\Lambda_K$ is not an additional structural index; it is the difference between the single-date partial sums of Section 4.1 evaluated at the two dates. It is simply the part of the exact change in $\log D$ recovered when closed walks up to order $K$ are included. As $K$ tends to infinity,

$$\Lambda_K(a,b) \to \delta_{a,b}.$$

Figure 4.2 reports this accumulation for the four main phases. The vertical axis expresses the partial sum as a fraction of the exact phase change. A value of one therefore means that the included feedback orders reconstruct the full observed change in $\log D$.

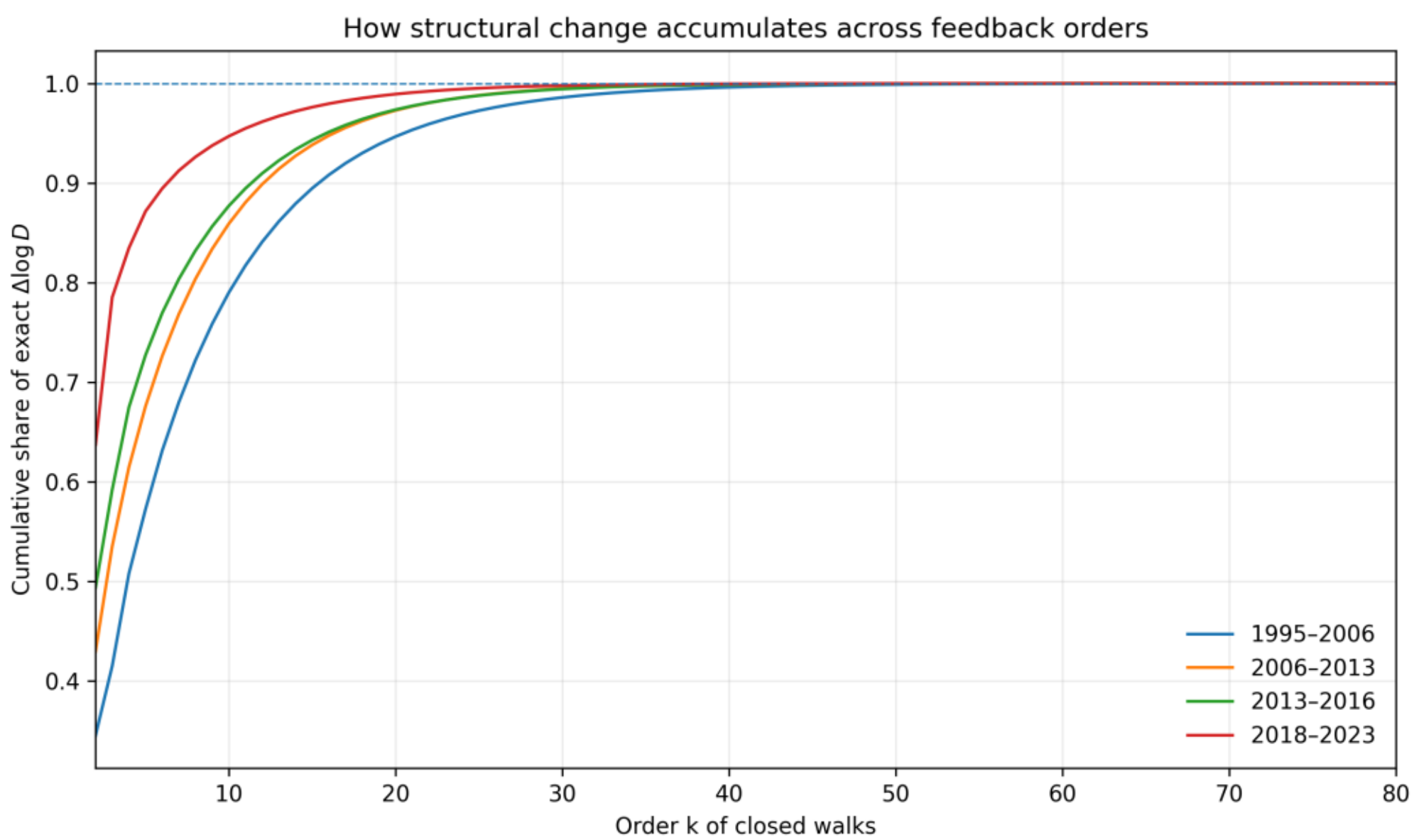

**Figure 4.2.** Cumulative reconstruction of structural change by the order of closed feedback walks.

*Source: author's calculations from BACI.*

The first result is that bilateral feedbacks are important but never sufficient. At order two,

$$-\frac{\Delta \mathrm{Tr}(B^2)}{2}$$

accounts for 34.5% of the 1995–2006 change, 43.0% of the 2006–2013 change, 49.4% of the 2013–2016 change, and 63.7% of the 2018–2023 change. The two-country reciprocal structures captured by $B^2$ therefore explain a substantial fraction of change, especially in the most recent phase, but the aggregate transformation cannot generally be reduced to bilateral reciprocity.

Adding orders three and four raises the cumulative share to 50.8% in 1995–2006, 61.5% in 2006–2013, 67.5% in 2013–2016 and 83.5% in 2018–2023. Thus, low-order feedbacks become increasingly informative as we move through the sample.

The contrast is even clearer when we ask how many orders are required to recover a given fraction of the exact phase change. In 1995–2006, the cumulative decomposition reaches 50% only at order 4, 80% at order 11, 90% at order 16 and 95% at order 21. In 2018–2023, by contrast, order 2 alone already exceeds 50%; 80% is reached by order 4, 90% by order 7 and 95% by order 11.

This does not mean that the WTW itself becomes a collection of shorter simple circuits. The trace terms count closed walks, not only elementary circuits, and higher orders include repetitions and combinations of shorter feedback structures. The result is more precise: *the temporal difference between two structures becomes progressively more concentrated in low-order recurrent mechanisms*. The early transformation of the WTW required substantial changes distributed over a wider range of feedback orders; the more recent net change is much more heavily carried by short feedbacks.

Table 4.2 provides the numerical decomposition.

| **Period** | $\Delta D$ | $\Delta \log D$ | $\Lambda_2$ | $\Lambda_4$ | $\Lambda_8$ | $\Lambda_{20}$ | $\Lambda_{50}$ |
|---|---|---|---|---|---|---|---|
| 1995–2006 | 0.093126 | 0.572443 | 0.197672 | 0.290685 | 0.413513 | 0.541884 | 0.571818 |
| 2006–2013 | -0.051746 | -0.277319 | -0.119232 | -0.170440 | -0.222953 | -0.269792 | -0.277261 |
| 2013–2016 | 0.019805 | 0.115394 | 0.057006 | 0.077854 | 0.096046 | 0.112350 | 0.115363 |
| 2018–2023 | 0.007784 | 0.041932 | 0.026731 | 0.034998 | 0.038840 | 0.041479 | 0.041928 |
| 1995–2023 | 0.069010 | 0.452683 | 0.173229 | 0.239523 | 0.328648 | 0.426594 | 0.452085 |

**Table 4.2.** Structural change reconstructed by increasing orders of closed walks.

Note. $\Lambda_K$ is the partial sum of the exact log-determinant decomposition through order $K$.

*Source: author's calculations from BACI.*

The long-run 1995–2023 comparison reinforces the same point. Bilateral feedbacks account for 38.3% of the exact change. Orders up to four account for 52.9%, orders up to eight for 72.6%, and orders up to twenty for 94.2%. By order fifty, 99.9% of the long-run change is reconstructed.

There are two implications. First, the EDT reading of structural change cannot be reduced to the evolution of direct reciprocal trade. A large part of the long-run transformation is carried by longer recurrent chains. Second, the very high orders required for numerical convergence should not be confused with economic importance. Almost all economically relevant variation is already recovered well before the formal stopping order. The adaptive calculation is nevertheless useful because it verifies that no unmeasured residual remains hidden in the decomposition.

The signs are also informative. During 1995–2006, the cumulative terms are positive: the relevant traces are lower in 2006 than in 1995, so feedback mass has been removed from the structure. During 2006–2013, the cumulative terms are negative: feedback mass is rebuilt. The trace decomposition therefore gives a direct structural meaning to the reversals visible in Figure 4.1.

### 4.5 The 2017 discontinuity: a localized feedback configuration, not a global regime shift

The aggregate series contains one observation that is too large and too rapidly reversed to be treated like the other movements. Between 2016 and 2017,

$$D_t: 0.181721 \rightarrow 0.149331,$$

so

$$\Delta D_{2016,2017} = -0.032390,$$

and

$$\delta_{2016,2017} = -0.196309.$$

The following year the movement is almost exactly reversed:

$$D_{2018} = 0.181764,$$

with

$$\delta_{2017,2018} = 0.196543.$$

A one-year V-shaped movement of this magnitude raises a natural question: does 2017 reveal a genuine global reorganization of trade feedbacks, or is it generated by a small number of exceptional flows that have a disproportionate structural effect after normalization?

The EDT framework allows this question to be answered without deleting the observation. We successively suppress product layers and recompute the determinant. The discontinuity is almost entirely associated with HS chapter 89, *Ships, boats and floating structures*. Figure 4.3 compares the original determinant with the determinant after suppressing HS89.

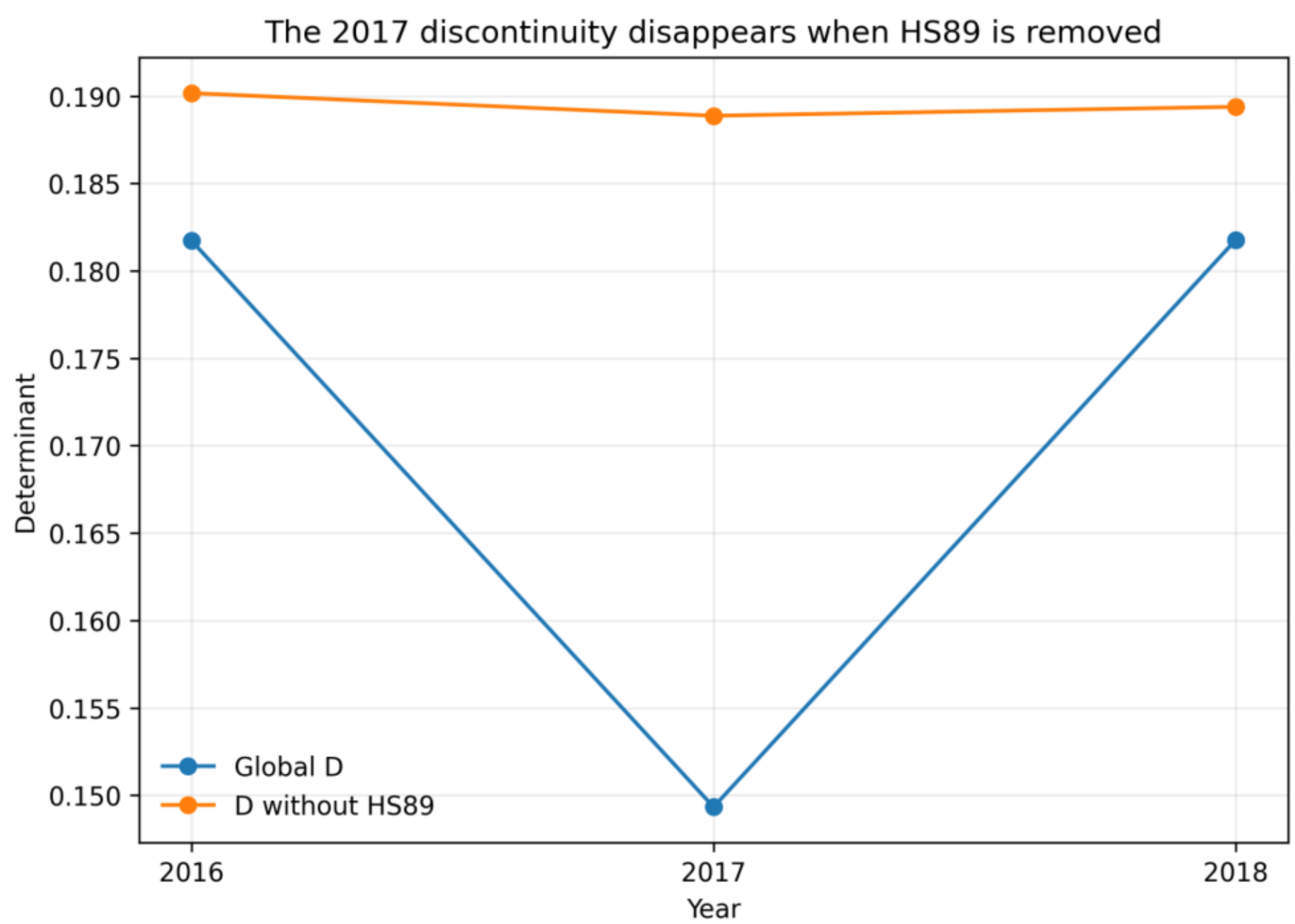

**Figure 4.3.** Diagnostic of the 2017 discontinuity. Removing HS89 almost completely eliminates the one-year fall in the aggregate determinant.

*Source: author's calculations from BACI.*

The contrast is striking. With all products included,

$$D_{2016} = 0.181721, \qquad D_{2017} = 0.149331, \qquad D_{2018} = 0.181764.$$

After suppressing HS89,

$$D_{-89,2016} = 0.190166,$$
$$D_{-89,2017} = 0.188872,$$

and

$$D_{-89,2018} = 0.189384.$$

The three years become almost continuous. The 2017 anomaly is therefore not a diffuse property of the world trade matrix. It is a localized structural configuration carried by one product chapter.

This result is particularly revealing because HS89 is not unusually large in terms of aggregate trade value in 2017. Its share of world trade in the BACI matrix is approximately 0.88%, lower than its 0.99% share in 2016. Yet its order-two trace changes from

$$\mathrm{Tr}[(B_{2016}^{89})^2] = 0.015403$$

to

$$\mathrm{Tr}[(B_{2017}^{89})^2] = 0.354305,$$

before returning to

$$\mathrm{Tr}[(B_{2018}^{89})^2] = 0.012951.$$

The episode is thus a clean illustration of the distinction between volume and structure. A product category representing less than one percent of raw world trade can strongly alter the global determinant if its flows close very heavy normalized feedbacks.

The audit can be pushed one level further. Within HS89, the HS6 category 890392 – motorboats other than outboard motorboats, for pleasure or sports, other than inflatable – contributes approximately 0.348050 to the 2017 order-two trace. This represents about 98.2% of the total HS89 order-two feedback in that year. The discontinuity is therefore not even a general property of ship trade.

At the country-pair level, two reciprocal relationships dominate. The Cayman Islands-Seychelles pair contributes 0.249741 to the HS89 order-two trace, or approximately 70.5%. The British Virgin Islands-Seychelles pair contributes 0.098064, or approximately 27.7%. Together they account for about 98.2% of the chapter's order-two feedback.

These facts lead to a precise empirical conclusion. The 2017 observation is *real in the data and real for the determinant*, but it should not be interpreted as evidence of a one-year global regime shift. It is the structural consequence of a highly concentrated reciprocal configuration. We therefore retain 2017 in the reported series — removing it ex ante would amount to editing the network according to the desired result — but we treat it separately when identifying medium-run phases.

The episode also demonstrates an important property of EDT diagnostics. The global determinant first detects an unusual change. Layer suppression identifies the product family responsible for it. The trace decomposition identifies the feedback order through which it operates. Disaggregation within the layer identifies the HS6 product, and bilateral inspection identifies the country pairs closing the feedback. The analysis moves from the global structure to the elementary flows without switching conceptual frameworks.

### 4.6 What the aggregate analysis establishes before opening the product layers

Four empirical results can be established at the aggregate level alone.

First, the WTW does not display a monotonic rise in circular structural interdependence. Over the full period, the determinant increases from 0.120537 to 0.189547 and $1 - D$ correspondingly falls. The long-run result is therefore one of weaker circular interdependence in 2023 than in 1995, notwithstanding the large expansion of international trade that occurred over the same historical period.

Second, this long-run result is generated by distinct structural phases. A strong relaxation of feedback between 1995 and 2006 is followed by a substantial rebuilding of feedback between 2006 and 2013, a renewed relaxation between 2013 and 2016, and a comparatively stable post-2018 period with sharp shorter movements in 2022 and 2023. The EDT periodization is therefore endogenous to the feedback structure rather than imposed from a chronology of external events.

Third, structural change is not confined to bilateral reciprocity. Order-two feedback is important, but longer closed walks account for a large fraction of the transformations, particularly in the earlier phases. The exact trace decomposition shows, however, that the more recent net change is increasingly concentrated in lower orders. This difference in the distribution of change across feedback lengths is itself a feature of structural transformation.

Fourth, large movements in the determinant need not correspond to large movements in aggregate trade values. The 2017 case shows the reverse possibility with unusual clarity: a narrow product-country configuration carrying less than one percent of raw trade can temporarily alter the aggregate determinant because it closes very strong normalized feedback loops. Conversely, a major volume shock such as 2020 can leave the normalized structural architecture comparatively stable.

These results already justify separating the structural analysis of the WTW from a purely volumetric reading of globalization. They do not yet tell us *which product categories* generate the medium-run transformations. Nor do they tell us whether the observed changes occur inside individual product networks or through the combination of different products within the same feedback mechanisms. Those are multilayer questions. Section 5 therefore opens the aggregate determinant by product layers and uses the exact decomposition developed in Section 2 to distinguish within-layer from genuinely multilayer structural change.

## 5. The multilayer transformation of the World Trade Web

Section 4 established that the structural interdependence of the World Trade Web (WTW) does not follow a monotonic trajectory. The determinant rises markedly between 1995 and 2006, falls between 2006 and 2013, partially recovers thereafter, and displays a short-lived anomaly in 2017. The aggregate analysis also showed that these movements are not confined to bilateral reciprocity: changes in longer closed walks account for a substantial fraction of the variation in $D_t$.

The aggregate matrix, however, does not reveal whether these changes take place *inside* product-specific trade networks or in the way different product networks combine. The multilayer construction developed in Section 2 makes this distinction exact. At every date,

$$\Phi_t = \sum_{g \in \mathcal{W}} \Phi_t^{(g)},$$

while the block-diagonal representation that keeps product layers structurally separate is

$$\Phi_t^{\diamond} = \oplus_{g \in \mathcal{W}} \Phi_t^{(g)}.$$

The corresponding determinants are

$$D_t = \det(I - \Phi_t)$$

and

$$D_t^{\diamond} = \det(I - \Phi_t^{\diamond}) = \prod_{g \in \mathcal{W}} \det\left(I - \Phi_t^{(g)}\right).$$

The comparison between these two objects is the starting point of this section. $D_t^{\diamond}$ allows feedback only within a product layer; $D_t$ additionally allows the product carried by an arc to change along a closed walk. The difference between them therefore reveals the structural importance of product combinations. We first examine this difference in levels, then decompose structural change exactly into within-layer and genuinely multilayer components, and finally identify the product categories and layer interactions through which the multilayer architecture is reorganized.

### 5.1 The aggregate WTW is structurally different from the collection of its product networks

The first empirical fact is the size of the gap between the separated-layer and aggregate representations. Table 5.1 reports selected years. Except for the exceptional 2017 episode examined in Section 4, $D_t^{\circ}$ remains in a narrow band around 0.91–0.93. By contrast, the determinant of the aggregate network lies between approximately 0.12 and 0.21 over most of the period.

| Year | $D_t$ | $D_t^{\circ}$ | $\kappa_t = D_t^{\circ} - D_t$ |
|---|---|---|---|
| 1995 | 0.1205 | 0.9267 | 0.8061 |
| 2000 | 0.1712 | 0.9131 | 0.7418 |
| 2006 | 0.2137 | 0.9294 | 0.7158 |
| 2009 | 0.1857 | 0.9304 | 0.7447 |
| 2013 | 0.1619 | 0.9127 | 0.7508 |
| 2016 | 0.1817 | 0.9198 | 0.7381 |
| 2018 | 0.1818 | 0.9186 | 0.7369 |
| 2020 | 0.1839 | 0.9250 | 0.7411 |
| 2022 | 0.2038 | 0.9180 | 0.7142 |
| 2023 | 0.1895 | 0.9204 | 0.7309 |

**Table 5.1.** Aggregate determinant, separated-layer determinant and multilayer heterogeneity, selected years.

Note. Differences and shares are computed from unrounded values; the last displayed digit may therefore differ from the difference of the displayed entries.

*Source: author's calculations from BACI.*

The interpretation is direct. Product-specific networks considered separately are only weakly circular relative to the aggregate trade system. Much of the circular interdependence observed in the WTW appears only when successive international flows are allowed to belong to different product categories. In other words, global trade is not structurally equivalent to a collection of independent commodity networks placed side by side. Its feedback architecture is produced by their combination.

The heterogeneity quantity of Lebert and Poindron (2025),

$$\kappa_t = D_t^{\circ} - D_t,$$

is correspondingly large throughout the period. This is a level difference, not a share: it should not be read as the percentage of interdependence attributable to multilayer feedback. Its role is to measure the determinant gap generated by aggregation of the product layers. The trace decomposition below provides the exact additive separation in log-determinant space.

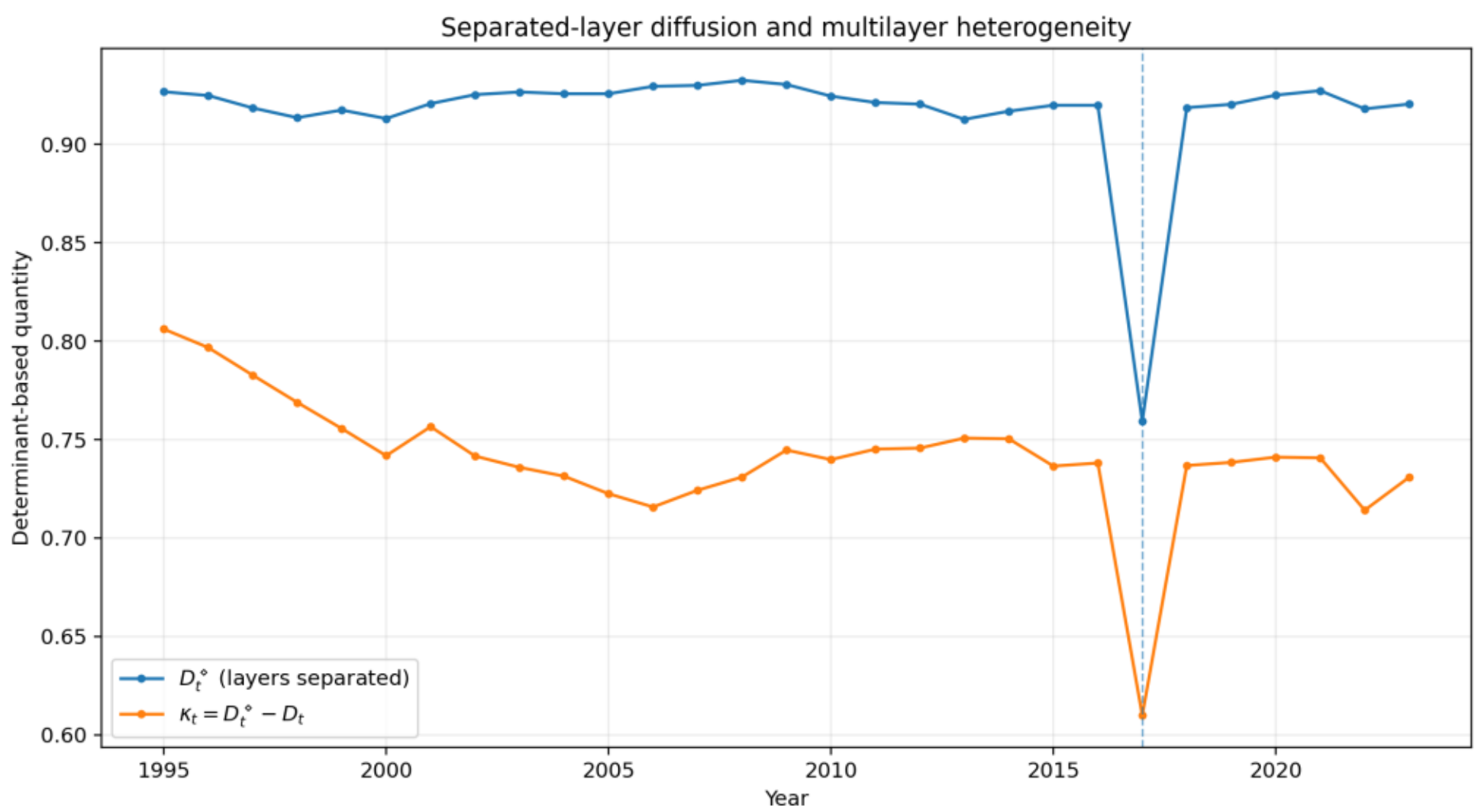

**Figure 5.1.** Separated-layer diffusion $D_t^{\circ}$ and multilayer heterogeneity $\kappa_t$.

*Source: author's calculations from BACI.*

Figure 5.1 also clarifies the 2017 anomaly. The separated-layer determinant falls from 0.9198 in 2016 to 0.7592 in 2017 before returning to 0.9186 in 2018. Section 4 showed that this movement originates overwhelmingly in HS89 and, within that layer, in HS890392. The fall in $\kappa_t$ in 2017 is therefore not evidence of a sudden disappearance of multilayer structure. It is the mechanical consequence of an exceptional increase in circularity *inside one layer*. This provides a useful empirical illustration of why $D_t^\diamond$ and $D_t$ must be distinguished.

### 5.2 Structural change is overwhelmingly a change in cross-product feedback

The level comparison can be converted into an exact temporal decomposition. From Section 2,

$$\log D_t = \log D_t^\diamond + (\log D_t - \log D_t^\diamond).$$

Between two dates $a$ and $b$,

$$\log D_b - \log D_a = (\log D_b^\diamond - \log D_a^\diamond) + [(\log D_b - \log D_b^\diamond) - (\log D_a - \log D_a^\diamond)].$$

We denote the first term as the within-layer or monolayer component and the second as the genuinely multilayer component. These labels do not introduce new indices; they identify the two exact terms of the equality.

Figure 5.2 reports the annual decomposition. With one conspicuous exception, the multilayer component lies very close to the total annual change in $\log D_t$. The exception is 2017–2018, where the HS89 event produces a large monolayer movement that reverses almost completely one year later.

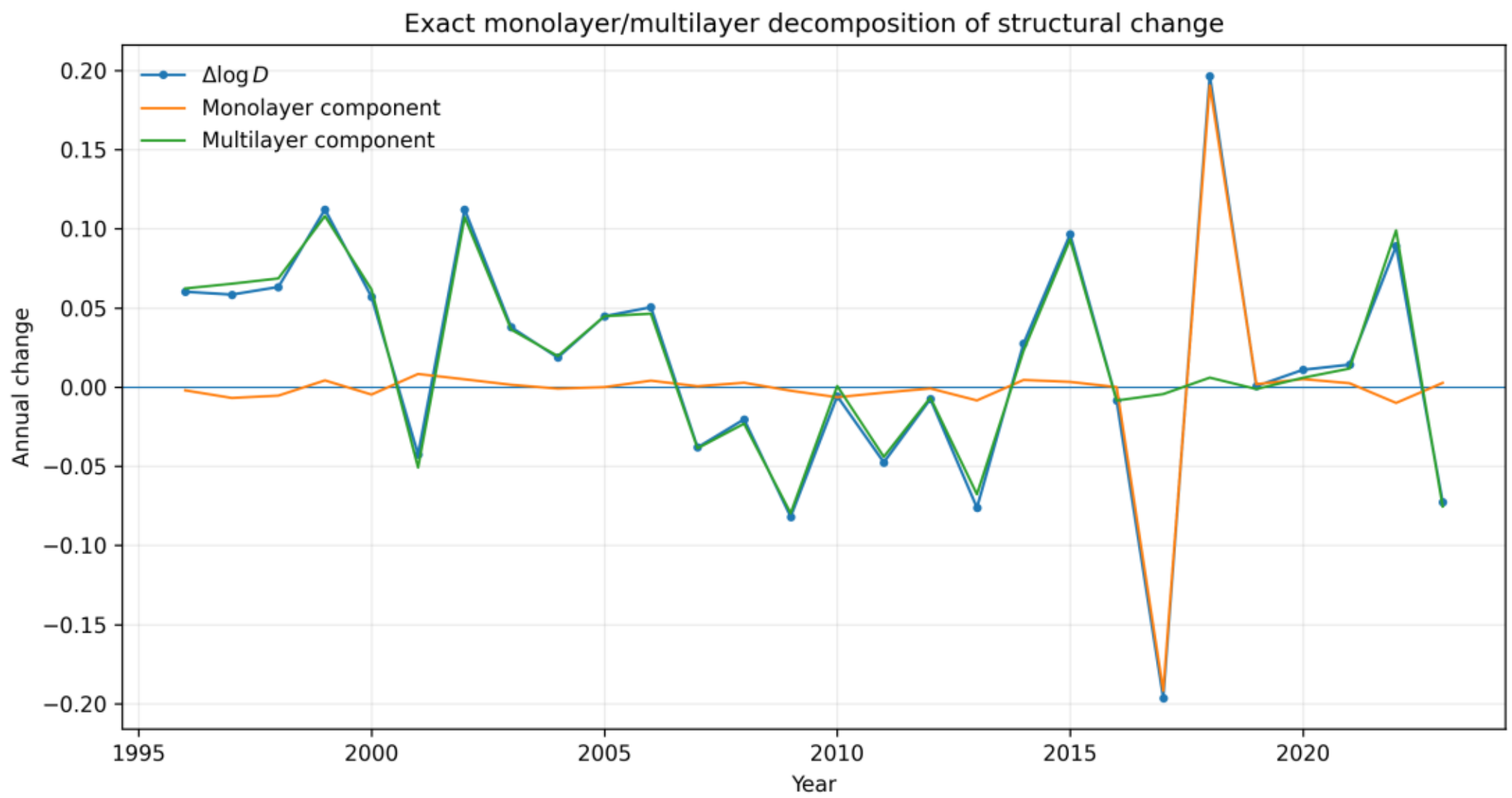

**Figure 5.2.** Exact annual decomposition of $\Delta\log D_t$ into within-layer and genuinely multilayer components.

*Source: author's calculations from BACI.*

The main structural phases identified in Section 4 make the result especially clear. Table 5.2 reports the exact decomposition over each phase.

| **Period** | *ΔD* | *Δ*log*D* | **Within-layer** | **Multilayer** |
|---|---|---|---|---|
| 1995–2006 | 0.0931 | 0.5724 | 0.0030 | 0.5695 |
| 2006–2013 | -0.0517 | -0.2773 | -0.0182 | -0.2591 |
| 2013–2016 | 0.0198 | 0.1154 | 0.0078 | 0.1076 |
| 2018–2023 | 0.0078 | 0.0419 | 0.0020 | 0.0399 |
| 1995–2023 | 0.0690 | 0.4527 | -0.0068 | 0.4594 |

**Table 5.2.** Exact within-layer and multilayer decomposition of structural change over the main phases.

*Source: author's calculations from BACI.*

Between 1995 and 2006, the aggregate log-determinant rises by 0.5724. The within-layer contribution is only 0.0030, whereas the multilayer contribution is 0.5695. More than 99% of the change therefore has the same origin: cross-product feedback becomes substantially weaker. The increase in diffusion observed during the first decade is not primarily due to product networks becoming less circular internally. It is due to a reorganization of the combinations through which different products close international feedback loops.

The reversal after 2006 has the same structural character. Between 2006 and 2013, $\log D$ falls by 0.2773. Within-layer feedback accounts for -0.0182, while the multilayer component accounts for -0.2591. The renewed rise in circular interdependence is again overwhelmingly a cross-product phenomenon.

The same conclusion holds for 2013–2016 and 2018–2023, for which the multilayer component accounts for approximately 93% and 95% of the total log-determinant movement respectively. The long-run comparison is even more revealing. From 1995 to 2023,

$$\Delta \log D = 0.4527,$$

but

$$\Delta \log D^{\text{within}} = -0.0068,$$

while

$$\Delta \log D^{\text{multi}} = 0.4594.$$

Taken separately, product layers become very slightly *more* circular over the period, not less. Their within-layer contribution would therefore have reduced diffusion. The aggregate WTW nevertheless becomes less circular because cross-product feedback weakens by a larger amount. The multilayer component not only explains the entire net change; it offsets a small movement in the opposite direction within individual product networks.

This result substantially changes the interpretation of structural globalization. The relevant transformation is not simply the strengthening or weakening of trade within separate commodity systems. It is the reorganization of the productive combinations embodied in reciprocal and longer international chains.

### 5.3 Multilayer feedback is already dominant at order two and becomes virtually exhaustive at longer orders

The previous decomposition aggregates all closed-walk lengths. The trace formulation makes it possible to open the multilayer component order by order. Recall that

$$\mathrm{Tr}\left(\Phi_t^k\right) = \sum_{g_1,\dots,g_k} \mathrm{Tr}\left(\Phi_t^{(g_1)} \cdots \Phi_t^{(g_k)}\right).$$

The monolayer part is

$$\sum_g \mathrm{Tr}\left[\left(\Phi_t^{(g)}\right)^k\right],$$

and the difference contains exactly the terms in which at least two layer labels differ. Since all matrices are non-negative, the ratio of the latter to the aggregate trace can be read directly as the fraction of the closed-walk mass of order $k$ that is genuinely multilayer.

| **Year** | $k=2$ | $k=3$ | $k=4$ | $k=5$ | $k=6$ | $k=7$ | $k=8$ |
|---|---|---|---|---|---|---|---|
| 1995 | 93.16% | 99.25% | 99.85% | 99.98% | 100.00% | 100.00% | 100.00% |
| 2006 | 91.98% | 99.01% | 99.85% | 99.98% | 100.00% | 100.00% | 100.00% |
| 2013 | 91.11% | 99.14% | 99.87% | 99.99% | 100.00% | 100.00% | 100.00% |
| 2018 | 91.18% | 99.03% | 99.84% | 99.98% | 100.00% | 100.00% | 100.00% |
| 2023 | 91.17% | 98.90% | 99.83% | 99.98% | 100.00% | 100.00% | 100.00% |

**Table 5.3.** Share of the aggregate closed-walk mass generated by genuinely multilayer terms, selected years and orders.

*Source: author's calculations from BACI.*

The result is striking even at the shortest possible feedback length (Table 5.3). At $k = 2$, between 91% and 93% of the trace is generated by terms of the form

$$\mathrm{Tr}\left(\Phi_t^{(g)}\Phi_t^{(h)}\right), \qquad g \neq h.$$

Economically, a two-country feedback usually does not consist of country $i$ exporting a product category to $j$ and $j$ returning the same category to $i$. It more often closes because the return flow belongs to another HS2 chapter. Bilateral reciprocity in the structural sense is therefore already predominantly cross-product.

At order three, the multilayer share rises to about 99%. From order four onward, the monolayer remainder becomes extremely small, and from order five it is practically negligible at the reported precision. Long closed feedbacks in world trade are thus almost necessarily multi-product. The multilayer character of the WTW is not a property that emerges only after many layers have been statistically aggregated. It is present in the elementary feedback mechanisms themselves and becomes stronger with path length.

This result also explains why the temporal decomposition in the previous subsection is so dominated by the multilayer term. Once most closed-walk mass is produced by combinations of product categories, changes in the architecture of those combinations have much more scope to alter $D_t$ than changes confined to one product layer.

### 5.4 Which product layers structure global feedback?

The preceding results establish that product combinations matter. They do not yet identify which layers have the largest marginal footprint in the aggregate feedback structure. For each HS2 layer $h$, we therefore use the EDT layer-suppression quantity

$$S_{h,t} = D_{-h,t} - D_t \geq 0.$$

$S_{h,t}$ measures the increase in diffusion produced by removing layer $h$ from the aggregate network. A large value means that the layer participates strongly in the feedback architecture of world trade.

A crucial qualification is required. The quantities $S_{h,t}$ are marginal effects and are *not additive shares*. The same closed feedback can involve several product layers, so removing either one may break it. The sum of $S_{h,t}$ therefore has no interpretation as a decomposition of $1 - D_t$. Interactions between layers are precisely what prevent such additivity. The purpose of $S_{h,t}$ is to rank and follow the structural footprint of individual layers while holding the remainder of the WTW fixed.

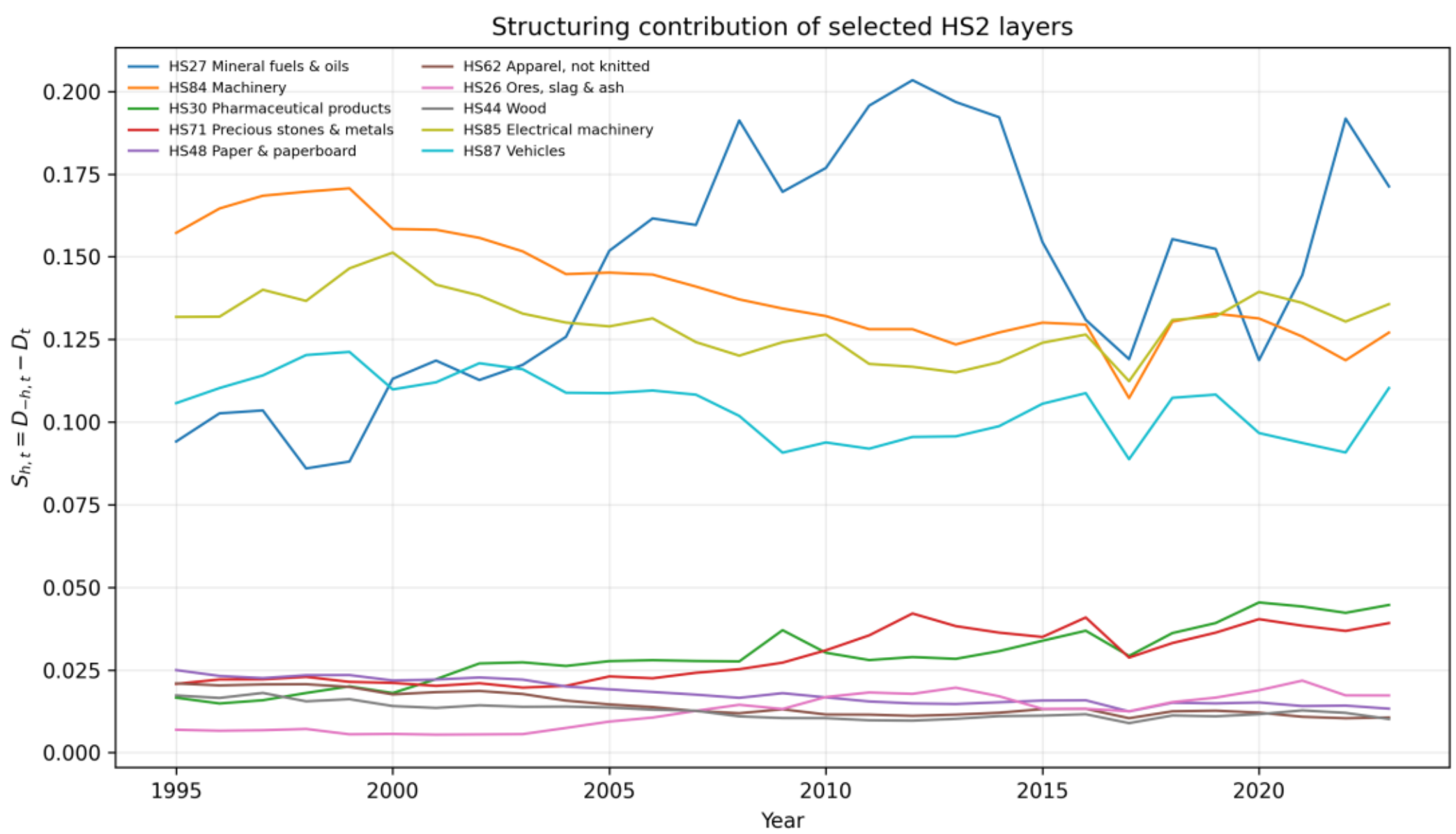

**Figure 5.3.** Marginal structuring contribution $S_{h,t} = D_{-h,t} - D_t$ for selected HS2 layers.

*Source: author's calculations from BACI.*

Figure 5.3 shows both persistence and substantial recomposition. Machinery (HS84), electrical machinery (HS85), vehicles (HS87) and mineral fuels (HS27) form the enduring upper group. Yet their relative positions change markedly. In 1995, machinery has the largest marginal structural footprint ($S_{84} = 0.1572$), followed by electrical machinery (0.1318), vehicles (0.1057), and mineral fuels (0.0941). By 2006, mineral fuels have become the leading layer ($S_{27} = 0.1616$). Their contribution continues to rise and reaches approximately 0.197 in 2013. In 2023 it remains the largest at 0.1713, ahead of electrical machinery (0.1357), machinery (0.1270), and vehicles (0.1103).

The long-run changes confirm that this is not simply a reshuffling among the four largest layers (Table 5.4).

| HS2 | Product category | Change in $S_h$, 1995–2023 |
|---|---|---|
| 27 | Mineral fuels & oils | +0.0772 |
| 84 | Machinery | -0.0302 |
| 30 | Pharmaceutical products | +0.0281 |
| 71 | Precious stones & metals | +0.0184 |
| 48 | Paper & paperboard | -0.0117 |
| 26 | Ores, slag & ash | +0.0104 |
| 62 | Apparel, not knitted | -0.0103 |
| 44 | Wood | -0.0071 |
| 52 | Cotton | -0.0065 |
| 72 | Iron & steel | -0.0065 |
| 29 | Organic chemicals | -0.0061 |
| 55 | Man-made staple fibres | -0.0049 |
| 54 | Man-made filaments | -0.0048 |
| 87 | Vehicles | +0.0046 |
| 12 | Oil seeds | +0.0044 |
| 33 | Essential oils, perfumery & cosmetics | +0.0041 |
| 85 | Electrical machinery | +0.0039 |
| 49 | Printed books & newspapers | -0.0039 |
| 41 | Raw hides & skins | -0.0038 |
| 19 | Preparations of cereals/flour/milk | +0.0035 |

**Table 5.4.** Largest long-run changes in the marginal structuring contribution of HS2 layers, 1995–2023.

*Source: author's calculations from BACI.*

Three features stand out. First, the manufacturing complex remains structurally central. Machinery, electrical machinery and vehicles are still among the four largest layers in 2023. The results therefore do not indicate a simple replacement of manufacturing by raw materials.

Second, however, energy acquires a substantially larger marginal role. HS27 records by far the largest increase in $S_h$. This means that removing energy trade breaks more of the aggregate feedback architecture in 2023 than it did in 1995. The result concerns structural position, not trade value: it does not say that fuels merely account for a larger share of world exports, but that their flow pattern has become more consequential for the closure of international feedbacks.

Third, several smaller layers move strongly in relative structural importance. Pharmaceuticals, precious stones and metals, and ores all gain. By contrast, paper, apparel, wood, cotton, several textile categories, and organic chemicals lose marginal structuring power. The long-run transformation therefore combines persistence at the top with a reorientation of the secondary layers through which the global system closes its feedback chains.

The contrast between HS30 and HS29 is particularly suggestive. Pharmaceuticals gain markedly in individual structuring contribution, while organic chemicals lose. The interaction analysis below shows that the relationship between the two also changes sign and becomes increasingly negative. The EDT therefore distinguishes a layer's own marginal structural role from its structural relationship with another layer.

### 5.5 Structural interactions between product layers

The marginal effect of one layer does not reveal whether the effects of two removals add, overlap, or reinforce one another. Section 2 defines their pairwise interaction as

$$\Gamma_t(g,h) = D_{-\{g,h\},t} - D_{-g,t} - D_{-h,t} + D_t.$$

This is a second-order discrete difference under layer suppression. When $\Gamma_t(g,h) > 0$, joint removal raises diffusion by more than the sum of separate removals. When $\Gamma_t(g,h) < 0$, it raises diffusion by less, so the separate effects overlap. Complementarity and redundancy refer here to these removal effects, as illustrated in Section 2.11, rather than directly to the productive roles of the goods.

As with $S_h$, $\Gamma$ should be interpreted structurally. A negative interaction does not establish technological substitutability in a production-function sense, nor does a positive interaction prove engineering complementarity. It identifies the sign of the interaction between two *trade-flow structures* within the EDT determinant. Economic interpretation must therefore remain consistent with the aggregation level and the nature of the data.

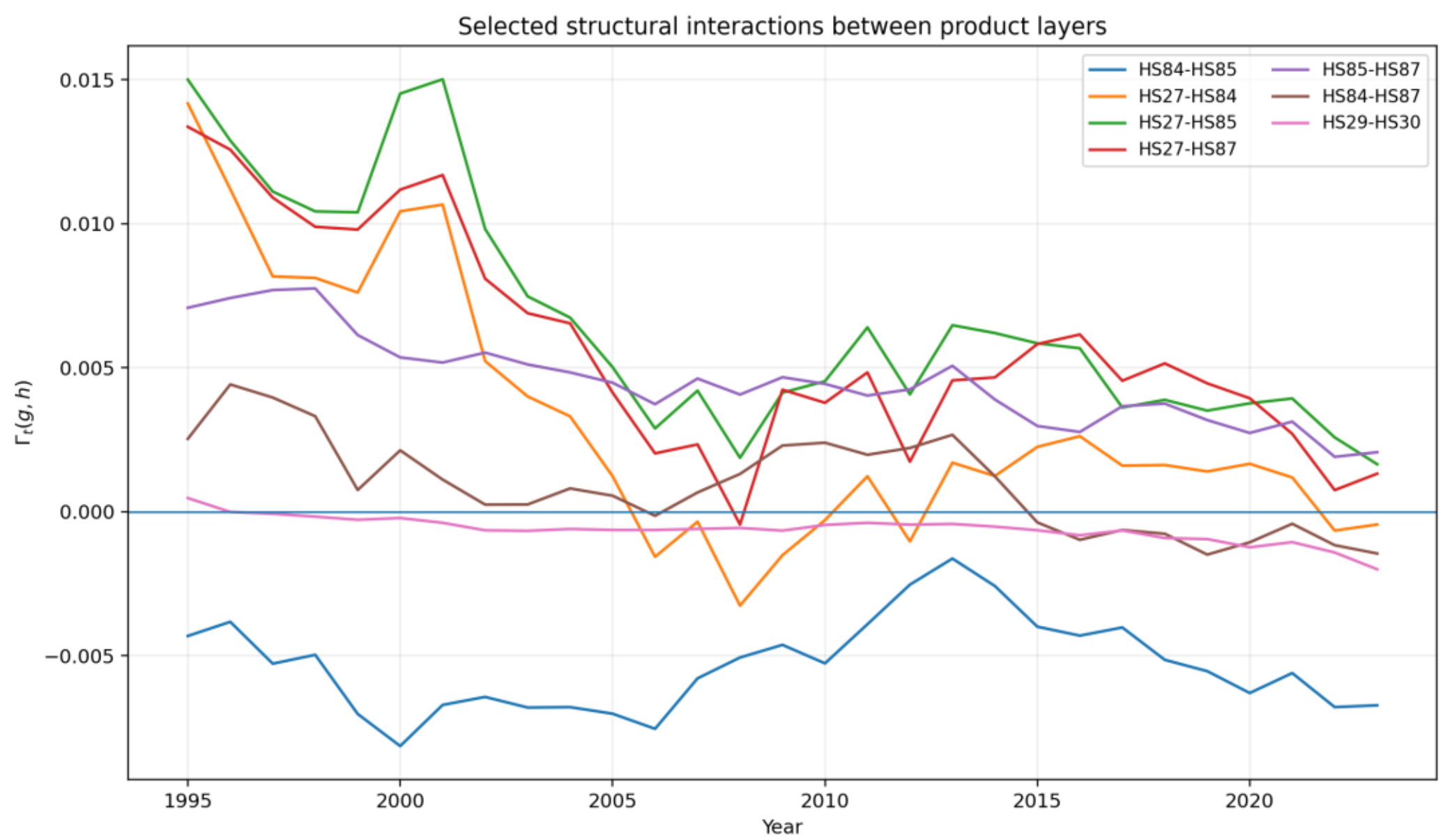

**Figure 5.4.** Selected pairwise structural interactions $\Gamma_t(g, h)$, 1995–2023.

*Source: author's calculations from BACI.*

The first major result concerns the relationships between energy and the manufacturing core (Figure 5.4). In 1995, the interactions of mineral fuels with machinery, electrical machinery and vehicles are all strongly positive:

$$\Gamma_{1995}(27,84) = 0.01418,$$

$$\Gamma_{1995}(27,85) = 0.01501,$$

and

$$\Gamma_{1995}(27,87) = 0.01337.$$

By 2023, the corresponding values are -0.00045, 0.00165 and 0.00132. The super-additivity of the energy–manufacturing removal effects therefore weakens sharply, and becomes a small overlap for machinery. This result is especially important when read together with the rise in $S_{27}$. Energy becomes more structurally important *individually* while its interaction with the traditional manufacturing core becomes much weaker. Its growing structural footprint is therefore not reducible to the specific pairwise relationships that characterized the mid-1990s.

The manufacturing complex itself also changes. The interaction between machinery and electrical machinery is negative throughout the whole period and becomes strongly negative again by 2023:

$$\Gamma_{2023}(84,85) = -0.00673,$$

the largest negative interaction among all 4,560 HS2 pairs in that year. Joint removal of electrical machinery and vehicles retains a positive interaction, but their interaction falls from 0.00708 in 1995 to 0.00206 in 2023. Machinery and vehicles move from a positive interaction in 1995 to a negative one in 2023.

A similar transformation occurs in chemicals. The interaction between organic chemicals and pharmaceuticals is slightly positive in 1995 but becomes negative from the late 1990s onward and reaches $-0.00200$ in 2023. This accompanies the opposite long-run movements in their marginal structuring contributions: HS30 rises sharply, whereas HS29 declines (Table 5.5).

| **Pair** | **1995** | **2006** | **2013** | **2016** | **2018** | **2023** | **1995–2023 change** |
|---|---|---|---|---|---|---|---|
| HS84–HS85 | -0.00432 | -0.00754 | -0.00163 | -0.00431 | -0.00515 | -0.00673 | -0.00241 |
| HS27–HS84 | 0.01418 | -0.00157 | 0.00170 | 0.00262 | 0.00162 | -0.00045 | -0.01463 |
| HS27–HS85 | 0.01501 | 0.00289 | 0.00647 | 0.00567 | 0.00388 | 0.00165 | -0.01336 |
| HS27–HS87 | 0.01337 | 0.00202 | 0.00456 | 0.00615 | 0.00515 | 0.00132 | -0.01205 |
| HS85–HS87 | 0.00708 | 0.00373 | 0.00507 | 0.00277 | 0.00375 | 0.00206 | -0.00502 |
| HS84–HS87 | 0.00253 | -0.00014 | 0.00267 | -0.00098 | -0.00077 | -0.00146 | -0.00398 |
| HS29–HS30 | 0.00047 | -0.00064 | -0.00043 | -0.00082 | -0.00092 | -0.00200 | -0.00247 |

**Table 5.5.** Selected structural interactions between product layers.

*Source: author's calculations from BACI.*

The full distribution moves in the same direction, although the simple number of positive and negative pairs should not be confused with a weighted interaction index. Of the 4,560 HS2 pairs, 390 have a negative $\Gamma$ in 1995, compared with 684 in 2023. The system therefore contains more pairs whose separate removal effects overlap at the end of the period. The magnitudes remain highly heterogeneous, however, so the sign count is reported only as a descriptive fact.

The strongest interactions in 2023 show that energy remains deeply involved in cross-layer structure despite the collapse of its old interactions with HS84–HS87. Four of the five largest positive interactions involve HS27 (Table 5.6).

| **Sign** | **Pair** | **Product layers** | $\Gamma_{2023}$ |
|---|---|---|---|
| Positive | HS27–HS39 | Mineral fuels & oils / Plastics | 0.00369 |
| Positive | HS27–HS72 | Mineral fuels & oils / Iron & steel | 0.00289 |
| Positive | HS85–HS87 | Electrical machinery / Vehicles | 0.00206 |
| Positive | HS27–HS30 | Mineral fuels & oils / Pharmaceutical products | 0.00202 |
| Positive | HS27–HS48 | Mineral fuels & oils / Paper & paperboard | 0.00176 |
| Negative | HS84–HS85 | Machinery / Electrical machinery | -0.00673 |
| Negative | HS85–HS90 | Electrical machinery / Optical, medical & precision instruments | -0.00231 |
| Negative | HS84–HS90 | Machinery / Optical, medical & precision instruments | -0.00204 |
| Negative | HS29–HS30 | Organic chemicals / Pharmaceutical products | -0.00200 |
| Negative | HS30–HS84 | Pharmaceutical products / Machinery | -0.00176 |

**Table 5.6.** Largest positive and negative pairwise structural interactions in 2023.

*Source: author's calculations from BACI.*

This pattern sharpens the interpretation of the layer-level results. The rise of HS27 is not associated with a generalized increase in super-additivity of removal effects with every industrial layer. Rather, the set of product categories with which energy contributes non-additively to the feedback architecture has itself been recomposed. Likewise, the rise of pharmaceuticals is accompanied by increasingly negative interactions with organic chemicals and machinery, while a positive relationship with energy appears among the largest interactions in 2023. These are structural statements about the organization of international flows, not claims about technological causality. Their economic meaning is revisited in Section 7, where the EDT interpretation is compared with the objects and purposes of established approaches to the World Trade Web.

### 5.6 What the multilayer decomposition adds to the aggregate history

The multilayer analysis changes the interpretation of the trajectory established in Section 4 in four ways.

First, the circularity of the aggregate WTW is overwhelmingly generated by combinations of product categories. The separated-layer determinant remains close to one, whereas aggregation produces a much lower determinant. The result is not an artifact of long paths: even two-step feedback is more than 90% multilayer, and feedback of order three or above is almost entirely so.

Second, the major temporal phases of the WTW are therefore phases of *cross-product reorganization*. The 1995–2006 weakening of structural interdependence, the 2006–2013 reversal, and the subsequent movements are not primarily produced by parallel changes within individual HS2 networks. They are produced by changes in the way different product networks connect to close international feedback chains. Over the full period, the within-layer component moves slightly in the opposite direction to the aggregate determinant; the entire net increase in diffusion comes from the multilayer component.

Third, the product composition of this feedback architecture is not stable. The manufacturing core remains important, but mineral fuels become the layer with the largest marginal structuring role. Pharmaceuticals, precious metals and ores also gain structural weight, while several mature material, paper and textile layers lose it. These changes concern the *position of the layers in the feedback system*, not their simple share in world trade.

Fourth, changes in individual structural importance do not imply parallel changes in pairwise interaction. The strongest example is energy: HS27 becomes much more structuring while its positive interactions with machinery, electrical machinery and vehicles collapse. The internal relationships of the manufacturing complex also become more redundant in several important pairs. The WTW is therefore transformed not only by a change in the importance of product layers but by a change in the *architecture of their interactions*.

Taken together, these results support a precise interpretation of structural change in the multilayer WTW. The central object is not the number of products traded, nor the density of each commodity network, nor the evolution of bilateral trade shares taken separately. It is the changing set of cross-product feedback configurations generated when product-specific directed flow matrices are superimposed. This is the level at which the EDT identifies the dominant long-run transformation of world trade between 1995 and 2023.

## 6. The geography of structural interdependence: partitioning the World Trade Web

The previous sections opened the World Trade Web (WTW) along two dimensions. Section 4 described the evolution of the aggregate EDT structure through the determinant and its closed-walk decomposition, while Section 5 showed that most of the structural change in world trade is generated by the way product layers combine. A third dimension remains to be examined: geography. Closed feedback can remain entirely inside a broad world region, or it can leave one region, pass through one or several others, and return. The theorem of the partition developed in Section 2 provides an exact way to separate these two forms of circularity without importing a community-detection algorithm, a modularity criterion, or an external measure of regionalization.

This section applies that theorem to a fixed geographical partition of the country set. The partition is deliberately exogenous to the observed trade network. Countries are grouped according to the continental regions of the United Nations M49 geographical classification: Africa, the Americas, Asia, Europe and Oceania. The purpose is not to discover an optimal partition of the WTW, but to ask a transparent structural question: how much of the feedback measured by EDT closes within these large regions, and how much requires trade relations that cross their boundaries?

The exercise yields three complementary results. First, a sizeable part of global circular interdependence is genuinely cross-regional, but this share is neither constant nor monotonic. Second, the geographic origin of structural change differs across the phases identified in Section 4: the movements before 2013 are mainly driven by changes in cross-regional feedback, whereas later movements are more mixed and can even combine weaker intra-regional feedback with stronger cross-regional feedback. Third, geography interacts with feedback length. Short reciprocal loops are predominantly internal to broad regions, whereas long closed walks become overwhelmingly transregional. This last result contrasts sharply with the product dimension, where genuinely multilayer feedback already dominates at order two.

### 6.1 A fixed geographical partition and its EDT implementation

Let the set of countries be partitioned into $P$ mutually exclusive geographical groups,

$$\mathcal{P} = (N_1, \dots, N_P),$$

with

$$\bigcup_{r=1}^{P} N_r = N, \qquad N_r \cap N_s = \varnothing \quad (r \neq s).$$

The empirical partition contains five economically meaningful macro-regions: Africa, the Americas, Asia, Europe and Oceania. The country mapping is kept fixed over 1995–2023 and therefore does not respond to changes in trade itself. The 238 nodes in the common country universe are distributed as follows: 58 in Africa, 50 in the Americas, 52 in Asia, 49 in Europe and 28 in Oceania. The French Southern Territories (ATF) form a one-node residual "Antarctic/Other" category required for an exhaustive mapping. Because a one-node block has no self-flow under the convention used throughout the article, its internal determinant is exactly one and its internal log-determinant is zero. It therefore does not affect the within-region results reported below. Four historical or non-ISO Comtrade/BACI identifiers are mapped explicitly to their natural continental region: S19 to Asia, R20 to Europe, ZA1 to Africa and PUS to Oceania.

For every year $t$, reorder the normalized aggregate trade matrix $\Phi_t$ by geographical group and write it in block form,

$$
\Phi_t = \begin{pmatrix} \Phi_{11,t} & \Phi_{12,t} & \cdots & \Phi_{1P,t} \\ \Phi_{21,t} & \Phi_{22,t} & \cdots & \Phi_{2P,t} \\ \vdots & \vdots & \ddots & \vdots \\ \Phi_{P1,t} & \Phi_{P2,t} & \cdots & \Phi_{PP,t} \end{pmatrix}.
$$

The window-induced structure keeps only within-region flows,

$$
\Phi_t^\circ = \operatorname{diag}\left(\Phi_{11,t}, \dots, \Phi_{PP,t}\right).
$$

Let

$$
D_{r,t} = \det\left(I - \Phi_{rr,t}\right)
$$

be the determinant of region $r$ after cross-region arcs are deleted, keeping all country margins at their annual global values. The remaining rows are not renormalized into an autonomous regional economy. Since $\Phi_t^\circ$ is block diagonal,

$$
D_t^\circ = \det(I - \Phi_t^\circ) = \prod_{r=1}^{P} D_{r,t}.
$$

The theorem of the partition implies

$$
\boxed{D_t^\circ \geq D_t.}
$$

All annual computations satisfy this inequality. The smallest numerical margins remain many orders of magnitude above the $10^{-10}$ tolerance used in the audit, and the identity is therefore treated as an empirical check on the construction rather than as an assumption imposed on the results.

For temporal decomposition, the log-determinant is more convenient because the block product becomes additive:

$$\log D_t^{\circ} = \sum_{r=1}^{P} \log D_{r,t}.$$

Add and subtract this term from the aggregate log-determinant:

$$\log D_t = \underbrace{\sum_{r=1}^{P} \log D_{r,t}}_{\text{feedback internal to regions}} + \underbrace{\left(\log D_t - \sum_{r=1}^{P} \log D_{r,t}\right)}_{\text{feedback crossing regional boundaries}}.$$

Define the second term only as a label for the exact residual,

$$C_t^{\text{geo}} = \log D_t - \log D_t^{\circ} \leq 0.$$

No new network index is introduced. $C_t^{\text{geo}}$ is simply the part of the EDT log-determinant that disappears when every inter-regional arc is suppressed. In the trace representation,

$$C_t^{\text{geo}} = -\sum_{k=2}^{\infty} \frac{\text{Tr}\left(\Phi_t^k\right) - \sum_r \text{Tr}\left(\Phi_{rr,t}^k\right)}{k}.$$

Hence it is generated exactly by closed walks that use at least one inter-regional arc. This formulation also makes the geographic decomposition comparable to the product decomposition of Section 5: one partitions the node set, the other partitions the arc set by product layer, but both open the same determinant through operations that are internal to EDT.

### 6.2 A substantial but changing part of global feedback crosses regional boundaries

Figure 6.1 compares the determinant of the full WTW with the determinant of the window-induced structure. The inequality $D_t^{\circ} \geq D_t$ is large in every year. Suppressing all inter-regional links raises diffusion substantially because it destroys feedback mechanisms whose closure requires more than one macro-region.

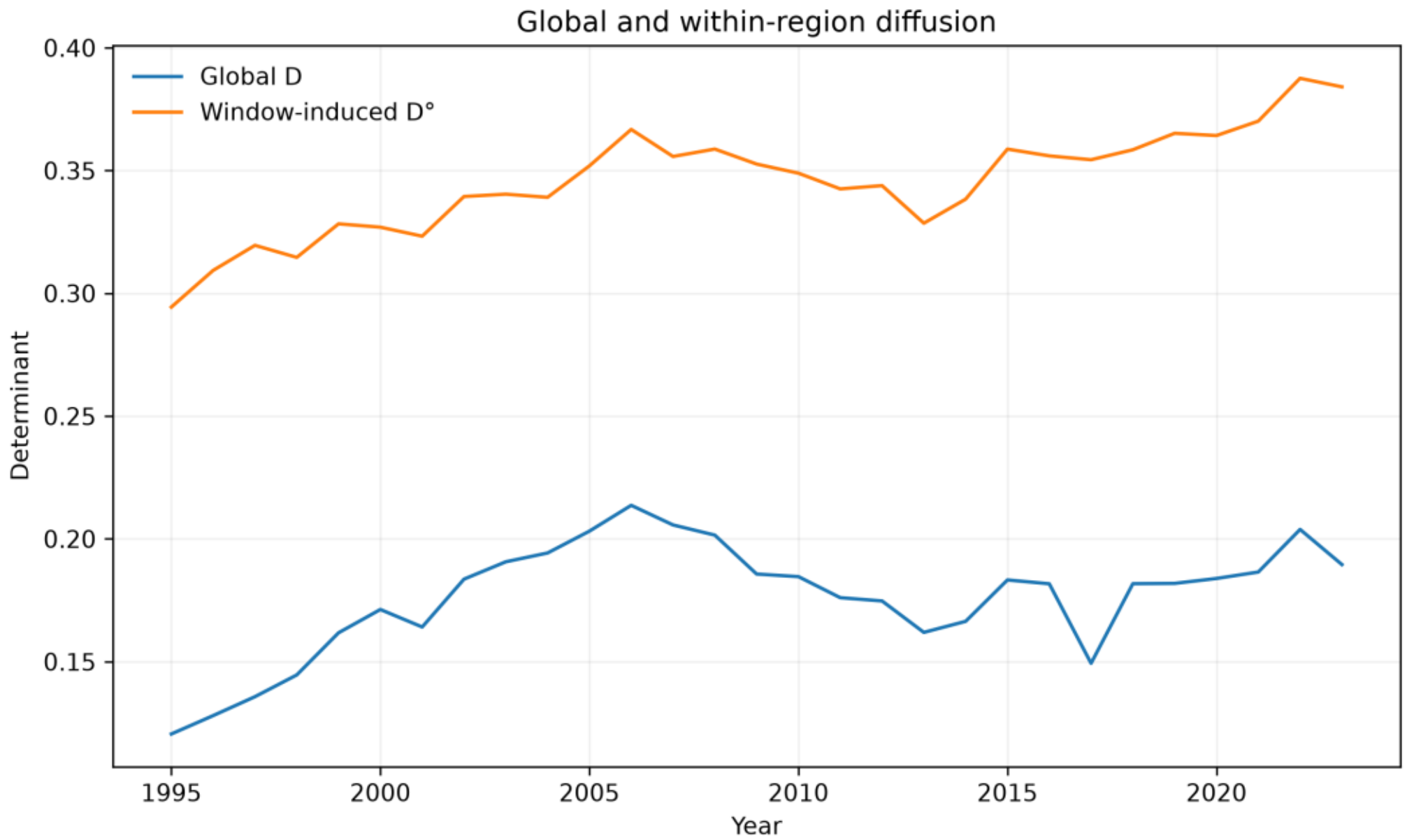

**Figure 6.1.** Aggregate diffusion and diffusion after suppressing all inter-regional flows. *Source: author's calculations from BACI.*

The level difference is already pronounced at the beginning of the period. In 1995,

$$D_{1995} = 0.1205, \qquad D^{\circ}_{1995} = 0.2944.$$

The determinant gap is therefore 0.1739. In 2023,

$$D_{2023} = 0.1895, \qquad D^{\circ}_{2023} = 0.3841,$$

and the gap rises to 0.1946. The theorem of the partition therefore identifies a persistent structural contribution of cross-regional feedback throughout the period.

The log-determinant decomposition is more informative than the level gap because it is exactly additive. Table 6.1 reports selected years. In 1995, the full log-determinant is $-2.1158$, of which $-1.2228$ comes from feedback internal to the regional blocks and $-0.8930$ from feedback that crosses at least one regional boundary. In absolute terms, the cross-regional component represents 42.2% of $|\log D|$. This share falls to 35.0% in 2006, then rises again. By 2023 it is 42.5%, almost exactly its initial level.

| Year | $D_t$ | $D_t^{\circ}$ | $D_t^{\circ} - D_t$ | $C_t^{\text{geo}}$ | **Cross-region share of $\lvert\log D_t\rvert$** |
|---|---|---|---|---|---|
| 1995 | 0.1205 | 0.2944 | 0.1739 | -0.8930 | 42.2% |
| 2000 | 0.1712 | 0.3269 | 0.1557 | -0.6467 | 36.6% |
| 2006 | 0.2137 | 0.3668 | 0.1531 | -0.5403 | 35.0% |
| 2013 | 0.1619 | 0.3286 | 0.1667 | -0.7077 | 38.9% |
| 2016 | 0.1817 | 0.3560 | 0.1743 | -0.6724 | 39.4% |
| 2017 | 0.1493 | 0.3544 | 0.2051 | -0.8643 | 45.5% |
| 2018 | 0.1818 | 0.3585 | 0.1767 | -0.6792 | 39.8% |
| 2022 | 0.2038 | 0.3876 | 0.1837 | -0.6426 | 40.4% |
| 2023 | 0.1895 | 0.3841 | 0.1946 | -0.7063 | 42.5% |

**Table 6.1.** Aggregate and window-induced determinants, selected years.

Note. Differences and shares are computed from unrounded values; the last displayed digit may therefore differ from the difference of the displayed entries.

*Source: author's calculations from BACI.*

Two points matter for interpretation. First, the WTW is not well described as a juxtaposition of continental trade systems linked only at the margin. A large fraction of the determinant-generating feedback necessarily crosses continental boundaries. Second, the importance of this cross-regional component is not monotonic. The rise in $D_t$ before 2006 is accompanied by a decline in the relative contribution of cross-regional feedback, whereas the subsequent strengthening of circular interdependence partly rebuilds that component.

This is a structural statement, not a statement about the value share of inter-regional trade. A region may exchange large trade volumes with the rest of the world without those flows forming strong closed feedback. Conversely, relatively smaller flows may matter structurally if they close recurrent chains linking otherwise separate parts of the system. The partition theorem isolates the latter phenomenon.

### 6.3 The geographical origin of structural change differs across phases

The static decomposition can be differenced exactly through time. Between dates $a$ and $b$,

$$\log D_b - \log D_a = (\log D_b^{\circ} - \log D_a^{\circ}) + \left(C_b^{\text{geo}} - C_a^{\text{geo}}\right).$$

The first term measures change in feedback that remains inside the fixed geographical regions. The second measures change in feedback that requires at least one regional boundary to be crossed. Figure 6.2 reports the annual decomposition.

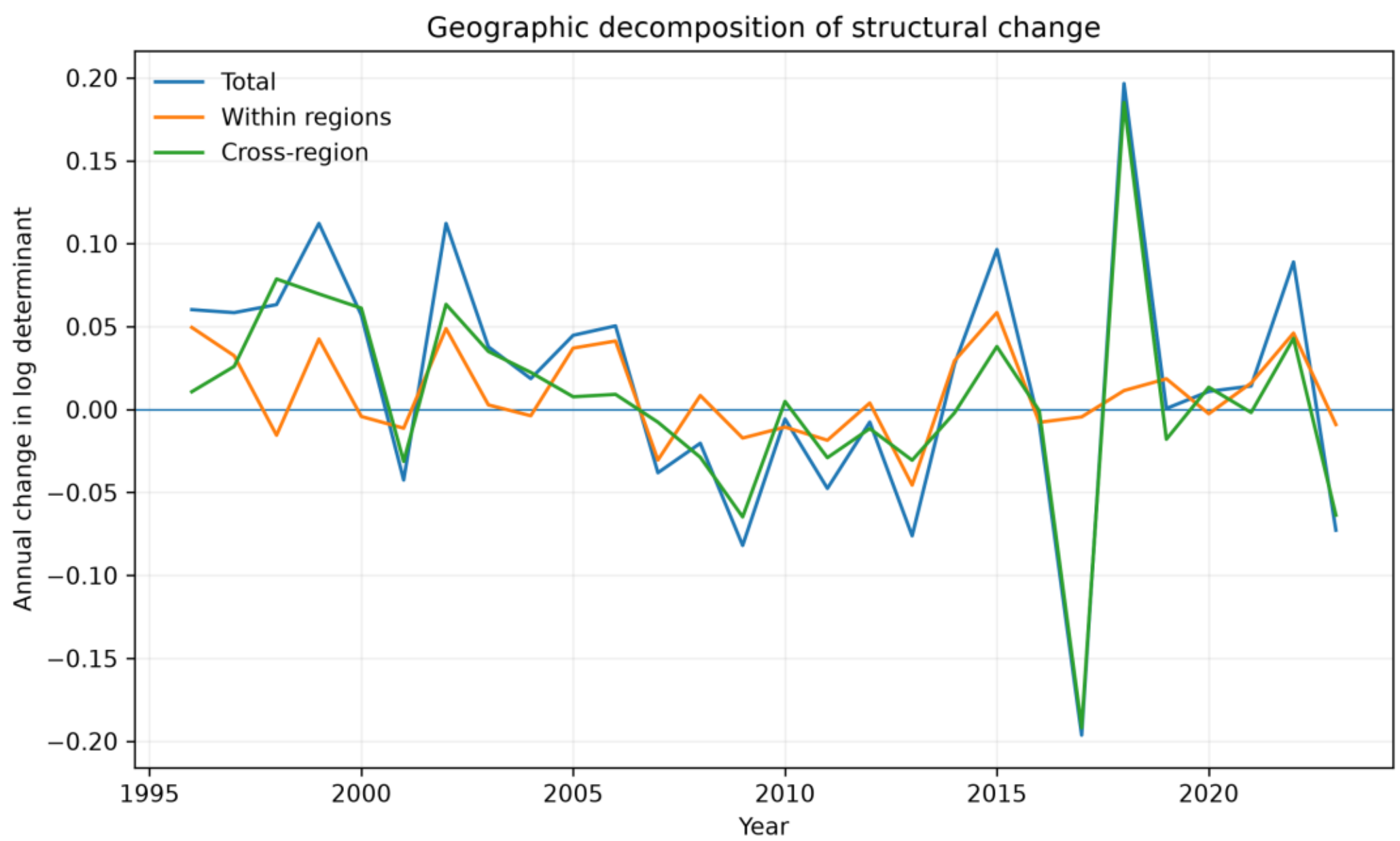

**Figure 6.2.** Exact annual decomposition of structural change into within-region and cross-region components.

*Source: author's calculations from BACI.*

The annual series immediately shows that the two geographical components need not move together. Some annual changes in the global determinant are produced largely inside regions; others are driven by the reorganization of feedback between them (Table 6.2). The major phases established in Section 4 make this distinction easier to interpret.

| Period | $\Delta \log D$ | Within regions | Cross regions | Within share of net change | Cross-region share of net change |
|---|---|---|---|---|---|
| 1995–2006 | 0.5724 | 0.2197 | 0.3527 | 38.4% | 61.6% |
| 2006–2013 | -0.2773 | -0.1099 | -0.1674 | 39.6% | 60.4% |
| 2013–2016 | 0.1154 | 0.0802 | 0.0352 | 69.5% | 30.5% |
| 2018–2023 | 0.0419 | 0.0690 | -0.0271 | 164.6% | -64.6% |
| 1995–2023 | 0.4527 | 0.2659 | 0.1868 | 58.7% | 41.3% |

**Table 6.2.** Geographic decomposition of the main structural phases.

*Source: author's calculations from BACI.*

The first phase, 1995–2006, is primarily a cross-regional transformation. The global log-determinant rises by 0.5724. Of this increase, 0.2197 is produced by weaker feedback inside the geographical regions and 0.3527 by weaker feedback crossing their boundaries. Thus 61.6% of the net movement is associated with the inter-regional component. The rise in EDT diffusion during the first decade is therefore not simply an internal reorganization of Europe, Asia or the Americas. A larger part comes from the loosening of circular chains that connect these regions to one another.

The reversal between 2006 and 2013 has almost the same geographical composition with the opposite sign. The global log-determinant falls by 0.2773. Internal regional feedback accounts for -0.1099, while the cross-regional component contributes -0.1674. Again, approximately 60% of the movement is cross-regional. The renewed strengthening of circular interdependence after 2006 therefore rebuilds transregional feedback as well as strengthening feedback inside several regions. The 2013–2016 recovery differs. Almost 70% of the increase in $\log D$ now comes from changes inside regions, with only 30.5% coming from the cross-regional component. The geographical mechanism of structural change has therefore shifted even though the sign of the aggregate movement resembles that observed before 2006.

The most instructive phase is 2018–2023. Within-region feedback weakens enough to raise the log-determinant by 0.0690. Cross-regional feedback moves in the opposite direction and lowers it by 0.0271. The net global change is therefore only +0.0419. Percentages above 100% or below zero in Table 6.2 are not anomalies: they indicate offsetting exact components. In this period the WTW becomes less circular inside broad regions while simultaneously becoming more circular across their boundaries. The two geographical dimensions partly cancel one another.

Over the full 1995–2023 interval, the aggregate increase of 0.4527 in $\log D$ is split into +0.2659 within regions and +0.1868 across regions. The long-run decline in circular interdependence is thus geographically broad-based. Approximately 59% of the net log-determinant movement is internal to the regions and 41% is cross-regional. This contrasts with the product decomposition of Section 5, where the long-run change is almost entirely generated by cross-product feedback. Geography and product composition therefore open different dimensions of the same structural transformation.

The exceptional 2017 movement provides a particularly useful illustration of the complementarity of the two decompositions. Section 5 showed that the 2016–2017 fall in $\log D$ is overwhelmingly *monolayer* in the product dimension because it is concentrated in HS89, and within HS89 in motorboats (HS890392). The geographical decomposition shows the opposite characteristic along the node dimension.

Between 2016 and 2017,

$$\Delta \log D = -0.1963,$$

of which only

$$\Delta \log D^{\text{within regions}} = -0.0044$$

comes from feedback contained inside the geographical blocks, while

$$\Delta C^{\text{geo}} = -0.1919$$

comes from cross-regional feedback. The reversal in 2018 is similarly cross-regional. At order two, the Africa–Americas contribution to $\mathrm{Tr}(\Phi^2)$ jumps from 0.0030 in 2016 to 0.3537 in 2017 before falling back to 0.0027 in 2018. This matches the earlier micro-diagnostic, in which reciprocal HS890392 flows between Seychelles on the African side and Cayman Islands or the British Virgin Islands on the American side generated almost the entire anomaly.

The two decompositions of the 2016–2017 change are almost mirror images. With the separated-layer determinants of Section 5.1, the product decomposition attributes approximately −0.1919 to within-layer change and −0.0044 to the multilayer component, whereas the geographical decomposition attributes −0.0044 to change inside regions and −0.1919 to cross-regional feedback. This is not an algebraic identity. Both decompositions isolate the same reciprocal HS890392 loop, which is at once confined to a single product layer and extended across a regional boundary, so that its effect of about −0.19 enters the within-layer component in one case and the cross-regional component in the other. The two small remainders measure different things, namely changes in cross-product feedback elsewhere in the network on the one hand and changes in feedback closing inside regions on the other, and their agreement at the reported precision is a coincidence. The 2017–2018 reversal shows that the symmetry is not mechanical: the product decomposition then gives approximately +0.1906 within layers and +0.0060 for the multilayer component, whereas the geographical decomposition gives +0.0114 within regions and +0.1851 across regions.

The episode is therefore structurally simple in one dimension and transregional in another. It is a single-product disturbance that creates an exceptional cross-continental reciprocal loop. This is precisely why the EDT partition operators are useful: product and geography do not compete as explanations; they locate the same feedback mechanism along different dimensions of the structure.

### 6.4 Which regions account for the internal component?

The window-induced determinant is the product of the regional determinants. Its evolution can therefore be decomposed exactly into the log-determinants of the individual regions,

$$\Delta \log D^{\circ} = \sum_{r} \Delta \log D_{r}.$$

Figure 6.3 reports the annual trajectories. Europe has by far the most negative regional log-determinant throughout the period, followed at some distance by the Americas and Asia. Africa and Oceania remain much closer to zero. This ordering should not be interpreted as a ranking of trade volume or economic importance. It indicates the intensity of circular feedback contained wholly inside each regional block under the common EDT normalization.

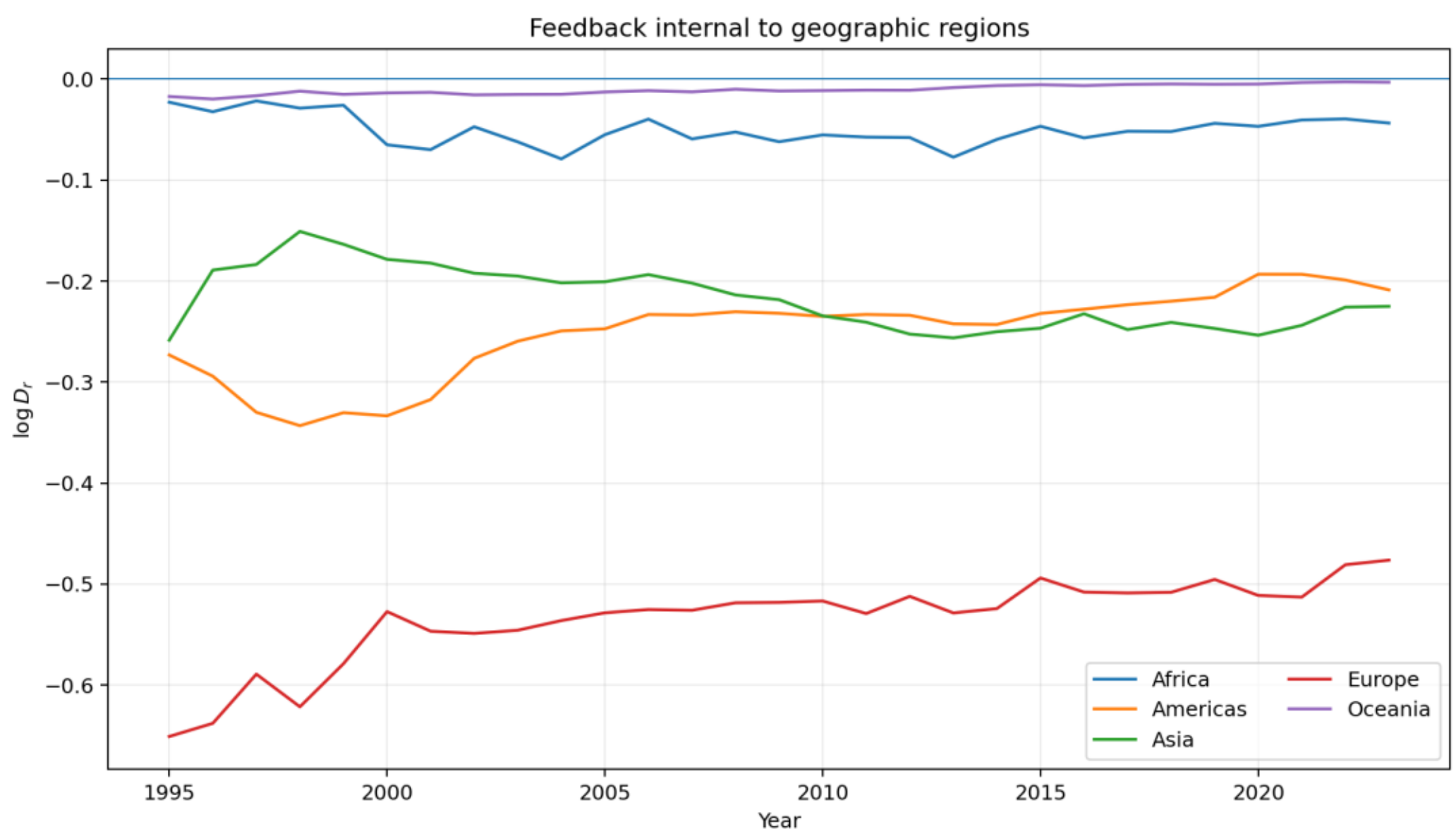

**Figure 6.3.** Feedback internal to the geographical regions, measured by regional log-determinants.

*Source: author's calculations from BACI.*

In 1995, Europe has

$$\log D_{Europe} = -0.6510,$$

compared with $-0.2731$ for the Americas and $-0.2585$ for Asia. By 2023, the corresponding values are $-0.4764$, -0.2087 and $-0.2251$. Europe therefore remains the region with the strongest internal circular feedback, but that feedback weakens substantially over the period. Africa moves in the opposite direction: its regional log-determinant falls from $-0.0229$ to $-0.0436$, indicating a modest strengthening of within-African circularity from a very low initial level.

| **Region** | $D_r$, **1995** | log$D_r$, **1995** | $D_r$, **2023** | log$D_r$, **2023** | $\Delta$log$D_r$, **1995–2023** | **Contribution to within-region net change** |
|---|---|---|---|---|---|---|
| Africa | 0.9774 | -0.0229 | 0.9574 | -0.0436 | -0.0207 | -7.8% |
| Americas | 0.7610 | -0.2731 | 0.8116 | -0.2087 | 0.0644 | 24.2% |
| Asia | 0.7722 | -0.2585 | 0.7985 | -0.2251 | 0.0335 | 12.6% |
| Europe | 0.5215 | -0.6510 | 0.6210 | -0.4764 | 0.1746 | 65.7% |
| Oceania | 0.9829 | -0.0172 | 0.9969 | -0.0031 | 0.0141 | 5.3% |

**Table 6.3.** Regional determinants and long-run change in internally closed feedback.

*Source: author's calculations from BACI.*

Europe accounts for about two thirds of the positive long-run change in the within-region component (Table 6.3). The Americas contribute about one quarter, Asia around one eighth, and Oceania a small positive amount. Africa offsets part of these movements because its internal circularity increases rather than declines. These percentages are additive because they refer to changes in regional log-determinants, not to marginal suppression effects.

The phase decomposition adds nuance. Between 1995 and 2006, Europe contributes $+0.1257$ to the $+0.2197$ within-region rise in log$D^{\circ}$, while Asia contributes $+0.0649$ and the Americas $+0.0401$. Africa moves slightly in the opposite direction ($-0.0168$). Between 2006 and 2013, the internal component reverses mainly because Asia ($-0.0626$) and Africa ($-0.0377$) become more circular; the Americas also contribute negatively ($-0.0093$), whereas Europe changes little ($-0.0034$). During 2013–2016 all five main regions move toward greater internal diffusion, and during 2018–2023 the internal relaxation is again broad-based, led by Europe.

The regional picture therefore does not reduce to a simple narrative of increasing or decreasing regionalization. Different regions contribute differently at different dates, and the cross-regional component may move in the same or the opposite direction. The value of the partition theorem is precisely that it keeps these components algebraically distinct.

### 6.5 Feedback length changes the meaning of geography

The trace formulation reveals a further regularity. For every order $k$, define the within-region trace mass

$$T_{k,t}^{\text{within}} = \sum_r \mathrm{Tr}\left(\Phi_{rr,t}^{k}\right)$$

and the cross-region remainder

$$T_{k,t}^{\text{cross}} = \mathrm{Tr}\left(\Phi_t^k\right) - \sum_r \mathrm{Tr}\left(\Phi_{rr,t}^{k}\right).$$

Because the matrices are non-negative,

$$0 \leq \frac{T_{k,t}^{\text{cross}}}{\mathrm{Tr}\left(\Phi_t^k\right)} \leq 1.$$

This ratio is not introduced as an independent structural index. It simply reports what fraction of the already-defined closed-walk mass of order $k$ contains at least one inter-regional transition.

Figure 6.4 shows a remarkably stable pattern. Geography matters very differently for short and long feedback. At order two, most reciprocal feedback remains inside the broad regions. In 1995, only 15.6% of $\mathrm{Tr}(\Phi^2)$ is cross-regional; the value is 14.8% in 2006 and 19.7% in 2023. At order three, the cross-regional share is already 45.7% in 1995 and 48.1% in 2023. It exceeds one half at order four and rises rapidly thereafter. In 2023, 75.5% of order-five feedback, 92.0% of order-eight feedback and 98.1% of order-twelve feedback cross regional boundaries.

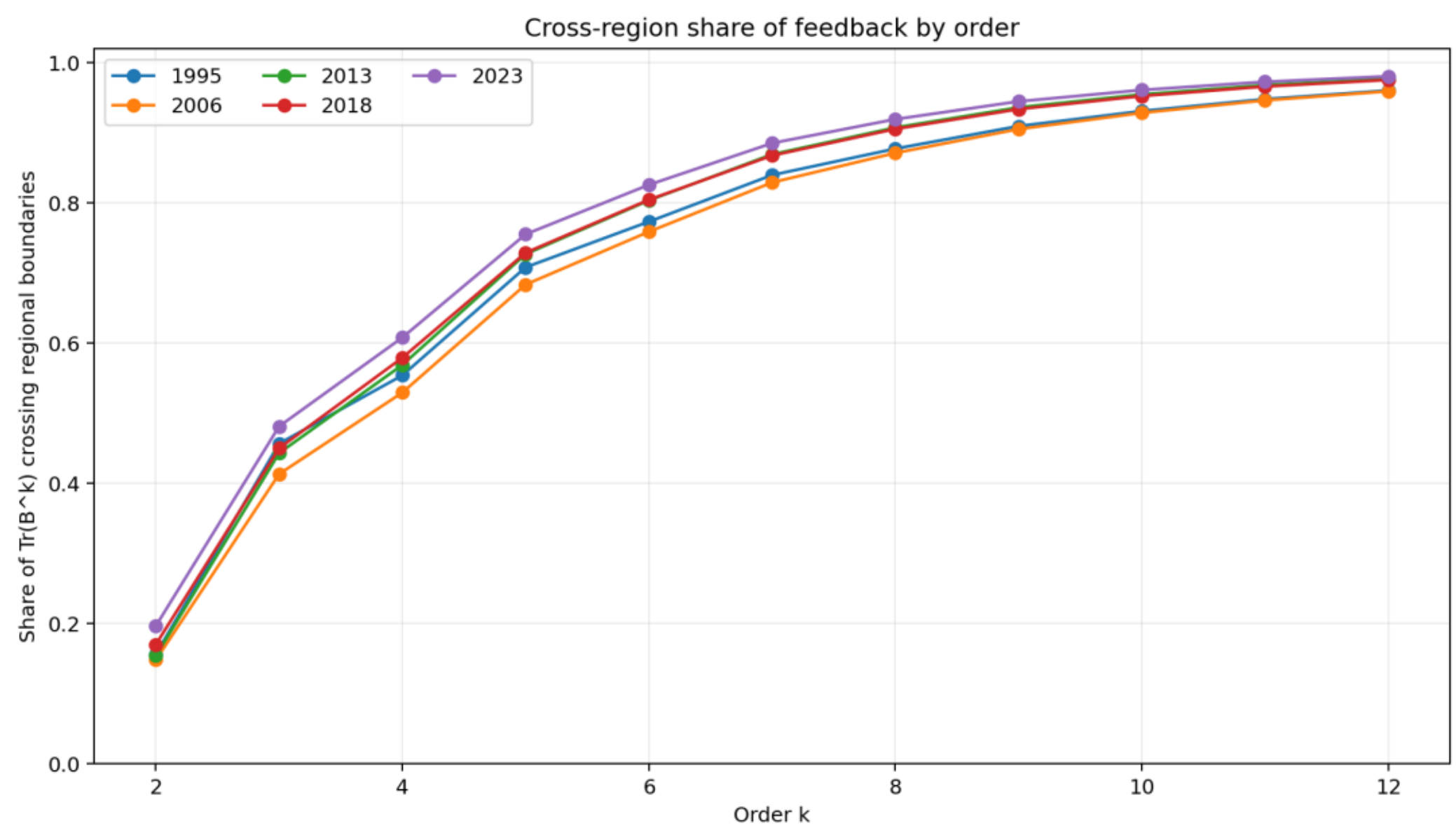

**Figure 6.4.** Share of closed-walk mass of each order that crosses at least one geographical boundary.

*Source: author's calculations from BACI.*

This result gives a precise structural meaning to the idea that the WTW becomes global through chains. A short reciprocal loop can remain geographically localized: two European countries, for example, may exchange flows in both directions without involving another continent. As the length of the closed walk grows, remaining inside one macro-region becomes progressively less likely. Long feedback is therefore almost necessarily transregional.

The comparison with Section 5 is important. In the product dimension, more than 91% of order-two feedback is already genuinely multilayer: reciprocal trade between two countries usually closes through different product categories. In the geographical dimension, only about 15–20% of order-two feedback crosses continents. Thus the WTW is *multilayer before it is transregional* at the shortest feedback length. Product heterogeneity is embedded in elementary bilateral reciprocity, whereas geographical globalization becomes progressively more visible as feedback chains lengthen.

The evolution over time is also informative. The cross-regional share at every order from two to twelve is higher in 2023 than in 2006. The difference is modest at the highest orders because the share is already close to one, but it is visible at the short and intermediate orders. This suggests that the recent WTW does not merely contain long global chains; even some of its shorter feedback structures have become more geographically dispersed.

### 6.6 The regional pairs supporting shortest cross-regional feedback

Order two has a special advantage for geographical interpretation because it can be decomposed exactly into unordered pairs of regions. For $r \neq s$, the contribution of the regional pair $(r, s)$ to the cross-region part of $\mathrm{Tr}(\Phi^2)$ is

$$2\mathrm{Tr}(\Phi_{rs}\Phi_{sr}).$$

Expanding the trace gives

$$2\sum_{i\in N_r}\sum_{j\in N_s}\phi_{ij}\,\phi_{ji}.$$

The term therefore measures reciprocal two-step feedback in which one country belongs to region $r$ and the other to region $s$. The pair contributions sum exactly to the cross-region order-two trace, an identity verified numerically in every year to machine precision.

Figure 6.5 shows the six economically largest or most dynamic regional dyads. Two pairs dominate the shortest transregional feedback throughout most of the period: Americas–Asia and Asia–Europe. Their relative importance, however, changes substantially.

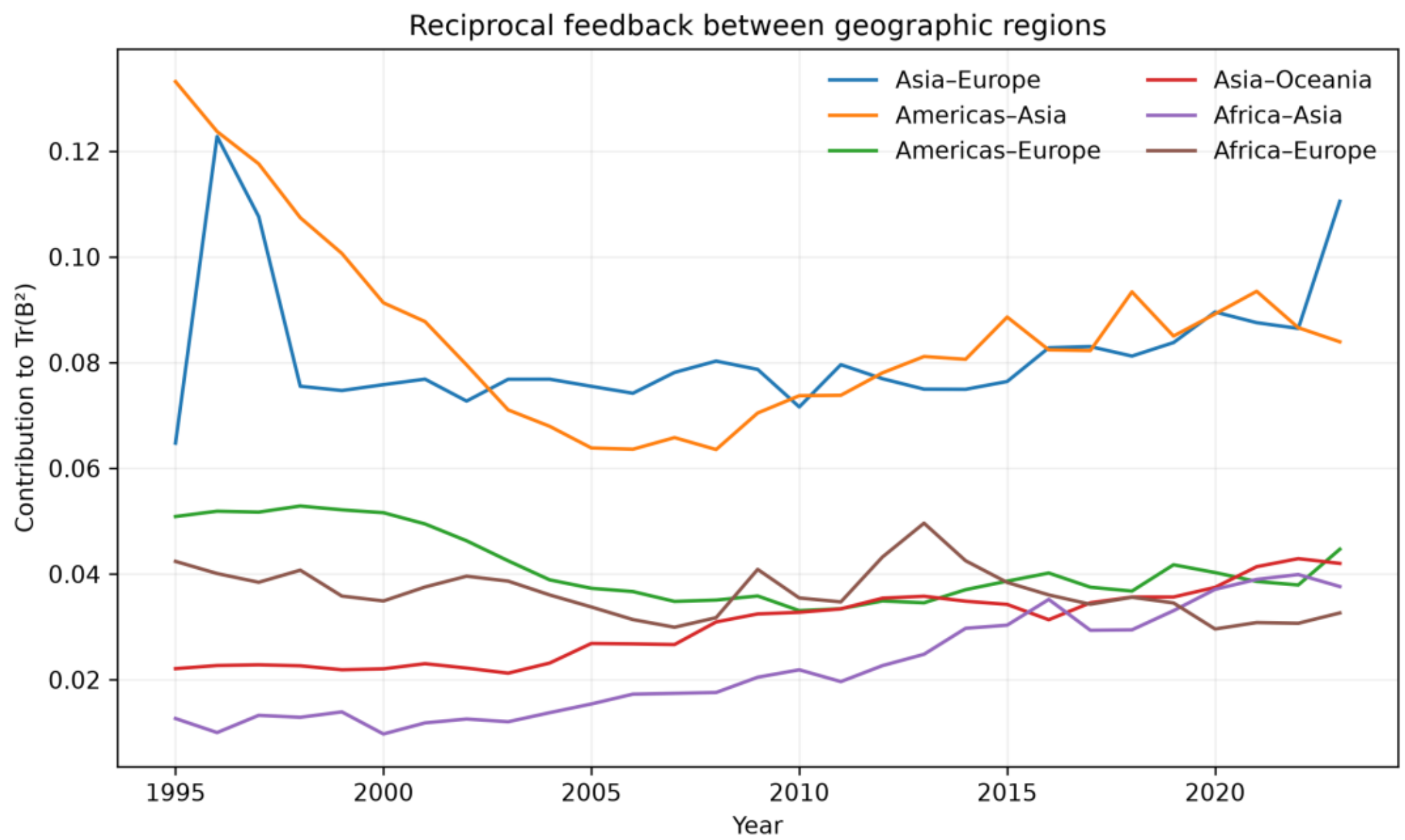

**Figure 6.5.** Reciprocal order-two feedback between selected geographical regions.

*Source: author's calculations from BACI.*

| **Regional pair** | **Contribution to cross-region** $\mathrm{Tr}(\Phi^2)$**, 1995** | **2023** | **Change** |
|---|---|---|---|
| Americas–Asia | 0.1332 | 0.0839 | -0.0492 |
| Asia–Europe | 0.0648 | 0.1105 | 0.0458 |
| Americas–Europe | 0.0509 | 0.0447 | -0.0062 |
| Africa–Europe | 0.0424 | 0.0326 | -0.0098 |
| Asia–Oceania | 0.0221 | 0.0420 | 0.0199 |
| Africa–Asia | 0.0127 | 0.0376 | 0.0250 |

**Table 6.4.** Long-run change in selected order-two regional pair contributions.

Note. Differences and shares are computed from unrounded values; the last displayed digit may therefore differ from the difference of the displayed entries.

*Source: author's calculations from BACI.*

In 1995, Americas–Asia alone accounts for 39.5% of all cross-regional order-two feedback. Asia–Europe accounts for 19.2%, Americas–Europe for 15.1% and Africa–Europe for 12.6%. By 2023 the ordering has changed. Asia–Europe becomes the largest pair, with 31.0% of the cross-regional order-two total, while Americas–Asia falls to 23.5%. Asia–Oceania and Africa–Asia both gain strongly, the latter rising from 3.8% to 10.6% of cross-regional order-two feedback.

The change is therefore not a generalized strengthening of every cross-regional link (Table 6.4). It is a recomposition. Reciprocal feedback between the Americas and Asia weakens relative to its very high initial level, while Europe–Asia, Africa–Asia and Asia–Oceania gain. At the same time, Africa–Europe and Americas–Europe decline modestly. The shortest cross-regional feedback architecture thus becomes more concentrated around several dyads involving Asia, but with different trajectories on the American, European, African and Oceanian sides.

This result should again be distinguished from bilateral trade shares. The quantity

$$2\mathrm{Tr}(\Phi_{rs}\Phi_{sr})$$

rewards mutual structural closure. A large one-way flow from one region to another does not generate a large value unless a return path exists in the opposite direction. The decomposition therefore identifies reciprocal dependence between regions, not simply the intensity of their gross trade.

**6.7 What the geographical partition adds to the structural picture**

The geographical application of the partition theorem adds a dimension that cannot be recovered from either the aggregate determinant alone or the product decomposition. Three conclusions are particularly relevant for the remainder of the article.

First, the distinction between internal and cross-regional feedback is quantitatively important. In levels, around 35–42% of the absolute aggregate log-determinant outside the isolated 2017 episode is generated by closed walks that cross regional boundaries. Over 1995–2023, 41% of the net increase in the global log-determinant comes from the weakening of this cross-regional component. The global WTW is therefore neither a purely regional system nor a homogeneous global one: both scales contribute materially to its circular architecture.

Second, the geographical composition of change is historically variable. Cross-regional feedback explains about 60% of the two major movements before 2013, but only about 30% of the 2013–2016 recovery. From 2018 to 2023, the two components move in opposite directions: internal regional feedback weakens while cross-regional feedback strengthens. The same aggregate movement in $D_t$ can therefore conceal different spatial reorganizations.

Third, geography and product composition reveal complementary aspects of the same feedback system. The 2017 anomaly is the clearest example: it is monolayer in product space and cross-regional in geographical space. More generally, Section 5 showed that elementary feedback is overwhelmingly cross-product, while the present section shows that elementary geographical feedback remains mainly within continents and becomes transregional as path length increases. The two decompositions therefore identify different mechanisms rather than redundant descriptions.

This distinction will be useful in the discussion section. Conventional WTW studies often characterize globalization through changes in density, bilateral intensity, centrality, communities or regional assortativity. EDT poses a different question: *where does feedback close?* The partition theorem answers it without first converting the network into communities or comparing the observed graph with a null model. The relevant boundary can be geographical, institutional or otherwise economically meaningful, and the effect of that boundary is measured directly through the same determinant that defines structural interdependence at the global level.

These geographical results complete the empirical opening of the determinant. Section 7 now places the EDT representation in a broader methodological perspective, comparing its formal object, propagation mechanism, and analytical purpose with established approaches to the World Trade Web, economic complexity, input-output analysis, global value chains, multilayer networks, and community detection.

## 7. Discussion: what kind of structure does EDT reveal in world trade?

The empirical results of the previous sections can now be placed in a broader methodological perspective. This comparison requires some care because the word *structure* does not designate the same object across the literatures that study international trade as a network, a country-product system, an input-output system, a set of global value chains, or a multilayer graph. A comparison limited to indicators would therefore be misleading. Degree, clustering, economic complexity, upstreamness, value-added content, modularity, and the EDT determinant do not provide competing numerical answers to a single question. They formalize different questions.

This distinction is the starting point of the discussion. The contribution of the Economic Dominance Theory (EDT) is not to add one more network statistic to the established empirical toolkit of the World Trade Web (WTW). It is to define structural dependence through a specific mechanism: directed influence is allowed to propagate through a normalized system of flows, and structure is identified with the architecture through which this influence circulates, returns, and closes into feedback. The determinant, the Ghosh inverse, the circuit expansion, the trace expansion, the theorem of the partition, and the multilayer interaction results are different views of that same object.

This is also why the appropriate comparison is not a horse race among measures. The relevant questions are more basic. What is the primitive object of each approach? Which relations are preserved when raw trade data are transformed into an analytical matrix? What kind of indirect effect is admitted? Does the formalism describe topology, capabilities, production accounting, position along a chain, or feedback dependence? What kinds of counterfactual operation are meaningful inside the theory? And, when time is introduced, what exactly is meant by structural change?

The discussion proceeds along these dimensions. It first contrasts EDT with the weighted-network tradition of the WTW, then locates it relative to path- and walk-based measures of indirect influence. It next distinguishes structural dependence from economic complexity, before turning to the closer algebraic relation with input-output analysis and the different substantive objective of global value-chain research. The multilayer formulation is then compared with general multilayer network theory and commodity-specific trade networks. The theorem of the partition provides the basis for a discussion of community detection and regionalization. A separate subsection considers the meaning of structural change through time. The final two subsections state the limitations of EDT and synthesize what is, and is not, distinctive about the framework.

### 7.1 From the topology of trade relations to a theory of structural dependence

The modern literature on the WTW has established a simple but consequential point: bilateral trade relations cannot be understood adequately by treating each dyad as if it were independent of the rest of the system. Network analysis makes the interdependence of relations explicit. A country's trade position depends not only on how much it exports or imports, but also on the partners to which it is connected, on the partners of those partners, and on the larger configuration in which these links are embedded. This idea motivates the binary and weighted analyses of connectivity, assortativity, clustering, centrality, reciprocity, core-periphery organization, and other higher-order properties developed in the WTW literature (Fagiolo et al., 2008, 2009, 2010; De Benedictis et al., 2013).

The distinction between binary and weighted representations was itself a major step. In a dense international trade network, the existence of a link and the economic importance of that link are very different pieces of information. Fagiolo et al. (2008) showed that the weighted WTW can display properties that differ substantially from those suggested by its binary projection. Their subsequent work studied the coevolution of weighted connectivity, clustering, assortativity, and centrality and emphasized the remarkable stability of many distributions of network statistics over time (Fagiolo et al., 2009, 2010). Squartini et al. (2011) pushed the point further by showing that weighted and binary constraints do not contain the same information about higher-order organization. These results are important for the present paper because they rule out a simplistic contrast in which network analysis would be reduced to counting links while EDT would somehow be the first method to take weights or indirect structure seriously. It is not.

The difference lies instead in the definition of the structural object. In a conventional network description, a weighted directed graph is first given and a family of descriptors is then computed on it. Different descriptors summarize different properties. Degree or strength describes immediate connectivity. Clustering describes local closure. Assortativity describes mixing between nodes with similar or dissimilar attributes. Centrality identifies positions that are important under a particular conception of flow or reachability. Community methods search for partitions satisfying a density or modularity criterion. The graph is common, but the structural meaning is distributed across a repertoire of measures.

EDT starts from a more restrictive but more integrated construction. Raw bilateral flows are transformed into a normalized matrix of directed coefficients, denoted here by $\Phi$. The normalization is not a cosmetic rescaling. It states what one unit of structural influence means and imposes a substantive rule about the capacity of a country to transmit influence. Under the trade-balance normalization used in this article, outgoing coefficients sum to one for balanced or surplus countries and to less than one for deficit countries. The latter therefore introduce leakage into the propagation mechanism. Once this matrix is fixed, the fundamental object is not simply the weighted graph but the linear system generated by

$$G = I - \Phi.$$

The distinction matters because the subsequent EDT quantities are not independent statistics chosen from a menu. They are linked algebraically. The inverse

$$G^{-1} = (I - \Phi)^{-1} = I + \Phi + \Phi^2 + \cdots$$

represents direct and indirect propagation. The determinant

$$D = \det(I - \Phi)$$

measures diffusion and, under the no-self-flow convention of this paper, $1 - D$ measures circular interdependence. The minors of the same matrix generate node-removal effects. Jacobi's formula gives arc sensitivities. The trace expansion decomposes the same determinant by feedback order. Partition and aggregation theorems describe what happens when parts of the same structure are disconnected or merged. The multilayer formulation preserves this algebra when product categories are distinguished.

This internal linkage is the first characteristic difference. EDT should therefore not be described as a “global” approach opposed to “local” network analysis. Such a contrast would be inaccurate. EDT has local objects. The Lantner centrality of country $i$ is

$$L_i = (I - \Phi)_{ii}^{-1} = \frac{D_{-i}}{D},$$

and the contribution of an arc, a node, or a product layer can be measured by determinant differences. What is distinctive is that these local or marginal quantities are defined *through the same global structural function*. Removing one node changes the determinant of the entire system, and the ratio of that modified determinant to the original determinant defines the node’s feedback role. Local position and global organization are not separate descriptive layers of the analysis.

A second difference concerns invariance and scale. Many WTW studies ask how to weight bilateral links in economically meaningful ways, and different weighting schemes can be appropriate for different questions. EDT makes one specific choice because the normalization must support a propagation interpretation. Multiplying every trade flow in a year by the same positive constant leaves $\Phi$ unchanged. Consequently, EDT structural change is not generated by a common expansion of nominal world trade. The result obtained in Section 4 is informative in precisely this sense. From 1995 to 2006, the diffusion determinant rose substantially, then fell until 2013, before entering a more mixed trajectory. A long period of globalization in the ordinary volume sense therefore does not translate into a monotonic movement of EDT interdependence. This finding does not contradict the WTW literature; it illustrates that the two approaches define different dimensions of structural evolution.

The distinction is particularly clear when one considers the determinant as a full-system function. Let $\lambda_1, \dots, \lambda_n$ denote the eigenvalues of $\Phi$. Then

$$D = \prod_{j=1}^{n} (1 - \lambda_j).$$

The determinant therefore depends on the full spectrum rather than on a dominant eigenvector alone. Yet the spectral expression is still not the main interpretation used here. Through the Leibniz expansion, $D$ is an inclusion-exclusion sum over families of vertex-disjoint circuits; through the logarithmic trace expansion, it is an accumulation of closed walks of every order. The same scalar can therefore be read combinatorially, spectrally, and dynamically. This multiplicity of exact representations is what turns $D$ from a generic network summary into the central structural function of EDT.

This also clarifies the relation between EDT and the empirical tradition represented by De Benedictis et al. (2013). Their BACI-based exposition is deliberately plural: it introduces the topology of the WTW and shows how standard binary and weighted network statistics can be computed and interpreted. The present article is narrower in one sense and broader in another. It is narrower because it deliberately restricts itself to one structural theory rather than surveying the network toolkit. It is broader because, once the EDT object is chosen, the same formalism is followed from the raw matrix through global structure, local suppression, product layers, geographical partitions, and temporal change. The aim is not a catalog of properties but a closed analytical system.

### 7.2 Paths, walks, and feedback: the mechanism behind a network measure

The preceding distinction becomes sharper when indirect influence is considered. Network measures differ not only because they use different formulas, but because those formulas embody different assumptions about how something can move through a network. Borgatti (2005) makes this point explicitly by distinguishing geodesics, paths, trails, and walks, and by showing that centrality measures are meaningful only relative to a plausible model of network flow.

Betweenness, for example, is naturally associated with processes that move through shortest paths or through paths selected according to a routing rule. Walk-based measures instead allow repeated visits to nodes and links and are appropriate for processes in which influence can recirculate.

EDT belongs unambiguously to the second family. Matrix multiplication implies that $(\Phi^k)_{ij}$ sums the weights of all directed walks of length $k$ from $i$ to $j$. The inverse therefore accumulates walks of every length. The diagonal element $(\Phi^k)_{ii}$ is generated only by walks that return to their starting node. Consequently,

$$L_i = 1 + \sum_{k=2}^{\infty} (\Phi^k)_{ii}$$

measures feedback returning to $i$, while

$$\mathrm{Tr}(\Phi^k) = \sum_i (\Phi^k)_{ii}$$

measures the total weight of closed walks of order $k$ in the system. This is a very different substantive object from the number of geodesics passing through a country.

This difference should not be overstated into a claim that walk-based influence is unique to EDT. Katz (1953) introduced a status measure that counts indirect connections with attenuation as path length grows. Bonacich (1987) developed a family of power and centrality measures in which the position of a node depends on the centrality of its neighbors, with a parameter controlling whether connections to central others increase or decrease power. Resolvent-based ideas are therefore part of a much larger family of network methods.

The comparison is nevertheless useful because it reveals what EDT adds. In Katz-type constructions, an attenuation parameter is introduced to guarantee convergence and to specify how strongly distant walks should count. In EDT, the series used in this paper is

$$I + \Phi + \Phi^2 + \cdots,$$

without an additional scalar attenuation parameter. Convergence follows instead from the construction of $\Phi$ and the condition $\rho(\Phi) < 1$. Effective attenuation is endogenous to the coefficients carried by the walks: the weight of a walk is the product of its successive bilateral coefficients, and leakage produced by the normalization prevents indefinite conservation of mass in the relevant strongly connected components. The theory therefore places the substantive burden on the definition of the weight matrix rather than on a free distance-decay parameter.

That point connects EDT to a broader methodological issue raised by Leenders (2002). In models of social influence, the weight matrix is not a neutral technical input. Different specifications of the same observed relations correspond to different theories of influence. The same principle applies here. A raw BACI flow matrix does not become a structural dependence matrix simply because it is treated as weighted and directed. The trade-balance normalization defines a mechanism: the capacity to transmit influence is bounded by the larger side of each country's external trade position, and a deficit appears as leakage. Other normalizations would represent other mechanisms. The EDT results are therefore conditional on an explicit theory of how observed trade values are transformed into influence coefficients.

The distinctive step comes after the resolvent. EDT does not stop at an individual walk-based centrality. It links walk propagation to system closure through the determinant. Under $\rho(\Phi) < 1$,

$$\log D = -\sum_{k=2}^{\infty} \frac{\mathrm{Tr}(\Phi^k)}{k}.$$

The object of interest is no longer "how central is country $i$?" but "how much closed feedback of each order is embedded in the entire system?" The factor $1/k$ performs a precise algebraic role in the logarithm of the determinant and avoids interpreting the trace as a simple count of independent circuits. Repeated traversals of shorter cycles are allowed because the trace counts closed walks, not only elementary cycles. This distinction was important in Section 2 and remains important in the comparison with network motifs or cycle counts. Network analysis can, of course, count directed cycles and motifs. EDT's claim is not that circuits were previously absent from graph theory. It is that circuits enter a particular algebra in which their combined contribution determines diffusion and in which structural interventions have monotonic determinant effects.

The difference is visible in the treatment of redundancy. A shortest-path betweenness measure can change drastically when one alternative geodesic appears, because traffic may be rerouted across equivalent shortest routes. In EDT, multiple return paths are not a nuisance to be resolved by a tie-breaking rule; they are part of the feedback architecture. If an influence can leave a node, circulate through several alternative chains, and return, those alternatives contribute to the corresponding powers of $\Phi$ and therefore to the determinant. Lebert and Poindron (2025) emphasize this distinction when relating EDT intermediacy to more familiar notions of betweenness: structural effects depend on circuits and redundancies, not only on the set of connecting paths.

This has a direct implication for interpretation. A country can have low geodesic betweenness yet matter for circular interdependence if it closes feedback loops that have alternative entrances and exits. Conversely, a country can lie on many shortest paths without being central to feedback if those paths do not close into recurrent structures. Neither statement implies that one measure is superior. They correspond to different flow models.

The same applies to arc deletion. In many graph applications, removing an edge is studied through connectivity, path length, efficiency, or robustness. In EDT, the effect of reducing an off-diagonal coefficient is evaluated on $D$. Because the matrices are non-negative and the relevant inverse is positive, the determinant moves monotonically with arc suppression under the conditions established in Section 2. This produces a well-defined comparative-static question: how much circular interdependence disappears when a relation is weakened or removed? The answer is internal to the same propagation model rather than imported from a separate robustness criterion.

The conceptual gain is therefore not that EDT has discovered indirect effects, walks, or centrality. Its contribution lies in *closing the chain of reasoning*: an economically specified matrix generates walks; closed walks generate the determinant; the determinant generates marginal structural effects; those effects extend to partitions and layers. This integrated sequence is the appropriate unit of comparison.

### 7.3 Economic complexity and structural dependence answer different questions

Economic complexity provides the most instructive contrast because it also starts from international trade data, treats products as central objects, and explicitly seeks structural information that is not visible in aggregate export volumes. The resemblance can therefore be misleading. In the canonical economic-complexity framework, countries are connected to the products in which they display revealed comparative advantage, yielding a bipartite country-product matrix. Diversity on the country side and ubiquity on the product side are combined recursively to infer the amount of productive knowledge or capabilities embodied in an economy (Hidalgo & Hausmann, 2009; Hausmann et al., 2014). The related product-space approach uses patterns of co-export to infer proximity among products and to study feasible paths of productive diversification (Hidalgo et al., 2007).

The basic data transformation already shows why EDT is not an alternative estimator of the same latent variable. In a country-product matrix, the destination of an export is normally not part of the primitive relation. What matters is whether country $i$ exports product $g$ with sufficient relative intensity. Bilateral partner information is aggregated away when the export basket is constructed. EDT does the opposite. The direction from exporter $i$ to importer $j$ is indispensable, because dependence is defined through chains that traverse countries. Product information is retained as a layer on those directed country-to-country arcs.

Schematically, the canonical complexity object can be represented as a matrix

$$M_{ig},$$

where $i$ indexes countries and $g$ products. The EDT multilayer object is instead a family

$$\{\Phi_{ij}^{(g)}\}_{g=1}^{\ell}.$$

The extra index is not a technical detail. It preserves the bilateral route through which each product moves. As a result, the questions differ from the outset.

Economic complexity asks what the composition of a country's productive or export basket reveals about the capabilities available to that country and about its opportunities for future diversification. Product-space proximity asks whether the capabilities required for one product are likely to overlap with those required for another. These are fundamentally developmental and productive questions. EDT asks how observed directed exchanges combine into recurrent chains of dependence. It is not trying to infer hidden productive capabilities from the basket. It is studying the structure through which influence propagates between observable poles.

The distinction is also visible in the status of products. In economic complexity, products carry information about productive sophistication and capability requirements. In the product space, two products are close when countries that export one also tend to export the other. In the multilayer EDT of this article, products are types of flow. Two product layers interact structurally when combining them changes the determinant by more or less than the sum of their separate effects. The pairwise interaction

$$\Gamma(g,h) = D_{-\{g,h\}} - D_{-g} - D_{-h} + D$$

therefore has no direct counterpart in product proximity. A positive or negative $\Gamma(g,h)$ is not evidence that two goods require similar capabilities. It says that their directed trade structures are complementary or redundant in forming global feedback.

The empirical results in Section 5 make the distinction concrete. HS84, HS85, HS87, HS27, or HS30 can be structurally important in EDT because removing those layers breaks many feedback mechanisms in the aggregate WTW. That does not imply that these chapters are the most complex products in the Hidalgo-Hausmann sense. Likewise, the rise of the structuring contribution of energy or pharmaceuticals is not a statement about the productive sophistication of countries specializing in those goods. It is a statement about the role of those flow layers in closing international circuits under the common EDT normalization.

Another difference concerns ranking. Economic-complexity research often produces country and product rankings because relative complexity is substantively central. EDT can generate rankings of Lantner centrality or layer structuring effects, but ranking is not its primary objective. The determinant is a property of the system, and local quantities are interpreted as contributions to that system. The natural EDT question is therefore less “which country is most complex?” than “how does the removal or modification of this country, arc, region, or product layer alter the feedback structure of the whole?”

The two approaches also have different relationships to development theory. Economic complexity is explicitly concerned with productive knowledge, diversification, and paths of structural transformation. The product space is designed to illuminate which new activities are reachable from a country’s current capabilities. The present EDT application does not model capability accumulation or technological learning. Its use of the term *structural change* is narrower: it refers to a change in the architecture of directed trade feedback between dates. A country can transform its productive capabilities without producing an immediate proportional change in $D$, and $D$ can change because bilateral trading patterns are reorganized even if the underlying capability set changes little.

These differences make the two frameworks potentially complementary rather than substitutable. Economic complexity can help explain *what countries are able to produce and diversify into*. EDT can help explain *how the resulting flows embed countries in systems of reciprocal dependence*. One could imagine a future design in which changes in complexity or product-space position are related to subsequent changes in EDT feedback architecture. Such an exercise would be substantively interesting precisely because the two variables represent different objects. Collapsing them into a single notion of “trade structure” would remove the reason for comparing them.

For the same reason, this article does not compute an Economic Complexity Index merely to correlate it with $D$ or with Lantner centrality. A high or low correlation would be difficult to interpret. The indices are not alternative noisy measurements of one latent construct. They answer different questions using different transformations of the same broad data domain.

### 7.4 EDT within the input-output tradition: algebraic proximity, substantive displacement

The relation with input-output analysis is closer. EDT inherits much of its matrix language from the input-output tradition, and the use of the Ghosh matrix is explicit. Ghosh (1958) developed an allocation-oriented counterpart to the Leontief model. The broader input-output literature uses systems of linear equations and matrix inverses to trace direct and indirect effects through interindustry relations, and modern treatments include multiplier analysis, linkage measures, decomposition methods, hypothetical extraction, inter-regional systems, and supply-side formulations (Miller & Blair, 2009).

At the algebraic level, the family resemblance is immediate. A Leontief inverse has the form

$$(I - A)^{-1},$$

while the object used here is

$$(I - \Phi)^{-1}.$$

Both represent repeated propagation through powers of a coefficient matrix. This common algebra is not accidental. Lantner's Economic Dominance Theory emerged from structural analysis of linear economic systems, and the determinant of the associated Ghosh-type matrix is central to the theory.

Yet the empirical matrix in this article should not be confused with a conventional input-output coefficient matrix. $\Phi$ is constructed from bilateral merchandise trade between countries, not from an industry-by-industry table of intermediate deliveries divided by sectoral output. Its normalization is based on the external trade balance of each country. The entries therefore represent structural trade coefficients under the EDT rule defined in Section 2, not technical production coefficients. This point matters because otherwise the reader might import a production-function interpretation that the data do not support.

A second clarification concerns the long-standing debate about the supply-driven Ghosh model. In the input-output literature, the causal interpretation of a Ghosh quantity model has been controversial, and Dietzenbacher (1997) famously argued for a reinterpretation as a price model. The present paper does not require a claim that an exogenous increase in primary inputs causes physical output changes according to a supply-driven Ghosh mechanism. "Ghosh matrix" is used here for the structural matrix $I - \Phi$ in the Lantner tradition. The empirical analysis is a structural analysis of normalized flows, not a supply-shock simulation. This distinction removes a possible source of confusion between the algebraic ancestry of EDT and the causal interpretation of standard input-output models.

The substantive displacement from conventional input-output analysis appears most clearly in the role of the determinant. Input-output analysis commonly focuses on the inverse because multipliers, total requirements, and linkage effects are the natural quantities of interest. EDT gives the determinant an independent structural interpretation. The determinant measures diffusion and, with zero diagonal flows, its gap from one measures circular interdependence. Through the circuit and trace expansions, it is interpreted as a function of feedback closure.

This changes the meaning of a structural intervention. Consider a hypothetical extraction. In input-output applications, removing a sector or a set of interindustry transactions is often used to estimate output loss or linkage importance. The counterfactual asks how much economic activity would be affected if a component were absent. EDT also performs extraction-like operations, but the outcome is different. Removing node $i$ produces $D_{-i}$; removing layer $g$ produces $D_{-g}$; removing inter-regional arcs produces $D^{\circ}$. The question is how much the *feedback architecture* changes, not how much output or value added disappears.

This distinction explains why an EDT contribution should not be read as a multiplier. If suppressing a product layer raises $D$ substantially, the layer is structurally important because it participates in many determinant-generating feedback configurations. That statement does not quantify the change in GDP, gross output, employment, or welfare caused by eliminating the corresponding trade. Behavioral substitution is also absent from the static suppression operation. The EDT counterfactual is structural and algebraic: it evaluates the network with specified arcs removed while other coefficients are held as observed.

At the same time, the input-output heritage gives EDT a stronger process interpretation than purely topological network descriptors. Powers of $\Phi$ have the same type of direct/indirect logic as powers of input-output coefficients. The difference is that this paper uses that propagation machinery to ask a question about recurrence and closure. One can summarize the displacement as follows: traditional multiplier analysis asks how an impulse propagates through a linear system; EDT asks how the system's architecture permits propagated influence to return and reinforce itself.

The trace representation makes the distinction exact. A multiplier is naturally read from the elements or sums of an inverse matrix. EDT's global structural object can instead be decomposed as

$$\log D = -\frac{\mathrm{Tr}(\Phi^2)}{2} - \frac{\mathrm{Tr}(\Phi^3)}{3} - \cdots.$$

Only closed walks enter because only diagonal elements of the powers contribute to the trace. Open production chains matter for propagation but do not directly contribute to this global feedback decomposition unless they participate in a closed return path.

This distinction also helps situate EDT relative to linkage analysis. Backward and forward linkages identify sectors whose input or output relations transmit shocks widely. EDT's node or layer intermediacy is related in spirit because both ask whether a component is structurally important. But EDT importance is specifically importance for circular interdependence. A component that sends influence broadly but does not contribute to closure may look important under a linkage measure and modest under a determinant-removal measure. Again, the methods need not agree because their finalities differ.

### 7.5 Global value chains: position along chains versus closure of chains

Research on global value chains (GVCs) is also rooted in input-output accounting but asks a different set of substantive questions. The key empirical problem is that gross trade values cross borders multiple times when production is internationally fragmented. Modern GVC methods therefore combine bilateral trade with input-output tables to distinguish domestic and foreign value added, identify double counting, and locate countries or industries along production sequences (Johnson & Noguera, 2012; Antràs et al., 2012; Koopman et al., 2014; Timmer et al., 2014).

The contrast with EDT can be stated in spatial terms. GVC analysis is primarily concerned with *position along a production chain*. Upstreamness measures the average distance of an industry from final use. Value-added decompositions ask where the value embodied in final demand was generated and how often gross flows cross borders before absorption. The analytical direction is longitudinal: from upstream inputs through successive production stages toward downstream final use.

EDT is instead concerned with *closure of chains into feedback*. A sequence

$$i \to j \to k$$

is an indirect trade path, but it contributes to the determinant's closed-walk decomposition only if the structure eventually returns, for example,

$$i \to j \to k \to i.$$

The distinction is conceptually important. A long and highly fragmented value chain can be structurally open in the EDT sense if it terminates without generating significant return influence.

A short set of reciprocal flows can generate strong EDT circularity without representing a long production chain.

The data requirements follow from the difference in purpose. GVC analysis requires information capable of separating intermediate from final uses and domestic from foreign value added. BACI provides detailed bilateral gross merchandise trade by product, but it does not by itself identify the domestic value-added content of each flow or the stage at which a good enters production. The present article therefore cannot and does not claim to measure vertical specialization, upstreamness, or trade in value added.

Conversely, a value-added accounting framework does not automatically produce the EDT object. Even with a perfect multi-regional input-output table, one would still have to specify which normalized directed coefficients represent structural influence and how feedback is to be evaluated. The two frameworks could be combined, for example by constructing EDT matrices from value-added flows rather than gross trade flows, but that would constitute a different empirical model.

The distinction also prevents a terminological misunderstanding about repeated walks. The log-determinant includes repeated traversals of a closed cycle at orders $2k$, $3k$, and so on. These mathematical repetitions are not the same thing as the "double counting" that GVC accounting removes from gross exports. GVC double counting refers to value that is embodied more than once in gross cross-border trade statistics. Repeated closed walks in the trace series are terms in the expansion of a structural function. They do not assert that the corresponding monetary value is produced repeatedly.

This difference in finality helps interpret one of the major empirical results of Section 5. The finding that almost all high-order feedback is cross-product and that long-run structural change is overwhelmingly generated by the multilayer component does not mean that production has necessarily become more vertically fragmented in the GVC sense. It means that closing international feedback loops increasingly requires combinations of different product layers. That can be related to production fragmentation, but the relation is an empirical hypothesis rather than an identity.

Similarly, the rise of HS27 or HS30 as structurally important layers does not imply that energy or pharmaceuticals have become more upstream or that their domestic value-added share has risen. The EDT claim is more precise and more limited: removing these layers alters the determinant more strongly because they are increasingly embedded in recurrent international trade structures.

The approaches are therefore complementary along a useful conceptual axis. GVC analysis tells us *where value is created and how production is sequenced*. EDT tells us *how observed flows form recurrent structures of dependence*. A complete account of contemporary globalization may require both, but conflating the two would make each less informative.

### 7.6 Multilayer trade networks: coexistence of layers versus structural interaction between layers

The multilayer comparison is especially important because the present article explicitly treats product categories as layers. General multilayer network theory was developed to represent systems in which the same or related entities participate in several types of relation, potentially with connections within and between layers (De Domenico et al., 2013; Kivelä et al., 2014). This literature generalizes degree, centrality, modularity, diffusion, and other network concepts to tensorial or supra-adjacency representations. In international trade, Barigozzi et al. (2010) showed that commodity-specific trade networks have heterogeneous topological properties and that aggregation can mask substantial differences across product networks.

EDT belongs to this broad multilayer family, but its multilayer object is deliberately specialized. Every layer contains the same set of country nodes, and a layer records bilateral trade in one product category. There are no explicit interlayer arcs connecting, for example, "France in machinery" to "France in chemicals." The product dimension is attached to the arcs between countries. The separated multilayer structure is therefore block diagonal, while the observed aggregate trade structure is obtained by summing layers.

The decisive condition is common normalization. If raw product matrices are $X^{(g)}$ and the aggregate matrix is

$$X = \sum_g X^{(g)},$$

then one common diagonal normalization matrix $Q$ is computed from aggregate trade, and

$$\Phi^{(g)} = Q^{-1} X^{(g)}, \qquad \Phi = Q^{-1} X.$$

Therefore

$$\boxed{\Phi = \sum_g \Phi^{(g)}.}$$

This identity is more than a convenient aggregation rule. It ensures that all layers are components of the same structural system and are expressed in the same unit of influence. If each product network were normalized independently, adding the layers would add matrices defined on different scales. The aggregate determinant would then no longer be interpretable as the result of fusing the separated product structures.

This leads to a second difference from a generic multilayer description: layer aggregation itself has a theorem. Let

$$D^{\diamond} = \prod_g \det\left(I - \Phi^{(g)}\right)$$

be the determinant of the block-diagonal structure in which every product layer remains separate, and let

$$D = \det\left(I - \sum_g \Phi^{(g)}\right)$$

be the determinant of aggregate trade. The multilayer EDT result gives

$$D^{\circ} \geq D.$$

The economic meaning follows from closed walks. In the separated structure, a walk that starts in product layer $g$ must remain in $g$. After aggregation, one leg of a closed walk can be carried by machinery, the next by energy, the next by chemicals, and so forth. Aggregation does not simply superimpose layers visually; it creates new possible closures from cross-product combinations. This is the source of the exact trace decomposition

$$\mathrm{Tr}(\Phi^k) = \sum_g \mathrm{Tr}\left[\left(\Phi^{(g)}\right)^k\right] + \text{cross-layer terms.}$$

The distinction between *having many layers* and *being structurally multilayer* is therefore important. A trade system can contain thousands of product categories while its feedback remains nearly monolayer if most closed walks use only one category. Conversely, a relatively small number of layers can generate strongly multilayer feedback if closure typically requires their combination.

The results of Section 5 show that the second situation describes the WTW studied here. At order two, more than nine-tenths of the feedback weight is already cross-layer. At order three the cross-layer share is about 99%, and at higher orders it approaches the whole trace. This empirical result is stronger than the descriptive statement that international trade contains many commodities. It says that reciprocal dependence is composed through product heterogeneity almost from the shortest possible feedback order.

The dynamic result is equally important. For any two dates,

$$\delta = \log \frac{D_1}{D_0} = \delta^{\text{within}} + \delta^{\text{multi}},$$

where

$$\delta^{\text{within}} = \log \frac{D_1^{\circ}}{D_0^{\circ}}$$

captures changes that occur inside individual product layers and

$$\delta^{\text{multi}} = (\log D_1 - \log D_1^{\circ}) - (\log D_0 - \log D_0^{\circ})$$

captures changes in cross-product closure. Over 1995–2023, the monolayer component is slightly negative while the multilayer component is positive and larger than the observed net increase in $\log D$. The long-run change in the aggregate structure is therefore not a weighted average of changes in separate commodity networks. It is primarily a change in the way commodity networks combine.

This is where EDT differs most sharply from a simple layer-by-layer comparison. Barigozzi et al. (2010) compare topological properties across commodity networks and use cross-commodity correlations to study their relationships. Such an approach asks whether layers have similar connectivity or weight patterns. EDT asks a different question: what new feedback appears *because the layers are aggregated*? Similarity and interaction are not the same. Two layers can have low correlation yet combine to close important circuits; two highly similar layers can be structurally redundant. The same contrast applies to the recent multilayer analysis of BACI by Li et al. (2026), which covers 1995–2024 at the product, chapter and section levels. Layers are compared through row-normalized random-walk transition profiles and a Jensen–Shannon divergence, product communities are obtained by spectral clustering, and country influence is assessed by PageRank perturbation. The objects are similarity between layers and positional influence; the closure of feedback across layers, which drives the results of Section 5, does not enter that framework.

The pairwise interaction $\Gamma(g, h)$ makes this distinction explicit. A negative value means that separate removal effects overlap; a positive value means that joint removal increases diffusion by more than the sum of separate removals. These are interactions of the removal capacity, not correlations between matrices or technological relationships between goods. In particular, two layers necessary for the same circuit can have a negative interaction because either removal breaks it (Lebert & Poindron, 2025).

The advantage of this specialization is interpretability. The disadvantage is that the model does not exploit all possibilities available in general multilayer network theory. It has no independently parameterized interlayer coupling, no layer-switching cost, and no supra-network dynamics other than those implied by aggregation of product-specific arcs. These are not omissions relative to the question asked in this article; they are restrictions that make the structural theorem possible. A more general multilayer model would answer broader questions but would require additional behavioral or structural assumptions.

### 7.7 Communities, regionalization, and the theorem of the partition

Community detection provides another case in which superficially similar operations have different purposes. In network science, a community is typically a set of nodes with denser or stronger internal connections than would be expected under some reference model. Methods based on edge betweenness, modularity maximization, spectral properties, or heuristics such as Louvain seek a partition that is revealed by the observed graph itself (Newman & Girvan, 2004; Newman, 2006; Blondel et al., 2008; Fortunato, 2010). The number, size, and composition of groups are outputs of the method.

The theorem of the partition in EDT begins from the opposite direction. A partition

$$\mathcal{P} = (N_1, \dots, N_P)$$

is supplied for substantive reasons. It may correspond to continents, trade agreements, institutional blocs, industries, or any other economically meaningful grouping. The question is not whether this is the “best” partition of the graph. The question is what part of circular interdependence disappears when links between those predefined groups are cut.

The window-induced matrix

$$\Phi^{\circ} = \mathrm{diag}(\Phi_{11}, \ldots, \Phi_{PP})$$

retains only within-group links. The theorem gives

$$D^{\circ} = \prod_{r} D_r \geq D.$$

The difference therefore has a direct structural interpretation: cross-boundary arcs create or reinforce feedback that lowers diffusion relative to the separated windows. In log-determinant form,

$$\log D - \sum_{r} \log D_r$$

is exactly the contribution of closed walks that cross at least one boundary of the partition.

This operation is conceptually different from modularity. Modularity asks whether observed within-group connectivity is unexpectedly dense relative to a null model and selects a partition to maximize that contrast. EDT asks how much determinant-generating feedback depends on connections between groups. A partition can have low modularity yet a large EDT cross-boundary effect if relatively few intergroup links close high-weight feedback loops. Conversely, a visually strong community partition can have a modest determinant effect if the intergroup links that are removed do not participate strongly in closed walks.

The distinction also explains why the geographical analysis in Section 6 deliberately used an exogenous partition. The aim was not to discover trade communities and then interpret them as continents. Countries were assigned to fixed broad geographical regions before the structural result was examined. This makes the counterfactual transparent: what happens to diffusion if all inter-regional connections are suppressed? The answer is a property of that economically chosen boundary system.

The empirical results show why this is useful. At order two, reciprocal feedback is predominantly within broad regions: only roughly one-sixth to one-fifth of the order-two trace crosses regional boundaries. As feedback order increases, however, the cross-regional share rises strongly and eventually dominates. Long closed walks almost inevitably leave one macro-region before returning. The result would be difficult to summarize by a single statement that the WTW is either "regionalized" or "globalized." It is regionalized at short feedback distances and transregional at long feedback distances.

This order dependence is exactly the kind of result for which a partition theorem is useful. A modularity score would summarize the strength of a partition, but it would not by itself say how regional boundaries enter feedback at successive walk lengths. EDT can write, for every $k$,

$$\mathrm{Tr}(\Phi^k) = \sum_r \mathrm{Tr}\,(\Phi_{rr}^k) + T_k^{\mathrm{cross}},$$

where the second term contains closed walks that cross at least one regional boundary. Geography can therefore be opened order by order without redefining the structural object.

The 2017 anomaly provides an even sharper illustration. In product space, the episode is overwhelmingly monolayer because it is concentrated in HS89 and, more specifically, motorboats. In geographical space, the same episode is overwhelmingly cross-regional because the normalized reciprocal flows responsible for it connect territories in different macro-regions. There is no contradiction. Product layers partition arcs by commodity; geographical windows partition nodes by location. A shock can therefore be simple along one dimension and complex along another. General community terminology would not capture this orthogonality as directly.

Node aggregation is a separate operation: countries within each group are replaced by a single pole, changing the paths that can be concatenated. Section 2.7 specifies a common-scale construction and shows why its determinant has no general ordering relative to the original one. The geographical results use the window operation, which retains country identities and original margins. They therefore measure feedback crossing regional boundaries without attributing new circuits to merged regional nodes.

None of this implies that EDT should replace community detection. If the research question is "what are the endogenous trading blocs in this network?", a community algorithm is the appropriate starting point. If the question is "how much of global circular dependence crosses the boundaries of a specified economic or geographical partition?", the EDT theorem addresses the question directly. The difference lies in finality, not in a contest over partitions.

### 7.8 Structural change without defining a distance between graphs

The temporal extension developed in this article is perhaps the point at which the distinctiveness of the EDT approach becomes most visible. Dynamic network analysis often proceeds by calculating a set of network statistics at successive dates and examining their evolution. Fagiolo et al. (2009, 2010), for example, study the coevolution and stability of weighted WTW properties. Another general strategy is to define a distance or similarity between two adjacency matrices or graphs and use that distance as an index of structural change. Both approaches are legitimate and can be very informative.

We study a particular dimension of temporal change. No distance between $\Phi_t$ and $\Phi_{t-1}$ is defined; instead, the signed change in the EDT diffusion index is

$$\Delta D_t = D_t - D_{t-1}.$$

The log form then opens this change through the same feedback decomposition used in the static theory:

$$\log D_t - \log D_{t-1} = -\sum_{k=2}^{\infty} \frac{\mathrm{Tr}\left(\Phi_t^k\right) - \mathrm{Tr}\left(\Phi_{t-1}^k\right)}{k}.$$

This identity allocates the net log-determinant change across feedback orders, including contributions of opposite signs. Exactness concerns this scalar decomposition. It does not imply that all changes in the network have been identified: different architectures can have identical determinants, and even identical trace sequences. The decomposition is an accounting of measured feedback change, not a causal explanation of its origin.

The same principle extends to products. Since

$$\log D_t = \log D_t^{\diamond} + (\log D_t - \log D_t^{\diamond}),$$

temporal change separates exactly into within-layer and cross-layer components. It extends again to geography through

$$\log D_t = \sum_r \log D_{r,t} + \left(\log D_t - \sum_r \log D_{r,t}\right).$$

A single observed change can therefore be opened along several theoretically defined dimensions without changing the dependent structural object. This is the main methodological advantage of treating structural change as a change in EDT rather than as an exogenous graph distance.

The empirical findings illustrate the gain. The rise of $D$ between 1995 and 2006 is not simply a period in which one network moved "far" from another. It is a period in which circular interdependence weakened, with a broad distribution across feedback orders and a dominant cross-product component. The reversal from 2006 to 2013 is again primarily cross-product, but the sign changes. The 2013–2016 movement is more concentrated in shorter feedback orders. From 2018 to 2023, within-region and cross-region changes partly offset each other even though the net movement in $D$ is modest. These are qualitatively different transformations that could have similar scalar distances under an unrelated graph metric.

The approach also separates structural change from pure trade growth. Because the normalization is invariant to multiplying all flows in a year by the same positive scalar, a common nominal expansion does not change $\Phi$. The method therefore asks whether *relative directed dependence* has changed, not whether world trade has become larger. This is particularly important in a 1995–2023 application spanning inflation, commodity-price cycles, and a large expansion of gross trade values.

Scale invariance does not, however, imply invariance to all price effects. Product-specific relative prices, bilateral price changes, and changes in the composition of nominal trade can alter $\Phi$. The EDT time series is therefore not a pure quantity index. It is a structural index built from observed trade values. This limitation should be kept in mind when interpreting energy-price shocks or exceptional commodity movements. The HS89 episode shows a related point: normalized coefficients can become structurally large even when the corresponding raw trade share is modest, especially for small economies or territories whose denominators are small.

Another advantage of the internal temporal decomposition is comparability when the node set changes. Adding an inactive node contributes an identity block to $I - \Phi$ and leaves the determinant and traces unchanged. Entry and exit therefore do not require an artificial balanced panel of active countries. This is a small technical property with a large empirical benefit: temporal change is not contaminated by mechanically changing matrix dimension when new reporting entities become active.

The temporal formulation is nevertheless descriptive rather than causal. If $D$ falls after a financial crisis, a trade war, a pandemic, or a commodity shock, the EDT decomposition identifies how the feedback architecture changed; it does not by itself establish that the event caused the change. Causal identification would require an additional research design. The strength of EDT is to provide a precise structural outcome that such a design could seek to explain.

In this article, structural change refers specifically to changes in normalized circular interdependence and its decomposed components. A zero aggregate change can conceal substantial reorganization. Explaining whether the identified movements arise from technology, policy, prices, comparative advantage, geopolitical shocks, or firm organization requires an additional research design.

### 7.9 What EDT does not do: scope conditions and limitations

A useful comparison must state not only what EDT reveals but also what it leaves unresolved. The framework is deliberately parsimonious. Its coherence comes from a small number of strong assumptions, and those assumptions delimit the claims that can be made from the results.

The first limitation is linearity. Propagation is represented through repeated multiplication by a fixed matrix $\Phi_t$ within each annual observation. The effect of a coefficient does not depend on the amount of influence already propagated, and the coefficients do not adjust endogenously as a walk unfolds. This is the same broad mathematical advantage and limitation that accompanies linear input-output systems and many network diffusion models. Nonlinear substitution, capacity constraints, congestion, strategic response, and price adjustment are outside the static EDT mechanism used here.

Second, the normalization is substantive rather than technical, but its components should be separated. Section 2.1 characterizes the maximum of exports and imports as the minimum common denominator under the substantive requirement that the country-level margin be independent of whether trade is represented in the export-oriented direction or in its import-oriented dual. Its algebraic relation to Balassa-type imbalance measures and to the corresponding Grubel–Lloyd-form complement locates this margin within a familiar family of trade statistics; that identity does not, by itself, select the maximum. The selection is conditional on orientation independence, boundedness on both sides of the country's external trade position, and minimality of the margin. Larger symmetric margins would also satisfy boundedness.

This does not make the choice innocuous. In the export-oriented representation used here, every balanced or surplus country has a row sum of exactly one, irrespective of its size, whereas a deficit country has a row sum equal to its export-import ratio and therefore strictly below one. A surplus economy exporting one trillion dollars and a balanced economy exporting one billion dollars can therefore emit the same total normalized mass. Country size enters only through the distribution of bilateral shares and through the coefficients that other countries direct toward it. The model measures relative structural exposure, not economic mass.

The common margin is symmetric in exports and imports; the leakage generated by the present model is not. The asymmetry appears only after choosing the export-oriented Ghosh representation. Balanced and surplus countries then have unit row sums, while deficit countries transmit only the fraction represented by their export-import ratio. This use also reverses the usual purpose of aggregate-imbalance adjustments in the intra-industry-trade literature: rather than purging the imbalance to isolate within-industry overlap, EDT retains it as structural information in the propagation mechanism. Sub-stochasticity is therefore a consequence of the selected margin together with the propagation orientation, not an ex post justification for an otherwise arbitrary denominator. Other margins or orientations would encode different theories of relative exposure.

The consequence, however, is that EDT structural importance is not an economic weight. A small economy whose reciprocal flows are large relative to its own external trade can receive large coefficients and can dominate the determinant. Section 4.5 documents exactly this mechanism for the 2017 HS89 episode. This is not a numerical error; it is a consequence of measuring structural exposure in relative rather than absolute terms.

Two responses follow, and they should be kept distinct. The first is diagnostic. Because the effect is mechanical, it is detectable, and layer suppression, trace decomposition and bilateral inspection can locate concentrated configurations without discarding them arbitrarily. The second is substantive robustness. Alternative trade margins would encode alternative theories of exposure; comparing them would be meaningful as a robustness exercise, but it would compare different weight-matrix specifications rather than provide a parameter-free correction to the present one.

Third, the empirical application uses gross merchandise trade. BACI provides an unusually rich bilateral-product dataset, but services are absent and domestic production relations are not observed. The resulting EDT structure is therefore a World Trade Web of internationally traded goods, not a complete model of the world economy. Domestic supply chains, service inputs, and non-traded activities can matter greatly for economic dependence while remaining invisible in the present matrices.

Fourth, gross trade values do not identify value-added origin. As discussed above, EDT should not be read as a GVC accounting framework. A product may cross borders several times, and BACI records the gross value of each recorded trade flow. The determinant summarizes structural feedback among those flows; it does not decompose the value embodied in them. A value-added EDT application would require multi-regional input-output data and a new decision about how value-added flows should be normalized.

Fifth, the framework is sensitive to relative coefficients, which is both a feature and a source of vulnerability. The 2017 HS89 episode demonstrates that a relatively small amount of raw trade can acquire a large structural weight when reciprocal flows are large relative to the external trade scale of the countries or territories involved. This is not a numerical error: it follows from the normalization. But it means that structural outliers must be diagnosed rather than automatically interpreted as global macroeconomic turning points. The suppression tools used in Sections 4 and 5 are therefore not optional embellishments. They are part of responsible EDT interpretation.

Sixth, the determinant is highly compressed. Distinct normalized matrices can have the same $D_t$, so a constant index does not establish structural stability. Even the complete trace sequence determines the spectrum rather than a unique network of labeled countries. Traces, minors, partitions and layers reveal complementary aspects of the feedback architecture; they do not make the scalar mapping injective. The empirical analysis therefore treats $D$ as an entry point into circular interdependence, not a complete descriptor of every trade relation.

Seventh, EDT does not provide statistical inference by itself. Given the observed matrix and the chosen normalization, $D$, the traces, and the removal effects are deterministic. Sampling variability, measurement error in trade values, uncertainty in mirror-data reconciliation, and sensitivity to classification are not represented by confidence intervals in the current analysis. Statistical uncertainty could be introduced through bootstrap, measurement-error models, or perturbation analysis, but those would be additions to the structural theory.

Eighth, structural suppression is not a behavioral counterfactual. When a layer is removed to compute $D_{-g}$, the remaining trade coefficients are not re-equilibrated. Countries do not substitute other products, reroute trade, alter prices, or enter new partnerships. The exercise answers a marginal structural question: how much of the observed feedback architecture is attributable to the presence of this layer? Policy simulation would require a model of adjustment. The same caution applies to node or arc removal.

Ninth, EDT does not discover causal direction from observed trade direction. An arc $i \to j$ records exports from $i$ to $j$ and is treated as a directed influence coefficient according to the theory. This orientation is substantively motivated, but it is not an econometric estimate showing that changes in $i$ cause changes in $j$. "Influence" in EDT is a structural propagation concept, not a causal effect in the potential-outcomes or structural-econometric sense.

Tenth, the current theory is most transparent for non-negative flow matrices. The monotonicity results used for suppression and partition rely on positivity. Signed relations, antagonistic effects, or flows with negative coefficients would require an extension of the theory and can alter determinant and convergence properties. The restriction is natural for merchandise trade values but matters when considering applications to financial positions, opinion influence, or technological relations in which signs may be meaningful.

Finally, EDT is not designed to answer every network question. If the objective is to identify shortest transport routes, optimize community modularity, infer productive capabilities, estimate value-added content, model endogenous network formation, or measure welfare, a specialized method is preferable. The framework is useful when the substantive question concerns propagation and circular dependence in a directed weighted system and when structural interventions can be meaningfully expressed as modifications of that system.

These limitations are not peripheral caveats. They define the research program implied by the paper. Several extensions follow naturally: value-added rather than gross-flow matrices; explicit services; alternative normalizations representing different theories of trade dependence; statistical perturbation of BACI flows; and causal designs in which EDT quantities become outcomes to be explained. What should be preserved across such extensions is not any particular empirical coefficient but the internal relation between the structural matrix, propagation, feedback, and structural modification.

### 7.10 An alternative theory of structural dependence, not an alternative toolbox

The comparisons above can now be summarized without reducing them to a league table of methods. The central lesson is that established approaches to international trade often use the word *structure* for different mathematical objects.

In WTW network analysis, structure is primarily the pattern of relations among countries and the distribution of topological or weighted properties over that graph. The principal questions concern connectivity, clustering, assortativity, centrality, reciprocity, communities, core-periphery organization, and the extent to which these properties are reproduced by suitable null models. This literature establishes that bilateral relations are embedded in a larger network and that binary and weighted descriptions contain different information.

In economic complexity, structure is the pattern linking countries to products and the latent productive capabilities inferred from diversification and ubiquity. The central questions concern productive knowledge, sophistication, diversification opportunities, and feasible movement through product space. Bilateral destination is secondary to the composition of the export basket.

In input-output and GVC approaches, structure is a system of production and accounting relations. The central questions concern requirements, multipliers, linkages, production sharing, value-added origin, and position relative to final demand. The inverse matrix is used to trace how production or value propagates through a system of intermediate transactions.

In general multilayer network theory, structure is a system of several relation types organized across layers, potentially with explicit interlayer coupling. The central questions concern how familiar network properties and processes change when layer identity is preserved rather than collapsed.

EDT takes a different but overlapping position. Its primitive object is a normalized directed flow matrix. Its mechanism is repeated propagation. Its central global property is closure into feedback. Its structural function is the determinant of $I - \Phi$, and its local, multilayer, partition, and temporal quantities are derived from modifications or decompositions of that same function. The natural question is therefore: *how is reciprocal dependence organized, and how does that organization change when parts of the structure are removed, separated, aggregated, or transformed through time?*

This formulation makes it possible to state more precisely what the present article contributes beyond the static multilayer model of Lebert and Poindron (2025). The first contribution is temporal. The equality

$$\log\frac{D_t}{D_{t-1}} = -\sum_{k=2}^{\infty}\frac{\mathrm{Tr}\left(\Phi_t^k\right) - \mathrm{Tr}\left(\Phi_{t-1}^k\right)}{k}$$

turns structural change into an exact change in feedback by order. The second contribution is dynamically multilayer. The identity

$$\delta = \delta^{\mathrm{within}} + \delta^{\mathrm{multi}}$$

separates changes occurring inside product networks from changes created by their combination. The third contribution is geographical. Applying the theorem of the partition through time separates changes internal to broad regions from changes in cross-regional feedback. These are not three unrelated indices. They are three decompositions of the same EDT structural function.

The empirical results make the distinction substantive. The WTW does not display a monotonic trajectory of circular interdependence over 1995–2023. Its main phases are generated overwhelmingly by changes in cross-product feedback rather than by changes inside isolated commodity networks. Yet the same network is more geographically localized at the shortest feedback order than at long orders: product heterogeneity enters closure earlier than transregional geography. The 2017 episode further shows that structural complexity is multidimensional, since the same anomaly is monolayer in product space but cross-regional in geographical space.

These results support a formulation that is deliberately narrower than the claim that EDT is "better" than established WTW methods:

> Different methods do not merely measure the same structure differently;
> they often define different structural objects.

Within EDT, that object can be summarized as follows:

> Structure is the architecture through which directed influences
> propagate, combine, and close into feedback.

This definition explains both the strengths and the boundaries of the approach. It explains why the determinant matters, why circuit and trace representations are central, why suppression and partition have theorem-based effects, why common normalization is essential in the multilayer model, and why temporal change can be decomposed without an external graph distance. It also explains why EDT does not directly measure capabilities, value-added content, community optimality, or welfare.

The most useful way to position EDT in the WTW literature is therefore as an additional *theory of structural dependence*, not as an alternative toolbox intended to replace network analysis, economic complexity, or global value-chain accounting. Those approaches can answer questions that EDT does not address. EDT contributes when the object of interest is the recurrent architecture of a directed flow system and when the researcher wants global structure, local contribution, layer interaction, geographical partition, and temporal change to remain mathematically connected.

The broader implication is methodological. International trade is sufficiently rich that no single representation can exhaust its structure. The value of a formalism lies partly in being explicit about what is preserved and what is discarded when raw flows are transformed into analytical objects. EDT preserves direction, bilateral destination, product identity, and the possibility of repeated propagation; it deliberately compresses other information. Its contribution is to show that, under those choices, a substantial part of the long-run transformation of world trade is not visible in isolated bilateral flows or isolated product layers. It resides in the changing architecture through which they close upon one another.

## 8. Conclusion

This article asked a deliberately structural question: how did the architecture of dependence in the World Trade Web change between 1995 and 2023 when dependence is defined not by trade volume, link density, or a collection of network descriptors, but by the propagation and closure of directed influence? The Economic Dominance Theory provides a precise answer to that question. Once bilateral trade flows are transformed into a normalized matrix of coefficients, indirect influence is generated by repeated matrix multiplication, circular interdependence is encoded by the determinant of the associated Ghosh matrix, and the determinant can be opened exactly through circuits, traces, suppression operations, partitions, and product layers.

The first conclusion is methodological. Changes in circular interdependence can be studied without defining a distance between graphs. The determinant summarizes this particular structural dimension at each date, and its temporal difference can be decomposed internally through the identity

$$\log D_{t_1} - \log D_{t_0} = -\sum_{k=2}^{\infty} \frac{\mathrm{Tr}\left(\Phi_{t_1}^{k}\right) - \mathrm{Tr}\left(\Phi_{t_0}^{k}\right)}{k}$$

which expresses an observed change in diffusion as an exact change in closed feedback by order. The multilayer and partition results extend the same logic. Product layers can be kept separate or aggregated, and geographical blocks can be isolated or reconnected, without changing the underlying structural criterion. The measured change in circular interdependence can therefore be decomposed by feedback length, by whether feedback is confined to one product or combines several products, and by whether it remains inside a geographical region or crosses regional boundaries. These are not three unrelated indicators. They are three openings of the same determinant-based object.

The second conclusion concerns the historical trajectory of the WTW. Trade globalization, understood as the long-run expansion of gross international exchange, is not equivalent to a monotonic increase in circular structural interdependence. The EDT determinant rises strongly from 1995 to 2006, falls from 2006 to 2013, rises again through 2016, and then moves within a comparatively narrower range after 2018, with visible movements in 2022 and 2023. The long-run determinant is higher in 2023 than in 1995. In the terminology of EDT, the world merchandise-trade system therefore exhibits greater diffusion and weaker circular interdependence at the end of the sample than at the beginning. But the important point is not the endpoint alone. It is the sequence of reversals through which that endpoint is reached.

This result illustrates the difference between the size of a system and its organization. A common proportional expansion of all flows leaves the normalized matrix unchanged. A historically important contraction of trade volumes can therefore have a modest effect on the EDT structure, while a comparatively small set of flows can have a large effect if it closes strong recurrent chains. The relatively muted structural signature of 2020 and the exceptional 2017 HS89 episode illustrate the two sides of this distinction. The former shows that a large volume shock need not imply an equally large change in normalized feedback architecture. The latter shows that structurally decisive feedback can arise from a narrow configuration with a small share of gross trade.

The third conclusion is that the structural transformation of world trade is not primarily a parallel transformation of independent commodity networks. It is a transformation in the way product networks combine. The separated-layer determinant remains close to one relative to the much lower determinant of aggregate trade, showing that much of the circularity of the WTW is created only when successive arcs of a closed walk are allowed to carry different products. This is already true at the shortest possible feedback length. More than 90% of the order-two trace is generated by genuinely cross-product terms; by order three the share is approximately 99%, and at higher orders it is virtually exhaustive.

The dynamic decomposition makes the result even stronger. The main phases identified in the aggregate series are overwhelmingly multilayer. From 1995 to 2006, almost the entire increase in the log-determinant is produced by weaker cross-product feedback. The 2006–2013 reversal is again predominantly cross-product. The same conclusion holds for the later phases. Over the full 1995–2023 interval, the within-layer component moves slightly toward *stronger* circularity. The aggregate network nevertheless becomes less circular because the multilayer component moves much further in the opposite direction. In other words, the long-run structural change would be misread if one studied the commodity layers independently and then inferred the aggregate history from them. The decisive object is the architecture created by their combination.

The composition of that architecture is itself changing. The persistence of machinery, electrical machinery, and vehicles among the most structuring layers shows that the WTW does not simply move from manufacturing toward raw materials. Yet mineral fuels acquire a much larger marginal role, while pharmaceuticals, precious stones and metals, and ores also gain. Several paper, wood, textile, and material categories lose structural weight. The pairwise interaction results add another layer of information. Energy becomes more structurally important individually even as its formerly strong positive interactions with major manufacturing layers weaken substantially. Separate removal effects inside the manufacturing complex and between organic chemicals and pharmaceuticals also overlap more in several cases. Structural importance and structural interaction are therefore distinct dimensions: a layer can become more consequential to global closure while changing the set of other layers with which it jointly performs that role.

The fourth conclusion is geographical. The WTW is neither a juxtaposition of continental systems nor a uniformly global structure. Both regional and transregional feedback matter. Outside the isolated 2017 episode, approximately 35–42% of the absolute aggregate log-determinant is associated with closed walks that cross at least one continental boundary. Over the full period, about 41% of the net increase in the log-determinant comes from the weakening of cross-regional feedback and about 59% from changes inside the broad regions. The balance changes substantially across subperiods. The two main movements before 2013 are predominantly cross-regional, whereas the 2013–2016 increase in diffusion is mainly internal to regions. From 2018 to 2023, the two components move in opposite directions: regional feedback weakens while cross-regional feedback strengthens.

Geography also depends strongly on feedback length. Short reciprocal feedback is mainly regional. Only about 15–20% of order-two closed-walk mass crosses continental boundaries. The cross-regional share rises quickly with $k$, passes one half at relatively low orders, and becomes overwhelming for long recurrent chains. Product and geography therefore enter structural globalization at different scales. Cross-product closure is already the rule for elementary two-step feedback; transregional closure becomes progressively dominant as chains lengthen. The formulation used in Section 6 – the WTW is multilayer before it is transregional – summarizes this difference compactly.

The 2017 episode shows why these decompositions are complementary rather than competing. The anomaly is concentrated within a single product chapter and ultimately within one HS6 category, which makes it essentially monolayer in product space. Yet the feedback is created by relations connecting Seychelles to Caribbean territories, which makes the same mechanism overwhelmingly cross-regional in geographical space. Product and geography identify different coordinates of one structural event. More generally, the capacity to move from a global determinant to feedback order, product layer, geographical block, and finally bilateral flow without changing analytical framework is one of the practical strengths of EDT.

The broader implication is that methods applied to the WTW should not be compared as if they were alternative estimators of one latent object called “structure.” Fagiolo and related weighted-network approaches describe topology, weighted organization, reciprocity, clustering, centrality, and other properties of the trade graph. Economic complexity infers productive capabilities and diversification opportunities from country-product patterns. Input-output and GVC approaches reconstruct production requirements, value-added propagation, and positions along production chains. General multilayer theory studies the consequences of preserving layer identity, while community methods seek partitions under explicit optimization criteria. EDT asks another question: how is recurrent dependence organized in a directed flow system, and how does that organization change when arcs, nodes, layers, or blocks are removed, separated, or recombined?

For that reason, the contribution of EDT is not another toolbox of unrelated network indicators. Its advantage is internal consistency. The normalization specifies a theory of influence. The Ghosh inverse represents propagation. The determinant represents system-level diffusion and circular interdependence. Minors and suppression operations identify local structural contributions. The trace expansion orders feedback by length. The multilayer algebra identifies cross-product closure and layer interactions. The partition theorem separates feedback by economically meaningful boundaries. The temporal extension makes changes in all of these objects comparable through time. The framework is restrictive, but the restrictions are what make the different levels of analysis mathematically connected.

Those restrictions also delimit the conclusions. The model is linear within each annual observation. The normalization is substantive rather than neutral. BACI records gross merchandise trade and excludes services and domestic production links. Gross flows do not identify value-added origin. The suppression of an observed layer or arc is a structural extraction, not a behavioral counterfactual in which agents substitute, reroute, or re-equilibrate. The determinant is deterministic given the observed matrix and does not by itself provide statistical uncertainty or causal identification. Relative prices and measurement peculiarities can alter normalized coefficients, and small economies can receive high structural weight when reciprocal flows are large relative to their external trade scale. These are not minor caveats; they define the scope within which the results should be read.

They also define a research agenda. One extension would apply the same formalism to multi-regional input-output data and value-added flows, allowing EDT feedback to be compared directly with GVC accounting. Another would incorporate services. Alternative normalizations could represent different theories of international dependence and provide a substantive robustness analysis rather than a purely numerical one. Statistical perturbation could assess sensitivity to measurement uncertainty in bilateral trade values. Causal research designs could use determinant-based quantities as outcomes to be explained by policy shocks, technological change, sanctions, or geopolitical fragmentation. The partition theorem could be applied to institutional or trade-bloc boundaries, while the multilayer framework could be extended beyond products to technological, financial, or knowledge-flow systems when the underlying coefficients remain compatible with the theory.

The central empirical lesson nevertheless stands without those extensions. Between 1995 and 2023, the structural history of world trade cannot be reduced to the growth of flows, to the evolution of separate commodity networks, or to a simple opposition between regionalization and globalization. The dominant long-run transformation lies in the changing combinations through which product-specific, directed bilateral relations close into recurrent feedback. Those combinations weakened overall, strengthened during important intermediate phases, shifted toward different product layers and interactions, and changed their geographical reach with feedback length. What changed was not only who traded with whom, or how much, but the architecture through which those relations returned upon one another.

That is the specific object EDT adds to the analysis of the World Trade Web: a theory connecting directed propagation to circular feedback, with an exact decomposition of changes in the circular component it measures.

## Annex A. A worked example: five countries, three product layers, two dates

This annex makes the main identities directly verifiable before they are applied to 238 countries and 96 layers. Exact circuit enumeration gives the determinant $D$. The trace series provides a separate numerical reconstruction. At order 200 its determinant approximation differs by about $1.5 \times 10^{-9}$; the log-determinant residual is reported in Section A.4. Thus agreement of the displayed decimals is distinguished from exact arithmetic and from convergence at machine precision.

### A.1 The raw data

Five countries, labeled $A$ to $E$, are assigned to two exogenous regions, $N_1 = \{A, B\}$ and $N_2 = \{C, D, E\}$. Three product layers are labeled 1, 2 and 3. Trade values are expressed in arbitrary units; by the scale invariance established in Section 2.1, only their ratios matter.

| Flow | Layer 1 | Layer 2 | Layer 3 |
|---|---|---|---|
| A → B | | | 40 |
| A → C | | 15 | |
| B → A | | 35 | |
| B → D | 10 | | |
| C → A | | 10 | |
| C → D | 30 | | |
| D → B | | | 12 |
| D → E | 20 | | |
| E → A | | | 8 |
| E → C | 25 | | |

**Table A.1.** Raw bilateral flows by product layer at date $t_0$.

Four structural configurations have been built into the example deliberately. The pair $A \rightleftarrows B$ closes as a two-step loop whose outward leg belongs to layer 3 and whose return leg belongs to layer 2: it is within-region and cross-layer. The pair $A \rightleftarrows C$ closes entirely inside layer 2 and crosses the regional boundary: it is cross-region and monolayer. The circuit $C \to D \to E \to C$ closes entirely inside layer 1 and entirely inside region $N_2$: it is within-region and monolayer, of order three. The pair $B \rightleftarrows D$ combines layers 1 and 3 and crosses the regional boundary.

Some circuits require all three layers: the four-step circuit from A through C and D to B and back to A is one example, as Table A.3 shows. The only three-country circuit, by contrast, lies entirely in layer 1. The latter fact explains the purely monolayer order-three trace in Section A.5; it does not extend to longer orders.

### A.2 Normalization and leakage

Aggregate exports, imports and the denominator $q_i = \max(E_i, M_i)$ are as follows.

| Country | $E_i$ | $M_i$ | $q_i$ | **Row sum of $\Phi$** | **Position** |
|---|---|---|---|---|---|
| A | 55 | 53 | 55 | 1.0000 | surplus |
| B | 45 | 52 | 52 | 0.8654 | deficit |
| C | 40 | 40 | 40 | 1.0000 | balanced |
| D | 32 | 40 | 40 | 0.8000 | deficit |
| E | 33 | 20 | 33 | 1.0000 | surplus |

**Table A.2.** Normalization of the toy structure at date $t_0$.

This table is the appropriate place to confront the substantive content of the normalization rather than to state it in passing. Three observations deserve to be made explicitly.

First, $A$, $C$ and $E$ have identical total outgoing influence, with row sums of exactly one, although their gross exports are 55, 40 and 33 respectively. Under this rule the capacity to transmit influence is not a function of size. A large surplus exporter and a small balanced exporter emit the same normalized mass, and size enters the structure only through the pattern of a country's bilateral shares and through the coefficients that others direct toward it.

Second, $B$ and $D$ leak. Their row sums, $45/52$ and $32/40$, are exactly their export-to-import ratios. External deficit is represented as structural leakage, and it is this leakage that delivers $\rho(\Phi) < 1$ in the strongly connected component.

Third, because $q_i$ is a country denominator computed from aggregate trade, a bilateral flow accounting for a large share of its origin country's margin receives a large coefficient. This is the mechanism behind the 2017 HS89 episode of Section 4.5, and it is visible here in miniature: $C \to D$ carries 30 units against a denominator of 40, so that $\phi_{CD} = 0.75$.

The resulting normalized aggregate matrix, with rows indexed by origin, is

$$\Phi = \begin{pmatrix} 0 & 0.727273 & 0.272727 & 0 & 0 \\ 0.673077 & 0 & 0 & 0.192308 & 0 \\ 0.250000 & 0 & 0 & 0.750000 & 0 \\ 0 & 0.300000 & 0 & 0 & 0.500000 \\ 0.242424 & 0 & 0.757576 & 0 & 0 \end{pmatrix}$$

with spectral radius $\rho(\Phi) = 0.9252 < 1$. The Neumann and logarithmic series therefore converge, as required by Section 2.1.

### A.3 The determinant by exact circuit enumeration

The structure admits exactly eight simple directed circuits and exactly two families of pairwise vertex-disjoint circuits. Both are enumerated in full below. This is the computation that cannot be performed on the empirical matrix and that is therefore indispensable here: it is the only place in the article where the alternating sign of the Leibniz expansion can be seen rather than asserted.

| **Circuit** | **Weight $\omega(c)$** | **Layers used** | **Regions** |
|---|---|---|---|
| A → B → A | 0.489510 | 3, 2 | $N_1$ |
| A → C → A | 0.068182 | 2, 2 | $N_1$–$N_2$ |
| B → D → B | 0.057692 | 1, 3 | $N_1$–$N_2$ |
| C → D → E → C | 0.284091 | 1, 1, 1 | $N_2$ |
| A → C → D → B → A | 0.041302 | 2, 1, 3, 2 | $N_1$–$N_2$ |
| A → B → D → E → A | 0.016953 | 3, 1, 1, 3 | $N_1$–$N_2$ |
| A → C → D → E → A | 0.024793 | 2, 1, 1, 3 | $N_1$–$N_2$ |
| A → B → D → E → C → A | 0.013244 | 3, 1, 1, 1, 2 | $N_1$–$N_2$ |
| **Sum** | **0.995768** | | |

**Table A.3.** Simple directed circuits of the toy structure and their weights.

| Family of two vertex-disjoint circuits | Product of weights |
|---|---|
| $\{A \rightleftarrows B\} \cup \{C \to D \to E \to C\}$ | 0.139065 |
| $\{A \rightleftarrows C\} \cup \{B \rightleftarrows D\}$ | 0.003934 |
| **Sum** | **0.142999** |

**Table A.4.** Families of vertex-disjoint circuits.

No family of three pairwise disjoint circuits exists on this support. Writing $\Sigma_1$ for the sum of Table A.3 and $\Sigma_2$ for the sum of Table A.4, the expansion of Section 2.4 gives

$$D = 1 - \Sigma_1 + \Sigma_2 = 1 - 0.995768 + 0.142999 = 0.147231.$$

This computation is the pedagogical center of the annex. Taken alone, $\Sigma_1$ would give $D = 0.004232$, which is wrong by a factor of thirty-five. A reader who has absorbed the informal idea that $1 - D$ is the total weight of the loops of the network is corrected here by an arithmetic that can be checked line by line. The two disjoint families enter with a positive sign because a permutation assigning two disjoint cycles has sign $(-1) \times (-1) = +1$. Circular interdependence is not additive over circuits, and the inclusion–exclusion structure of the determinant is not a technicality.

### A.4 The same determinant by the trace series

The logarithmic identity of Section 2.5 gives $\log D = -\sum_{k \geq 2} \mathrm{Tr}\,(\Phi^k)/k = -1.915755$. Writing $\Lambda_K$ for the partial sum truncated at order $K$, in the notation of Section 4.1, the reconstruction proceeds as follows.

| $K$ | $\Lambda_K$ |
|---|---|
| 2 | -0.615385 |
| 3 | -0.899476 |
| 4 | -1.167940 |
| 6 | -1.383665 |
| 10 | -1.619164 |
| 20 | -1.828928 |
| 50 | -1.911623 |
| 100 | -1.915709 |
| 200 | -1.915755 |

**Table A.5.** Reconstruction of $\log D$ by increasing orders of closed walks.

Two lessons transfer directly to Section 4.1. The first is that the order-two trace is verifiable by hand. Summing $2\phi_{ij}\phi_{ji}$ over unordered pairs gives $2(0.727273)(0.673077) = 0.979021$ for the pair $A, B$, then $2(0.272727)(0.250000) = 0.136364$ for $A, C$ and $2(0.192308)(0.300000) = 0.115385$ for $B, D$, whence $\mathrm{Tr}(\Phi^2) = 1.230769$.

The second lesson concerns truncation. With $\rho(\Phi) = 0.9252$, the absolute log residual is about 0.00004636 at order 100 and 0.00000001028 at order 200. Six displayed decimals agree by order 200, but the series is still truncated and its residual exceeds the empirical residuals reported in Section 4.1. This illustrates slow convergence near a unit spectral radius, not dominance of long economic chains: order twenty already accounts for 95.5% of the final log magnitude.

### A.5 The multilayer decomposition

The layer determinants and the determinant of the block-diagonal structure in which no closed walk may change product are as follows.

| Determinant | Value |
|---|---|
| $D_1$ | 0.715909 |
| $D_2$ | 0.931818 |
| $D_3$ | 1.000000 |
| $D^{\circ} = \prod_g D_g$ | 0.667097 |
| $D$ | 0.147231 |
| $\kappa = D^{\circ} - D$ | 0.519866 |

**Table A.6.** Layer determinants, separated-layer determinant and heterogeneity at $\mathrm{t}_0$.

Layer 3 has determinant exactly equal to one: taken alone it contains no circuit whatever. It is nevertheless structurally consequential, as Section A.6 shows. This single line disposes of the intuition that a layer matters in proportion to its own internal circularity.

In log-determinant space the separation is exact and additive,

$$\log D = \log D^{\circ} + (\log D - \log D^{\circ}), \qquad -1.915755 = -0.404820 + (-1.510935).$$

The order-by-order composition of the closed-walk mass is reported next.

| $k$ | $\mathrm{Tr}(\Phi^k)$ | **Within-layer** | **Multilayer share** |
|---|---|---|---|
| 2 | 1.230769 | 0.136364 | 88.9% |
| 3 | 0.852273 | 0.852273 | 0.0% |
| 4 | 1.073857 | 0.009298 | 99.1% |
| 5 | 0.245020 | 0.000000 | 100.0% |
| 6 | 1.000330 | 0.242757 | 75.7% |

**Table A.7.** Within-layer and genuinely multilayer composition of the trace at $\mathrm{t}_0$.

The order-three row is the point of the table. Here the entire order-three mass is monolayer, because the only three-cycle, $C \to D \to E \to C$, lies wholly inside layer 1. In the empirical WTW of Section 5.3 the corresponding figure is approximately 99% multilayer. The contrast should be read carefully: the empirical dominance of cross-product feedback is a finding about world trade, not a consequence of the algebra. A toy structure in which it fails is the cleanest way to establish that point.

A short exercise makes the same argument from the other direction. Keep every bilateral flow unchanged but reassign the arcs to layers, so that layer 1 contains $A \to B$, $C \to D$, $D \to E$ and $E \to C$, layer 2 contains $B \to A$, $A \to C$ and $C \to A$, and layer 3 contains $B \to D$, $D \to B$ and $E \to A$. The aggregate matrix $\Phi$, and therefore $D = 0.147231$, are unchanged. But $D^{\diamond}$ becomes $0.628611$, $\kappa$ becomes $0.481380$, and all three pairwise interactions turn negative. The product dimension is therefore not recoverable from the aggregate structure. It is genuinely additional information, and it is the attribution of arcs to layers that carries it.

### A.6 Layer suppression and pairwise interaction

The marginal structuring contribution of each layer, $S_g = D_{-g} - D$, is reported below.

| **Layer** | $D_{-g}$ | $S_g$ |
|---|---|---|
| 1 | 0.442308 | 0.295077 |
| 2 | 0.641264 | 0.494033 |
| 3 | 0.647728 | 0.500497 |

**Table A.8.** Marginal structuring contribution of the product layers at $t_0$.

Layer 3, whose own determinant equals one, has the largest marginal contribution of the three. Removing it destroys the outward leg of $A \rightleftarrows B$, the return leg of $B \rightleftarrows D$, and the arc $E \to A$, and therefore breaks the heaviest circuits of the system. The quantity $S_g$ measures participation in the aggregate feedback architecture; it measures neither internal circularity nor trade share. The example also makes the non-additivity immediate: $S_1 + S_2 + S_3 = 1.2896$, which exceeds $1 - D = 0.8528$. The $S_g$ are not shares and decompose nothing.

The pairwise interactions $\Gamma(g,h) = D_{-\{g,h\}} - D_{-g} - D_{-h} + D$ are reported together with the second-order approximation $-\mathrm{Tr}(\Phi_g \Phi_h)$ derived in Section 2.11.

| **Pair** | $\Gamma(g,h)$ | $-\mathrm{Tr}(\Phi_g \Phi_h)$ | **Removal effects** |
|---|---|---|---|
| (1, 2) | 0.063659 | 0.000000 | super-additive |
| (1, 3) | -0.010986 | -0.057692 | overlapping |
| (2, 3) | -0.425852 | -0.489510 | strong overlap |

**Table A.9.** Pairwise structural interactions and their second-order approximation at $t_0$.

The pair $(1,2)$ shares no two-step cross-layer loop, so $\mathrm{Tr}(\Phi_1\Phi_2) = 0$ exactly. The quadratic term therefore gives $\Gamma \approx 0$, but the exact interaction depends on higher orders. Both disjoint families in Table A.4, including $\{A \rightleftarrows B\} \cup \{C \to D \to E \to C\}$, contribute positively, totaling 0.142999. The three simple circuits using all three layers contribute negatively, totaling 0.079340. Their net contribution is 0.063659. The pair $(2,3)$ carries the heaviest bilateral loop: the quadratic term $-0.489510$ is close to the exact value $-0.425852$, although it does not include all contributions. Positive and negative interactions thus reflect different balances of determinant terms.

### A.7 The geographical partition

With $N_1 = \{A, B\}$ and $N_2 = \{C, D, E\}$, the window-induced structure retains only within-region arcs and yields $D_{N_1} = 0.510490$ and $D_{N_2} = 0.715909$, hence

$$D^\circ = D_{N_1} D_{N_2} = 0.365464 \geq D = 0.147231,$$

which verifies the theorem of the partition. The residual is $C^{\mathrm{geo}} = \log D - \log D^\circ = -0.909168$, or 47.5% of the absolute magnitude of $|\log D|$. The cross-regional share varies with order: 20.5% at $k = 2$, 0.0% at $k = 3$, and 55.4% at $k = 4$. This profile is not monotonic, because the only three-country circuit lies inside $N_2$. The broadly increasing empirical profile in Section 6.5 is therefore a result about the WTW, not an implication of the partition identity.

Reading Sections A.5 and A.7 side by side gives the orthogonality result of Section 6.3 in its simplest form. The circuit $C \to D \to E \to C$ is monolayer and intra-regional. The circuit $A \rightleftarrows B$ is multilayer and intra-regional. The circuit $A \rightleftarrows C$ is monolayer and cross-regional. Product layers partition arcs by commodity while geographical windows partition nodes by location, so that a disturbance may be simple along one axis and complex along the other, exactly as in the 2017 episode.

### A.8 A second date and the exact dynamic decomposition

At date $t_1$ three flows change and nothing else: $C \to D$ in layer 1 falls from 30 to 18, $B \to A$ in layer 2 rises from 35 to 50, and $A \to B$ in layer 3 rises from 40 to 52. The monolayer circuit of region $N_2$ is thereby weakened while the cross-layer bilateral loop $A \rightleftarrows B$ is strengthened.

| **Indices** | $t_0$ | $t_1$ |
|---|---|---|
| $\rho(\Phi)$ | 0.9252 | 0.9316 |
| $D$ | 0.147231 | 0.129919 |
| $\log D$ | -1.915755 | -2.040841 |
| $D^{\diamond}$ | 0.667097 | 0.743535 |
| $D^{\circ}$ | 0.365464 | 0.316798 |
| $S_3$ | 0.500497 | 0.601865 |
| $\Gamma(1,2)$ | 0.063659 | 0.072259 |

**Table A.10.** The toy structure at two dates.

The contribution of feedback order $k$ to the change is $-\left[\mathrm{Tr}\left(\Phi_1^k\right) - \mathrm{Tr}\left(\Phi_0^k\right)\right]/k$, which equals $-0.095783$ at $k = 2$, $+0.071023$ at $k = 3$ and $-0.043409$ at $k = 4$. The orders do not move together, and order three moves against the total. The corresponding partial sums of the change, in the notation of Section 4.4, are $\Lambda_2 = -0.095783$, $\Lambda_3 = -0.024760$, $\Lambda_4 = -0.068169$, $\Lambda_{20} = -0.105435$, $\Lambda_{50} = -0.122887$ and $\Lambda_{200} = -0.125086$. The reader can therefore watch the construction of Table 4.2 happen on ten lines of arithmetic.

The exact separation into within-layer and genuinely multilayer components gives

$$\Delta\log D = -0.125086, \qquad \Delta\log D^{\mathrm{within}} = +0.108480, \qquad \Delta\log D^{\mathrm{multi}} = -0.233566,$$

the within-layer term decomposing further as $+0.094588$ for layer 1, $+0.013892$ for layer 2 and 0 for layer 3. The two components have opposite signs, and the within-layer component is of the same order of magnitude as the total. Taken separately, the product networks became less circular; the aggregate system nevertheless became more circular, because cross-product closure strengthened by more than twice as much. This reproduces the offsetting-components mechanism of Section 5.2, with the directions reversed relative to 1995–2023, and it makes concrete why a layer-by-layer analysis followed by recomposition would have reversed the sign of the conclusion.

The geographical separation gives $\Delta \log D^{\text{within regions}} = -0.142904$ and $\Delta \log D^{\text{cross regions}} = +0.017818$. The internal component strengthens circularity while the cross-regional component weakens it. Their opposition reproduces the logic of the 2018–2023 comparison in Section 6.3, although the directions are reversed. Dividing components of opposite signs by their net total yields one share above 100% and another below zero; these are signed contributions, not proportions of disjoint positive quantities.

*A note on the normalization.* Replace $q_i = \max(E_i, M_i)$ by $q_i = E_i$ and recompute the row sums of $\Phi$. Every row now sums to exactly one, the leakage of $B$ and $D$ disappears, and the spectral condition $\rho(\Phi) < 1$ is no longer guaranteed by the argument of Section 2.1: an irreducible row-stochastic matrix has $\rho = 1$, the Neumann series diverges and $D = 0$. This exercise is worth doing before reading Section 7.9 because it separates two roles of the denominator. Section 2.1 first selects the maximum as the minimum common margin that is independent of whether trade is represented in the export-oriented direction or in its import-oriented dual. Choosing the export total alone instead makes the scale orientation-specific: the import-oriented dual would naturally require the import total. Once export-oriented Ghosh propagation is selected, that export-only denominator also makes the matrix row-stochastic; in an irreducible structure the spectral radius is one, so the Neumann series and the log-determinant expansion do not converge. By contrast, the common maximum makes deficit rows sub-stochastic. The orientation-independent measurement requirement and the convergence consequence are therefore distinct, but compatible.

### A.9 Reproduction code

The complete example is reproduced by the following program.

```
import numpy as np, itertools, math

n = 5
flows = {                                   # layer: [(from, to, value)]
    1: [(2,3,30), (3,4,20), (4,2,25), (1,3,10)],
    2: [(1,0,35), (0,2,15), (2,0,10)],
    3: [(0,1,40), (3,1,12), (4,0,8)],
```

```
}

X = {g: np.zeros((n,n)) for g in flows}
for g, lst in flows.items():
    for i, j, v in lst:
        X[g][i, j] += v

Xa  = sum(X.values())
q   = np.maximum(Xa.sum(1), Xa.sum(0))      # q_i = max(E_i, M_i)
Phi = {g: (X[g].T / q).T for g in X}
P   = (Xa.T / q).T

det = lambda M: np.linalg.det(np.eye(M.shape[0]) - M)

D  = det(P)
Dd = np.prod([det(Phi[g]) for g in Phi])
print("D =", D, " D_diamond =", Dd, " kappa =", Dd - D)

tr = [np.trace(np.linalg.matrix_power(P, k)) for k in range(1, 201)]
print("log residual =", abs(math.log(D) + sum(tr[k-1]/k for k in range(2,
201))))

for g in Phi:
    print("S", g, det(P - Phi[g]) - D)
for g, h in itertools.combinations(Phi, 2):
    print("Gamma", g, h,
          det(P - Phi[g] - Phi[h]) - det(P - Phi[g]) - det(P - Phi[h]) + D)

N1, N2 = [0,1], [2,3,4]
Dcirc = det(P[np.ix_(N1,N1)]) * det(P[np.ix_(N2,N2)])
print("D_circ =", Dcirc, " C_geo =", math.log(D) - math.log(Dcirc))
```

To obtain date $t_1$, replace the value 30 by 18 in layer 1, 35 by 50 in layer 2, and 40 by 52 in layer 3.